\documentclass[compress,numbers]{els-mrw} 

\usepackage{amsmath,amssymb,amsfonts,amsthm,makeidx,graphicx}
\usepackage{txfonts}
\usepackage{helvet}
\usepackage{hyperref}

\newcommand{\be}{\begin{equation}}
\newcommand{\ee}{\end{equation}}
\newcommand{\bea}{\begin{eqnarray}}
\newcommand{\eea}{\end{eqnarray}}

\newcommand{\gagamma}{g_{a\gamma}}

\begin{document}


\chapter{Experiments to test the hypothesis for solar and dark matter axions}\label{chap1} 



\author[1]{Babette D\"obrich}%
\author[2]{Igor G. Irastorza}%


\address[1]{\orgname{Max-Planck-Institut f\"ur Physik (Werner-Heisenberg-Institut)}, 
\orgaddress{Boltzmannstr. 8, 85748 Garching bei München, Germany}}
\address[2]{\orgname{Centro de Astropartículas y Física de Altas Energías (CAPA)}, \orgdiv{Universidad de Zaragoza}, \orgaddress{C/ Pedro Cerbuna, 12, 50009, Zaragoza, Spain}}

\articletag{Chapter Article tagline: update of previous edition, reprint.}

\maketitle

\begin{abstract}[Abstract]
 We present a pedagogical introduction to the direct search of axions as dark matter, as well as to searches for solar axions.
 The plethora of experimental searches exploit the axion's coupling to two photons: They attempt to convert the axion dark matter
 to photons in a resonator placed in an external magnetic field or convert solar axions into X-rays in a magnet pointing towards the sun. We give a basic introduction to this concept, its many variants and to searches that exploit the axion's other couplings. We also speculate about potentially transformative developments for such searches in the near-term future. 
\end{abstract}

\begin{keywords}
 	axion\sep dark matter\sep experiment \sep haloscope \sep bsm \sep helioscope
\end{keywords}

\begin{figure}[ht]
	\centering
	\includegraphics[height=5.1cm]{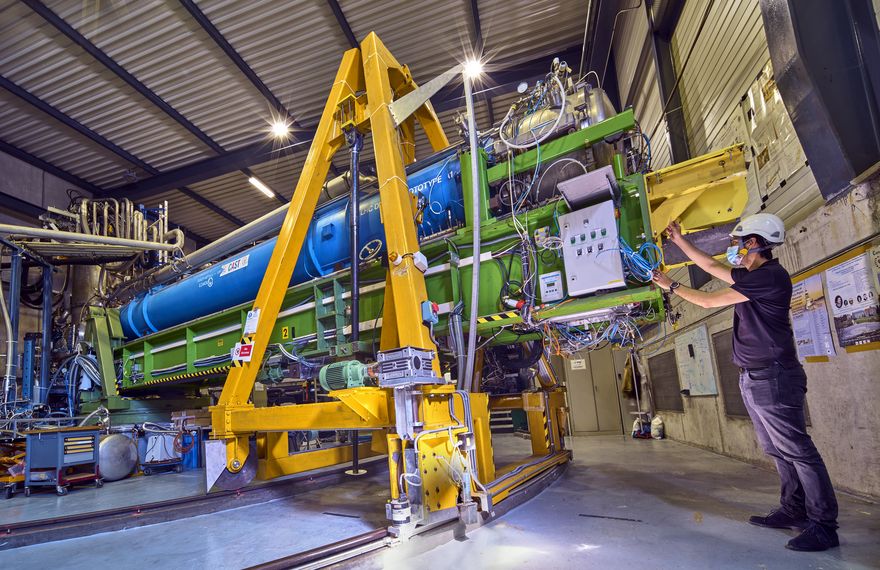}
    \hspace{0.5cm}
    \includegraphics[height=5.1cm]{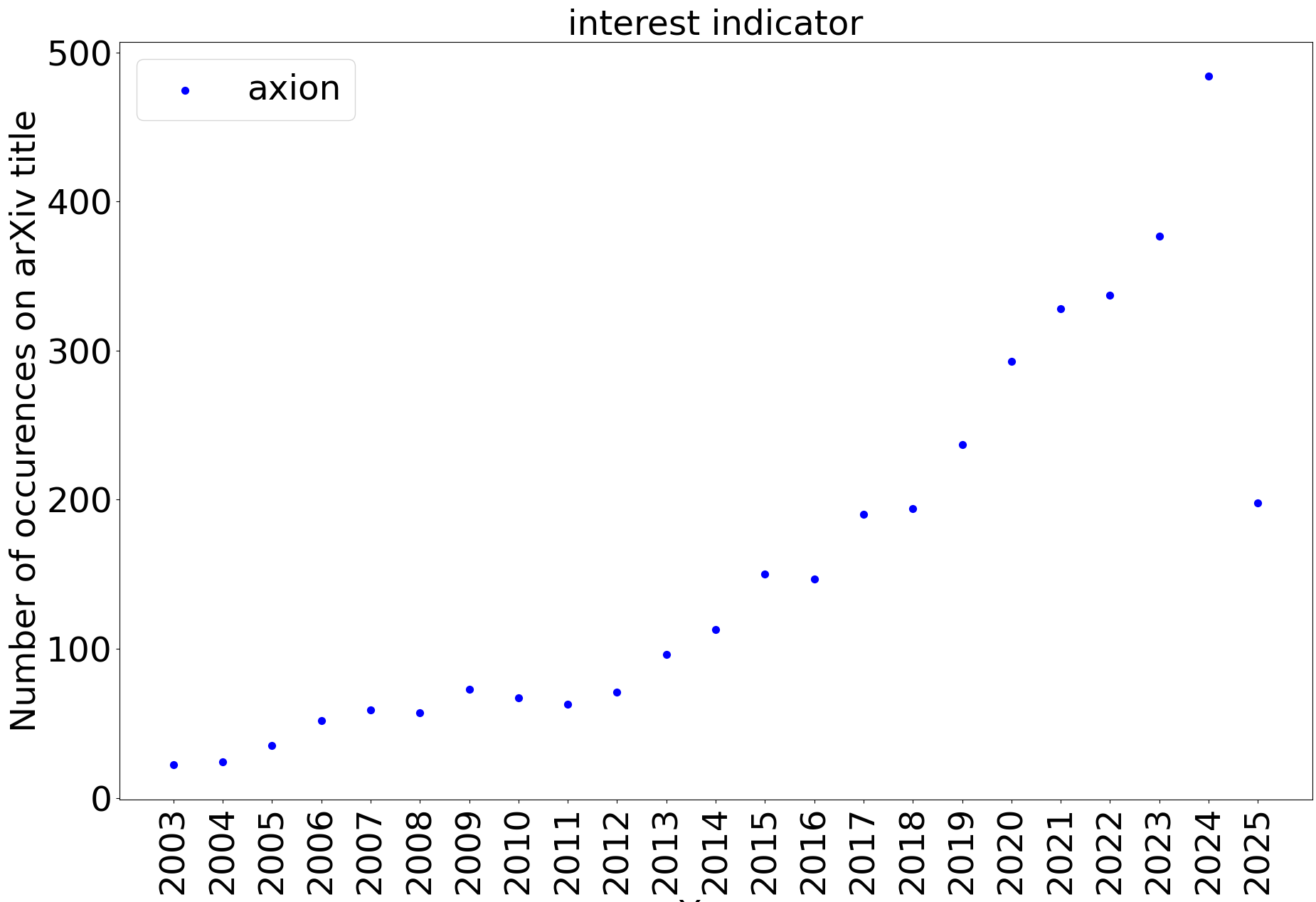}
	\caption{
Left:
 A picture \cite{Maximilien:2767149} showing  a scientist at work at the former CAST experiment at CERN. CAST is the world's so-far most sensitive solar axion search (Helioscope): Over the course of 20 years it searched for solar axions in various configurations by tracking the sun with a LHC test-magnet (blue tube in the picture). At the end of its lifetime, CAST hosted several setups that aimed to search directly for dark matter axions employing a radio-frequency resonator (Haloscope). Thus, this picture  features a Haloscope and Helioscope at once! \newline
 Right:
 History of axions on the arxiv, as an `interest indicator' in axions over the years. The plot counts the number of occurrences of the term `axion' in the title of arxiv preprints over the years (2025 only included up to April)
 }
	\label{fig:titlepage}
\end{figure}

\begin{glossary}[Nomenclature]
	\begin{tabular}{@{}lp{34pc}@{}}
		 SG & data filtering method proposed by A. Savitzky and M. J. E. Golay \\
    KSVZ &  axion model of Kim-Shifman-Vainshtein-Zakharov \\
    DFSZ & axion model of Dine-Fischler-Srednicki-Zhitnitskii \\
    vanilla axion & axion that fits KSVZ or DFSZ models \\
    SM &  Standard Model of particle physics \\
    BSM & Theories that go `beyond' the Standard Model of particle physics \\
    CDM & cold dark matter \\
    RF & Radio-frequency \\
	\end{tabular}
\end{glossary}

\section*{Objectives}
\begin{itemize}
	\item We provide an introduction to axion haloscopes and helioscopes, i.e. searches for axions as dark matter and solar axions. Concepts that were both suggested in Pierre Sikivie's seminal paper of 1983
	\item We describe the generic setup and the basic technology that drives these experiments and provide an idea on how the data in solar and dark matter axion experiments can be searched for a signal
	\item We outline that with helioscopes and haloscopes physics scenarios can be tested that are outside of their main scope 
	\item We attempt a view into the crystal ball to understand what these experiments might have in stock and where young students can gain expertise to enter the field 
\end{itemize}

\section{History of the experiments}\label{history}

This introductory text concerns itself with the experimental search for axions, particularly their possible role in understanding dark matter \cite{Cirelli:2024ssz,Balazs:2024uyj} and the possibility of observing them from the sun. 
To motivate that spending your time on axions is actually worth it, we refer
the reader to introductions on the cosmology of axions \cite{OHare:2024nmr}, the theory of axions \cite{DiLuzio:2020wdo,Sikivie:2024isv,Yu:2023gdq}, as well as a number of other introductory texts to the topic \cite{Irastorza:2021tdu} and its current experimental landscape \cite{Irastorza:2018dyq,Adams:2022pbo,Baryakhtar:2025jwh}. 
In the following we only provide a concise overview of the experiments' history.

Axions are hypothetical particles proposed in the 1970s to solve the ``strong CP problem'' in quantum chromodynamics (QCD), a puzzling feature of the Standard Model of particle physics~\cite{Peccei:1977hh,Peccei:1977ur,Weinberg:1977ma,Wilczek:1977pj}. Uncharted parameter space that allows the axion to solve the strong CP problem, predicts them to be very light-weight and weakly coupled. In addition to their role in particle physics, axions are also considered promising candidates for dark matter, as was pointed out in later papers~\cite{Preskill:1982cy,Abbott:1982af,Dine:1982ah}.

The ideas behind an axion haloscope as well as the axion helioscope were first proposed by Pierre Sikivie in 1983 \cite{Sikivie:1983ip}. Sikivie suggested that axions could convert into photons in the presence of a strong magnetic field, via the ``Primakoff effect''. 
His idea laid the foundation for most modern axion searches:

\vspace{0.1cm}
\begin{itemize}
\item In a haloscope experiment, a resonanting cavity is placed inside a strong magnetic field, which enhances the conversion of axions into detectable microwave photons. This approach targets axions with very low masses around the $\mu$eV scale and larger. The first practical haloscope experiment was built at Brookhaven National Laboratory in the late 1980s \cite{Wuensch:1989sa}, but it was not sensitive enough to detect `vanilla' axions. Significant progress was made in the 1990s with the launch of the Axion Dark Matter eXperiment (ADMX), the most advanced haloscope experiment to date. ADMX started its operations in the late 1990s and has undergone several upgrades to improve its sensitivity. By the 2000s, it had set the first significant constraints on axion masses and coupling strengths in the $\mu$eV range.
\item Helioscopes were conceived to detect axions produced in the Sun's core, where axions can be created via the Primakoff effect. These axions could then be converted back into photons under a strong magnetic magnet pointed at the Sun. After a first attempt at Brookhaven~\cite{Lazarus:1992ry} (just few hours of data with a static magnet), the first movable-magnet axion helioscope, SUMICO, was built in Tokyo in the 1990s~\cite{Moriyama:1998kd,Inoue:2002qy} . This work was succeeded by the development of the CERN Axion Solar Telescope (CAST), which has been in operation in 2003-2021. For this, CAST employed a ``repurposed'' Large Hadron Collider accelerator magnet prototype.
\end{itemize}
\vspace{0.1cm}

The past decade has seen significant advancements in haloscope experiments, driven by improved technology and steeply increasing interest in axion dark matter (cf. Fig. \ref{fig:titlepage}). 
ADMX, with multiple upgrades, became sensitive enough to probe axion models predicted by theories of QCD axions.
New experiments in various places around the globe, described below, have joined the search, each exploring different axion mass ranges and improving on cavity design, cryogenics, and microwave photon detection methods.

For helioscopes, the next-generation instrument will be the International Axion Observatory (IAXO) and its stepping-stone BabyIAXO. 
IAXO will feature a custom-made magnet instead of re-using an accelerator magnet, thereby increasing sensitivity by orders of magnitude compared to CAST. BabyIAXO and IAXO will expand the search to previously inaccessible regions of axion parameter space.

\section{Generic experiments description}\label{generic} 

Axions are expected to be very light particles and therefore signals of their existence are not expected to be seen easily at accelerators. More generic ALPs may evade the constraints on the axion mass and relatively massive ALP models are still viable. In this case they can still be searched for, e.g. at accelerators.
We will briefly comment on this possibility in Section \ref{sec:complementarity}. 
Here, we will refer to detection strategies for \textit{low energy} axions and ALPs, where low energy means $m_a \lesssim 1$~eV. The search for such low energy axions represents a particular experimental field that requires very specific combinations of know-hows, some of them not present in typical HEP groups\footnote{See the complementarity section \ref{sec:complementarity} for exceptions.}, and therefore requiring cross-disciplinary technology transfer. They include, among others, high-field magnets, super-conductivity, RF techniques, X-ray optics  \& astronomy, low background detection, low radioactivity techniques, quantum sensors, atomic physics, ultra-low temperature cryogenics etc. Their effective interplay with axion particle physicists is an important challenge in itself, that will be conveyed in the following sections.

We describe in the following the strategies to \textit{directly} detect axions in haloscopes and helioscopes. Other direct laboratory searches such as light-shining-through walls are discussed elsewhere\footnote{Experiments looking for axions or axion-induced effects entirely in the laboratory are light-shining-through-wall (LSW), laser polarization or new long-range macroscopic force experiments. However, due to the smallness of axion interactions, these experiments often fall short of reaching sensitivity to axions (unless in very non-conventional models), but could detect more generic ALPs.} What is meant here is that axions could be produced in the laboratory itself, or coming from other natural sources (in contrast to indirect detection, or detection of signatures of axions in cosmology or astrophysics like the ones described in~\cite{DiLuzio:2020wdo}). The most relevant natural sources of axions are the Sun and the dark matter halo. It is useful to categorize the different experimental approaches according to the source of axions used:

\begin{itemize}
\item Experiments attempting the detection of the very axions that may constitute our local dark matter galactic halo, often called ``axion haloscopes''~\footnote{The name \textit{axion haloscopes} (as in the case of \textit{axion helioscopes}) was coined by P. Sikivie in his seminal paper~\cite{Sikivie:1983ip} in which the --now widely spread-- magnetized RF-cavity approach was first proposed. Nowadays many variations of this method, or altogether new approaches, are being followed. The name \textit{axion haloscope} is sometimes used extensively for any technique looking for DM axions, and other times restricted to the conventional ``Sikivie haloscopes''.}
\item Experiments searching for axions emitted by the Sun and detected at terrestrial detectors, or ``axion helioscopes''.
\end{itemize}

\noindent
Haloscopes and helioscopes take advantage from the enormous flux of axions expected from extraterrestrial sources. Because of this, they are the only techniques having reached sensitivity down to QCD axion couplings in the low mass range of axion models. Haloscopes rely on the assumption that the 100\% of the dark matter is in the form of axions, and the dark matter density value assumed in searches follows a community convention. In the case of a subdominant axion component within dark matter their sensitivity should be rescaled accordingly. Helioscopes rely on the Sun emitting axions, but in its most conservative channel (being Primakoff conversion of solar plasma photons into axions) this is a relatively robust prediction of most models, relying only on the presence of the $\gagamma$ coupling. 

As will be shown in the following, most (but not all) of the axion detection strategies rely on the axion-photon coupling $\gagamma$. This is due to the fact that this coupling is generically present in most axion models, as well as that coherence effects with the electromagnetic field are easy to exploit to increase experimental sensitivity. Figs.~\ref{fig:helioscopes} and \ref{fig:haloscopes} show the overall panorama of experimental and observational bounds on the $\gagamma$-$m_a$ plane. 

\subsection{Dark matter experiments}
\label{sec:DMexps}

If our galactic dark matter halo is totally made of axions, the number density of these particles around us would be huge. The density of local dark matter is measured to be about $\rho \sim 0.4-0.6$ GeV/cm$^3$~\cite{deSalas:2020hbh}, which means that we would expect an axion number density of the order of:

\begin{equation}
n_a \sim \rho_a / m_a \sim 5 \times 10^{13}~\left(\frac{10\,\mu\mathrm{eV}}{m_a}\right) \mathrm{axions/cm}^3
\label{eq:axion_DM_density}
\end{equation}

These DM axions would be non-relativistic particles, with a typical velocity given by the virial velocity inside our galaxy, $\sim$300~km/s (that is, the axions get their velocity mainly by falling in the galactic potential well). The precise velocity distribution around this value is however dependent on assumptions on how the Milky Way dark matter halo formed. The typical approach in experiments is to follow the Standard Halo Model (SHM), which comes from the simplistic assumption that the halo is a thermalized pressure-less self-gravitating sphere of particles. The velocity distribution of the SHM at the Earth is given by a Maxwellian distribution truncated at the galactic escape velocity ($\sim$600~km/s). A more recent estimation from N-body simulations provides a more precise shape to the velocity distribution:

\begin{equation}
f(v)\propto  \left(v^2\right)^\gamma \exp\left(\frac{v^2}{2\sigma_v^2}\right)^\beta
\label{eq:axion_veldistri}
\end{equation}

\noindent with $\gamma$, $\beta$ and $\sigma_v$ being fitting constants given in~\cite{Lentz:2017aay}. In any case, the typical dispersion velocity is around $\sigma_v\sim10^{-3}$. This can be considered an upper limit on the velocity dispersion of DM particles, but particular models may predict finer phase-space substructure. Perhaps the most extreme case is the infall self-similar model developed in~\cite{Sikivie:1995dp,Chakrabarty:2020qgm}, in which phase-space substructure is created in the galactic infalling of dark matter axions. In~\cite{OHare:2023rtm}, spatial miniclusters and streams have a cosmological origin, as a consequence of axion dark matter being born after inflation. In both cases, theses models predict that a substantial fraction of the DM axions in the form of a few velocity streams with much lower values of $\sigma_v$.

In any case, this means that the population of DM axions is better described collectively by a coherent classical field (rather than a ``gas'' of particles, like the case of WIMPs). The field is coherent over lengths approximately equal to the de~Broglie wavelength:

\begin{equation}
\lambda_c \lesssim \frac{\pi/2}{m_a \sigma_v}\sim 200 \left(\frac{m_a}{\rm 10~\mu eV}\right)^{-1} {\rm m} .
\label{eq:coherencelenght}
\end{equation}

\noindent which, for typical axion masses, is well beyond the size of the experiments (and even larger coherence lengths are expected for models with low dispersion streams). The field can then be considered spatially constant in the local region of our experiment and oscillating with a well defined frequency close to the axion mass $\nu_a \sim m_a$. The spread in frequency $\delta \nu_a $ around this value reflects the above-mentioned velocity dispersion, and corresponds to $\delta \nu_a / \nu_a = 10^{-6}$. Its inverse is the axion quality factor $Q_a \sim 10^6$ (once more, for models with low dispersion streams, this peak in frequency is expected to have substructure with much lower $\delta \nu_a$). This is a very important feature for experiments as it will allow to exploit coherent techniques to enhance the signal strength at detection.

\subsection{Conventional haloscopes}

\begin{figure}[t] \centering
\includegraphics[width=0.7\textwidth, trim={0 1cm 0 1cm}, clip ]{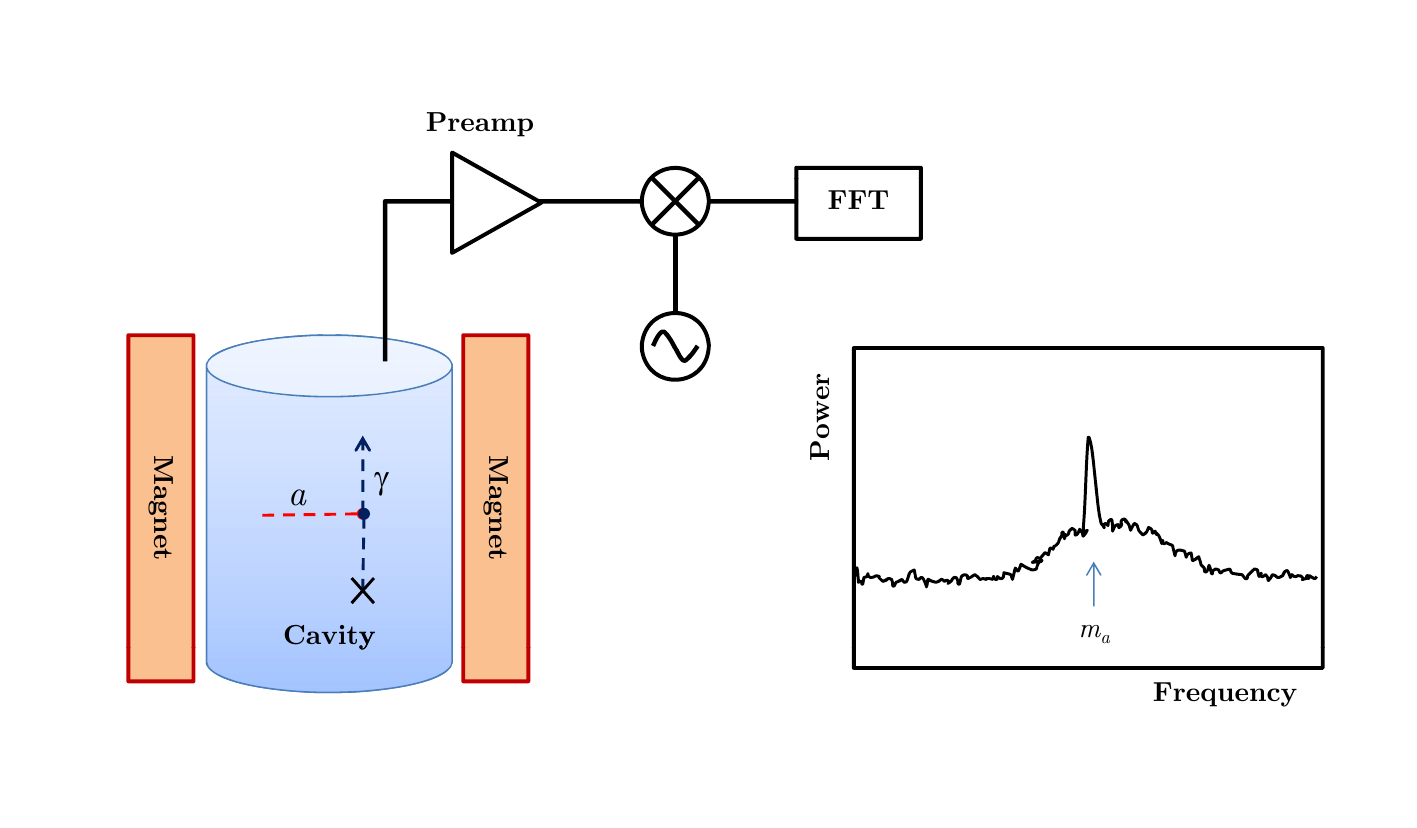}\hspace{2pc}%
\caption{\label{fig:haloscope_sketch} Conceptual arrangement
of an axion haloscope. If $m_a$ is within 1/$Q$ of the resonant frequency of the cavity, the axion will show as a narrow peak in the power spectrum extracted from the cavity. Figure taken from~\cite{Irastorza:2018dyq}}
\end{figure}

The conventional axion haloscope technique~\cite{Sikivie:1983ip} involves a high quality factor $Q$ microwave cavity inside a magnet, where $Q$ can be of order $10^{5}$ or even higher, using superconducting cavity coatings (see also Section \ref{sec:superconductors})). By virtue of the Primakoff conversion, DM axions produce photons in the magnetic field. If the resonant frequency of the cavity matches that of the axion field, the conversion is enhanced by a factor $Q$, and the resulting photons appear as an excited mode of the cavity. This power can then be extracted from the cavity via a suitable port connected to a radio-frequency (RF) detection chain with a low-noise amplifier. The power $P_s$ of such a signal can be calculated to be:

\begin{equation}
P_s  = \kappa \frac{Q}{m_a} \gagamma^2 B^2 |{\cal G}_m|^2  V \rho_a \
\label{eq:haloscope_signal}
\end{equation}

\noindent where $\kappa$ is the coupling of the cavity to the port, $B$ the external magnetic field, $V$ the volume of the cavity, and ${\cal G}_m$ a geometric factor accounting for the mode overlap between the given cavity mode electric field \textbf{E} and the external magnetic field \textbf{B}:

\begin{equation}
|{\cal G}_m|^2  = \frac{\left(\int dV \textbf{E}\cdot \textbf{B}\right)^2 }{V |\textbf{B}|^2\int dV \epsilon \textbf{E}^2},
\label{eq:geomtric_factor}
\end{equation}

\noindent where the integral is over the entire volume $V$ of the cavity, $\epsilon$ being the dielectric constant. The modes with higher ${\cal G}_m$ are the ones whose electric field is better aligned with the external magnetic field. For example, for a cylindrical cavity and a \textbf{B} field along the cylindrical axis, the TM$_{0n0}$ modes are the ones that couple with the axion, and the fundamental TM$_{010}$ mode provides the larger geometric factor $|{\cal G}_{\rm TM_{010}}|^2\sim 0.69$.

The signal expression in Eq.~(\ref{eq:haloscope_signal}) is only valid if the axion frequency matches the resonant frequency of the cavity within the very narrow cavity bandwidth $\sim m_a/Q$. Given that $m_a$ is not known, in order to scan a meaningful range of axion masses the cavity must be tunable in frequency. This is normally achieved by the implementation of precisely movable pieces that change the geometry of the cavity (e.g. movable rods). The experimental protocol involves a scanning procedure that devotes a small exposure time in each of the frequency points, then moving to the next one, and so forth. Covering a wide mass range can thus be time-consuming and thus be an experimental challenge.

Figure ~\ref{fig:haloscope_sketch} shows a sketch of the concept of the axion haloscope. A putative signal would appear as a narrow peak at the frequency corresponding to $m_a$ and with an intensity corresponding to Eq. (\ref{eq:haloscope_signal}). The capability of seeing such a signal will depend also on the level of noise and the exposure time. In absence of systematic effects, the longer the integration time, the smaller the noise fluctuations and the higher the signal-to-noise ratio. More on the analysis procedure is found in section~\ref{sec:analysis}. In general, the figure of merit $F_{\rm halo}$ of an axion haloscope can be defined as proportional to the time needed to scan a fixed mass range down to a given signal-to-noise ratio. This shows the main parameter dependencies:

\begin{equation}
F_{\rm halo}  \propto  \rho_a^2 g_{a\gamma}^4 m_a^2 B^4 V^2 T_{\rm sys}^{-2} |{\cal G}|^4 Q
\label{eq:haloscope_fom}
\end{equation}

\noindent where $T_{\rm sys}$ is the effective noise temperature of the detector. Typically the noise in these detectors come from thermal photons and therefore it is driven by the physical temperature of the system $T_{\rm phys}$. In reality $T_{\rm sys}$ includes additional components due to e.g. amplifier noise, $T_{\rm sys} = T_{\rm phys} + T_{\rm amp}$. Eq. (\ref{eq:haloscope_fom}) is useful to see the relative importance of each of the experimental parameters. Note the dependency with $\sim Q$ (instead of $Q^2$ naively expected from Eq.~\ref{eq:haloscope_signal}), which is due to the fact that improving $Q$ increases the signal strength but reduces the bandwidth of a single frequency point. This increases the total number of steps needed to scan a given mass range. Note also that improvement in $Q$ naively contributes to $F_{\rm halo}$ only as long as $Q<Q_a$ (however, see also \cite{Kim:2020kfo}), that is, the axion peak must be contained in the cavity bandwidth, otherwise some signal will be lost.

Experimental implementations of the axion haloscope paradigm have been led by the ADMX  collaboration, which has pioneered essential technologies such as high-quality factor (high-$Q$) resonant cavities within strong magnetic fields, quantum-limited RF amplification, and sub-Kelvin cryogenic environments. The central ADMX setup consists of a cylindrical copper-plated cavity of approximately 60~cm diameter and 1~m length, housed inside an $\sim$8~T superconducting solenoidal magnet. Frequency tuning is achieved through the precise positioning of dielectric and/or metallic tuning rods inside the cavity.
Over the years, ADMX has progressively improved its sensitivity to QCD axion dark matter in the $\mu$eV mass range. Currently, ADMX has successfully excluded axion-photon couplings in the 2.66--4.2~$\mu$eV range at sensitivity down to the DFSZ benchmark~\cite{Du:2018uak,Braine:2019fqb,ADMX:2021nhd}, a range recently extended to  5.41~$\mu$eV~\cite{ADMX:2025vom}.

The Center for Axion and Precision Physics Research (CAPP) has emerged as a competitive force in the search for axion dark matter, advancing a series of haloscope experiments with cutting-edge technologies. The CAPP-PACE and CAPP18T experiments have achieved sensitivities near the quantum noise limit in the $\mu$eV mass range, probing axion-photon couplings below the DFSZ benchmark in targeted mass bands~\cite{CAPP:2023pace,CAPP:2023capp18t}. More recently, the CAPP-MAX experiment has extended its reach into the 1.0--1.5~GHz frequency range, corresponding to axion masses around 4--5~$\mu$eV, delivering results that rival those of ADMX in terms of both sensitivity and mass coverage~\cite{CAPP:2024max}. These advances were enabled by the use of high-field magnets (up to 18~T), dilution refrigeration systems, and quantum-limited Josephson Parametric Amplifiers (JPAs). With continued development of high-frequency cavity designs and broadband scanning strategies, CAPP is well-positioned to explore higher-mass axion territory and provide complementary constraints to other leading experiments.

\begin{figure}[t]
\centering
\includegraphics[scale=0.8]{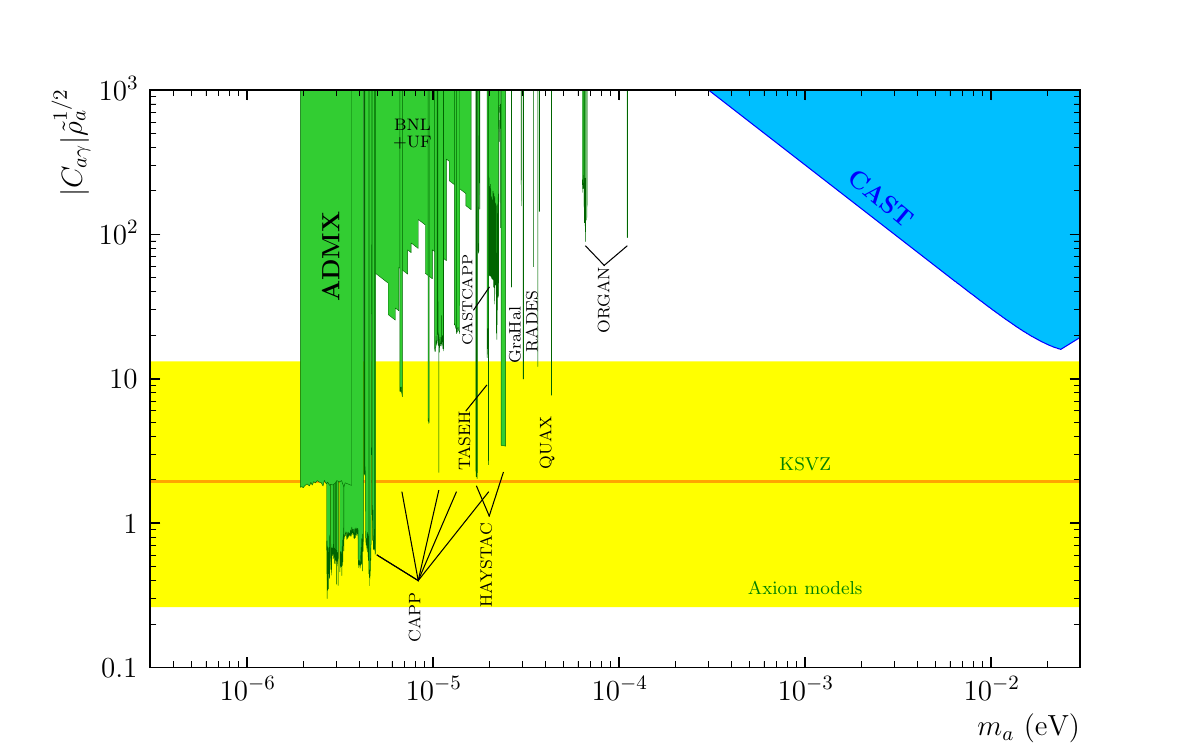}
\caption{Zoom-in of the region of parameters where most axion dark matter experiments are active (in green). The $y$-axis shows the adimensional coupling $C_{a\gamma} \propto \gagamma / m_a$ (scaled with the local axion DM density relative to the total DM density, $\tilde{\rho}_a = \rho_a/\rho_{DM}$, assumed unity in this plot, to stress that these experiments produce bounds that are dependent on the assumed fraction of DM in the form of axions). Thus the yellow region, where the conventional QCD axion models are, appears now as a horizontal band, but is the same yellow band shown in the other plots of this review.}
\label{fig:haloscopes}
\end{figure}

In recent years, a growing number of experimental initiatives have emerged, some of them implementing variations of the haloscope concept, or altogether novel detection concepts, making this subfield one of the most rapidly changing in the axion experimental landscape. 
Figure~\ref{fig:haloscopes} shows the current situation, with a number of new players accompanying ADMX and CAPP in the quest to cover different axion mass ranges.

Going to axion masses above $\sim10~\mu$eV is challenging for the conventional axion haloscopes concept. Higher frequencies imply lower volumes and correspondingly lower signals and sensitivity. This could be in part compensated by enhancing other experimental parameters (more intense magnetic fields, higher $Q$ factors, noise reduction, etc.). Recent years have seen a surge in haloscope experiments targeting those challenges. Techniques like superconducting coatings to increase $Q$~\cite{Alesini:2019ajt,Ahyoune:2024klt}, higher-order modes~\cite{Boutan:2018uoc}, and multi-cavity systems~\cite{Melcon:2020xvj,CAST:2020rlf,Jeong:2020cwz} are being explored to overcome volume limits, see also Sect.~\ref{sec:highmass}. Noise reduction by cooling the system to cryogenic temperatures hits technological and fundamental limits that can be overcome by novel quantum technologies, as will be explained in section~\ref{sec_quantum}. Substantially higher frequencies require novel concepts departing from the Sikivie haloscope, like the spherical dish or the dielectric haloscope concepts (section~\ref{sec:highmass}).

Lower frequencies imply proportionally larger cavity volumes and thus bigger, more expensive, magnets, but otherwise they are technically feasible. In addition, concepts like the lumped element and pick-up coil are being explores to expand the low-mass fronted, something that is explained in section \ref{sec:lowmass}.

\subsection{Solar axion experiments}
\label{sec:solar}

\begin{figure}[t]
\centering
\includegraphics[scale=0.8]{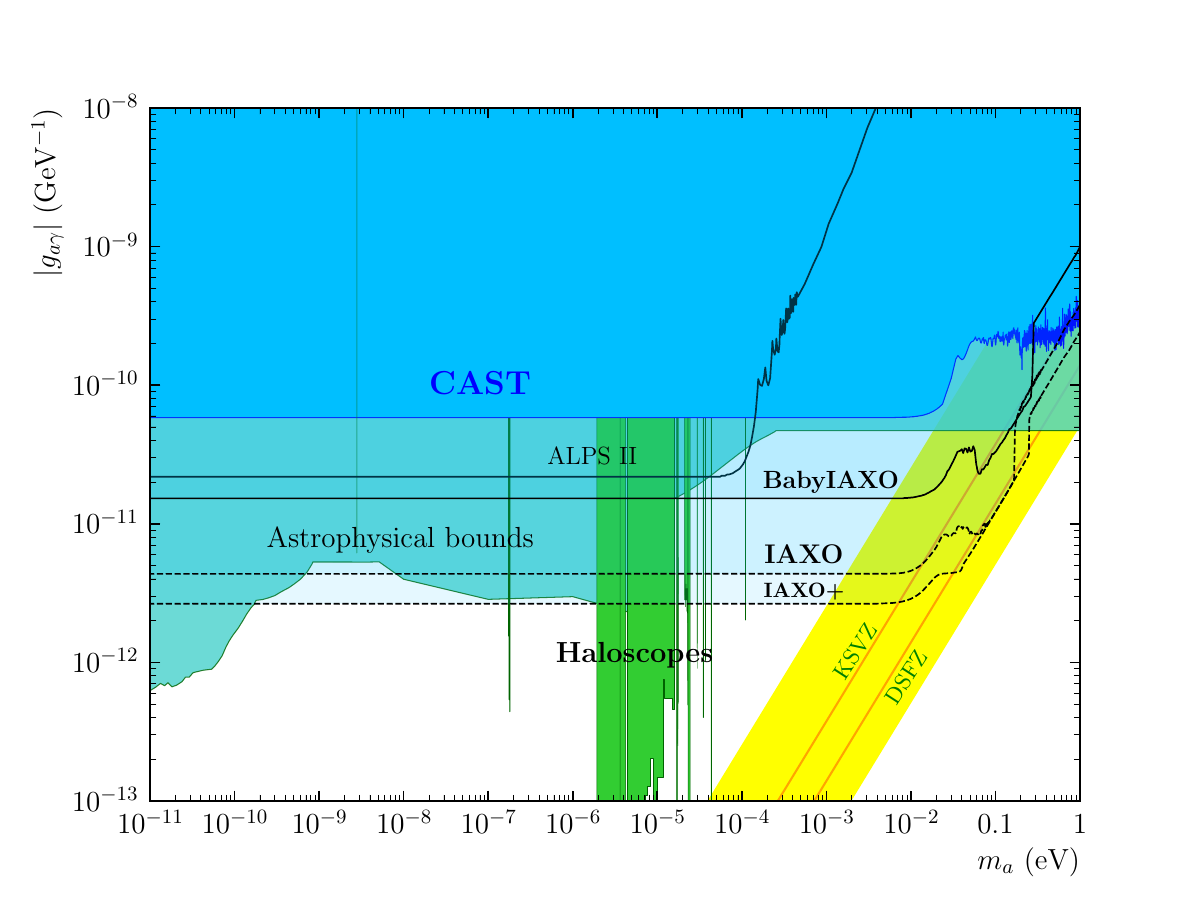}
\caption{Excluded regions and sensitivity prospects in the ($\gagamma$, $m_a$) plane, with a focus in the $\gagamma$ range relevant for helioscopes. Most relevant is the area excluded by CAST, as well as prospects from future helioscopes like BabyIAXO and IAXO (for the latter two scenarios, nominal and enhanced, or IAXO+, are considered), as well as the LSW experiment ALPS-II. Also the envelope of various astrophysical bounds, particularly relevant at low masses, is included. 
}
\label{fig:helioscopes}
\end{figure}

If axions exist, they would be produced in large quantities in the solar interior. The most important channel are Primakoff solar axions. They are a robust prediction by virtually any axion model, only requiring a non-zero $\gagamma$ and relying on well-known solar physics. Axions coupled to electrons offer additional production channels. Once produced, axions get out of the Sun unimpeded and travel to the Earth, offering a great opportunity for direct detection in terrestrial experiments. The leading technique to detect solar axions are axion helioscopes~\cite{Sikivie:1983ip}, one of the oldest concepts used to search for axions. Axion helioscopes (see Figure~\ref{fig:helioscopes}) are sensitive to a given $\gagamma$ in a very wide mass range, and after several past generations of helioscopes, the experimental efforts are now directed to increase the scale and thus push sensitivity to lower $\gagamma$ values. Contrary to the scenario described in the haloscope frontier, with a plethora of relatively small, sometimes table-top, experiments, most of the helioscope community has coalesced into a single collaboration, IAXO, to face the challenges to build a large scale next-generation helioscope. 


\begin{figure}[t]
\centering
\includegraphics[width=0.49\textwidth]{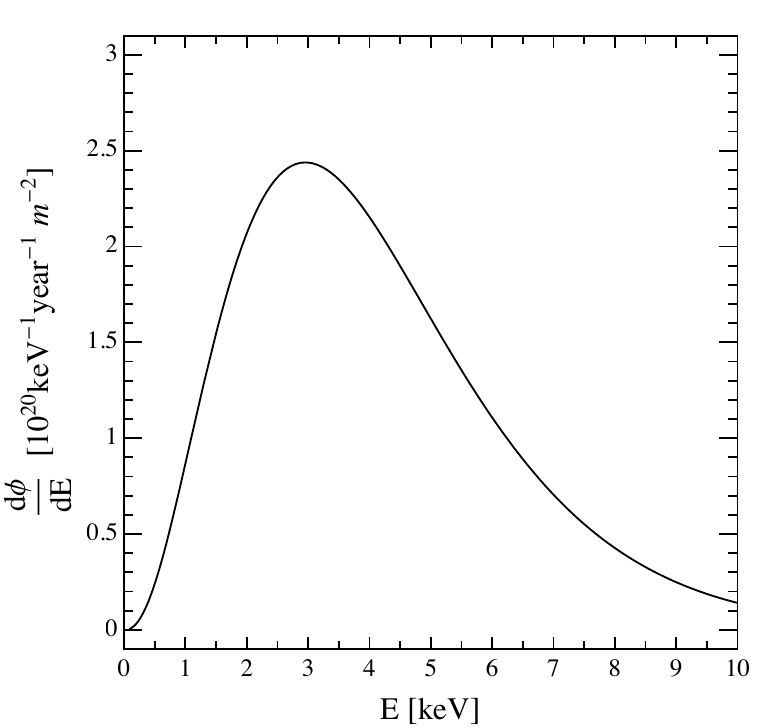}
\includegraphics[width=0.49\textwidth]{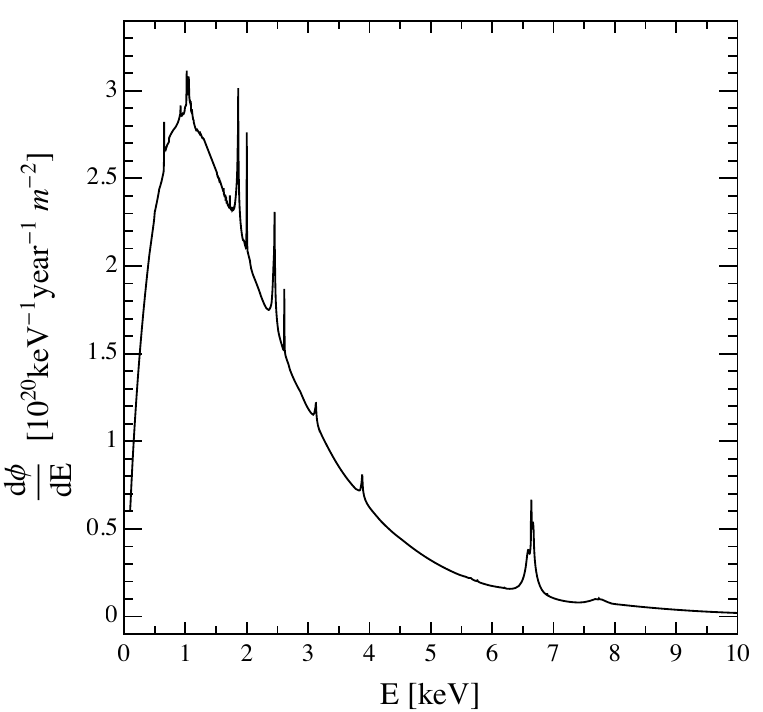}
\caption{\label{fig:axion_flux}Solar axion flux spectra at Earth by different production mechanisms. On the left, the most generic situation in which only the Primakoff conversion of plasma photons into axions is assumed. On the right the spectrum originating from the ABC processes~\cite{Redondo:2013wwa,Barth:2013sma}. The illustrative values of the coupling constants chosen are $\gagamma = 10^{-12}$~GeV$^{-1}$ and $g_{ae} = 10^{-13}$. Plots from~\cite{Irastorza:1567109}.}
\end{figure}

Photons from the solar plasma would convert into axions in the Coulomb fields of charged particles via the Primakoff axion-photon conversion. The produced axions have energies reflecting the typical solar core photon energies, i.e. around $\sim3$~keV. Therefore they are relativistic and the predicted flux is independent on $m_a$ (as long as $m_a \lesssim$~keV, which is the case for the QCD axion). A useful analytic approximation to the differential flux of Primakoff solar axions at Earth, accurate to less than 1\% in the 1--11 keV range, is given by~\cite{Andriamonje:2007ew}:

\begin{equation}
  \frac{{\rm d}\Phi_{\rm a}}{{\rm d}E}=6.02\times 10^{10} 
  \left(\frac{\gagamma}{10^{-10}\rm GeV^{-1}}\right)^2
  \,E^{2.481}e^{-E/1.205}\,
  \frac{1}{ {\rm cm}^{2}~{\rm s}~{\rm keV}}\ ,
\label{eq:primakoffflux}
\end{equation}

\noindent where $E$ is the axion energy expressed in keV. This Primakoff spectrum is shown in Fig.~\ref{fig:axion_flux} (left). As seen, it peaks at $\sim$3 keV and exponentially decreases for higher energies. Once the existence of a non-zero $\gagamma$ is assumed, the prediction of this axion flux is very robust, as the solar interior is well-known. A recent study of the uncertainties~\cite{Hoof:2021mld} confirms a statistical uncertainty at the percent level, although the number of axions emitted in helioseismological solar models is systematically larger by about 5\% compared to photospheric models. At energies below $\sim$keV the uncertainties are larger as other processes can contribute. Recent works have studied other interesting solar axion production channels that have not been exploited experimentally yet. On one side, axions can also be produced in the large scale magnetic field of the Sun. In particular, longitudinal or transversal plasmons can resonantly convert, leading to different detectable populations at sub-keV energies, with a dependence on the particular magnetic field profile of the Sun (and, for the case of transverse plasmons, on the ALP mass)~\cite{Caputo:2020quz,Guarini:2020hps,OHare:2020wum}.

In non-hadronic models axions couple with electrons at tree level. This coupling allows for additional mechanisms of axion production in the Sun~\cite{Redondo:2013wwa}, 
namely, atomic axio-recombination and axion-deexcitation, axio-Bremsstrahlung in electron-ion  or electron-electron collisions and Compton scattering with emission of an axion. They are sometimes collectively named ``ABC'' axions.

The spectral distribution of ABC solar axions is shown on the right of Figure~\ref{fig:axion_flux}. Albeit the relative strength of ABC and Primakoff fluxes depends on the particular values of the $g_{ae}$ and $\gagamma$ couplings, and therefore on the details of the axion model being considered, for non-hadronic models the ABC flux tends to dominate. Although all processes contribute substantially, free-free processes (bremsstrahlung) constitute the most important component, and are responsible for the fact that ABC axions are of somewhat lower energies than Primakoff axions, with a spectral maximum around $\sim$1~keV. This is because the axio-bremsstrahlung cross-section increases for lower energies and, in the hot solar core, electrons are more abundant than photons, and their energies are high with respect to atomic orbitals. In addition, the axio-de-excitation process is responsible for the presence of several narrow peaks, each one associated with different atomic transitions of the species present in the solar core. These two features would be of crucial importance in the case of a positive detection to confirm an axion discovery.

For the sake of completeness, we should mention that the existence of axion-nucleon couplings $g_{aN}$ also allows for additional mechanisms of axion production in the Sun. These emissions are mono-energetic and are associated with particular nuclear reactions in the solar core. Some examples of the emissions that have been searched for experimentally are: 14.4 keV axions emitted in the M1 transition of Fe-57 nuclei~\cite{DiLuzio:2021qct}, MeV axions from $^7$Li and D($p,\gamma$)$^3$He nuclear transitions or Tm$^{169}$ (see~\cite{Irastorza:2018dyq} for details and references).

\subsection{Axion helioscopes}

\begin{figure}[t] \centering
\includegraphics[width=\textwidth]{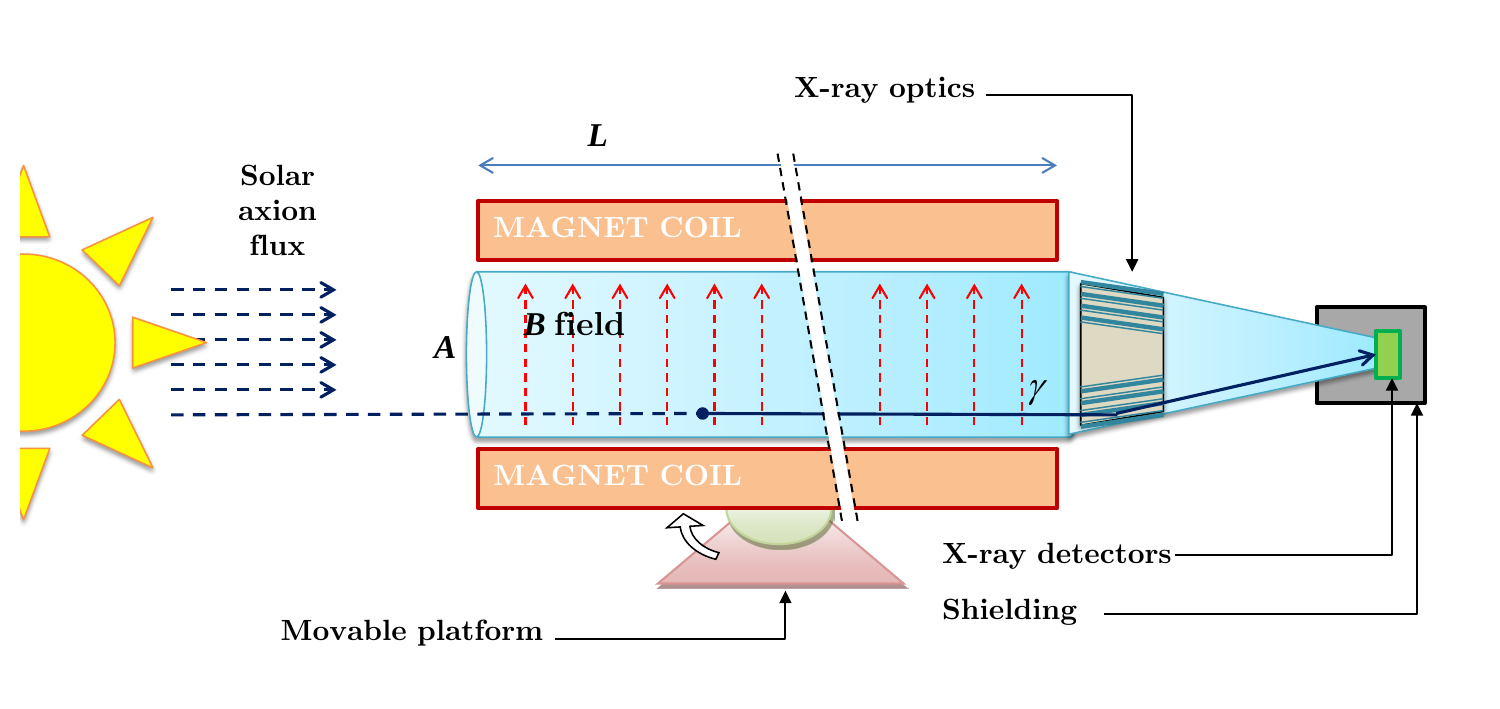}\hspace{2pc}%
\caption{\label{fig:helioscope_sketch} Conceptual arrangement
of an enhanced axion helioscope with X-ray focalization. Solar axions are converted into photons by the transverse magnetic field inside the bore of a powerful magnet. The resulting quasi-parallel beam of photons of cross sectional area $A$ is concentrated by an appropriate X-ray optics onto a small spot area $a$ in a low background detector. Figure taken from~\cite{Irastorza:2011gs}.}
\end{figure}

The axion helioscope detection concept~\cite{Sikivie:1983ip} is based on the conversion of solar axions into photons within a strong laboratory magnetic field. The converted photons retain the energy of the incoming axions and thus lie in the X-ray range, allowing their detection at the far end of the magnet when it is aligned with the Sun (see Fig.~\ref{fig:helioscope_sketch}).

The probability for axion-to-photon conversion in a homogeneous magnetic field of strength $B$ and length $L$ is given by~\cite{Sikivie:1983ip,Zioutas:2004hi,Andriamonje:2007ew}:

\begin{equation}
{\cal P}(a\to \gamma) = 2.6 \times 10^{-17} \left(\frac{\gagamma}{10^{-10}\rm~GeV^{-1}}\right)^2 \left(\frac{B}{10 \mathrm{\ T}}\right)^2
\left(\frac{L}{10 \mathrm{\ m}}\right)^2
\mathcal{F}(qL)
\label{eq:helio_conversion_prob}
\end{equation}

\noindent where $\mathcal{F}(qL)$ is a form factor that accounts for the loss of coherence:

\begin{equation}
\mathcal{F} = \left(\frac{2}{qL}\right)^2 \sin^2\left(\frac{qL}{2}\right),
\label{eq:formfactor}
\end{equation}

\noindent with $q = m_a^2 / 2E_a$ representing the momentum transfer, i.e., the mismatch between the photon and axion wave vectors, and $E_a$ the axion energy.
In the coherent regime, where $qL \ll 1$, the form factor approaches unity, $\mathcal{F}(qL) \rightarrow 1$, maximizing the conversion probability. However, as $qL$ increases, $\mathcal{F}(qL)$ decreases, leading to a suppression in the conversion rate.
For typical helioscope parameters (magnet lengths of order $\sim$10~m) and solar axion energies, this loss of coherence becomes significant for axion masses around $m_a \sim 0.01$eV. Consequently, the sensitivity of an axion helioscope remains flat for masses below this threshold, as illustrated in Figure\ref{fig:helioscopes}.

Figure~\ref{fig:helioscope_sketch} shows the typical configuration of axion helioscopes. Due to the dependencies expressed in Eq.~(\ref{eq:helio_conversion_prob}), dipole-like layouts for the magnet are preferred—that is, relatively long (in the Sun's direction) with a magnetic field in the transverse direction. The magnet is placed on a moving platform that allows it to point at the Sun and track it for long periods. At the end of the magnet opposite the Sun, the detection line(s) are placed.

In modern optically enhanced versions of helioscopes, X-ray optics are placed just at the end of the magnet bore to focus the almost-parallel beam of photons from axion conversion into small focal spots. This allows the use of relatively large magnet transverse areas while keeping a relatively small detector, thereby increasing the signal-to-noise ratio. X-ray optics are built following techniques developed for X-ray astronomy missions, based on the high reflectivity of X-rays when they strike a mirror at a small grazing angle. These optics resemble a collection of conical mirrors, nested one inside another, covering the entire magnet area.

The X-ray detectors are placed at the focal points of the optics and need only be slightly larger than the focal spot size ($\sim$cm$^2$). The presence of solar axions would manifest as an excess of counts over the detector background, the latter being measured either in the detector area outside the signal spot or during periods when the magnet is not pointing at the Sun. The detector should be energy-resolving and pixelated so that the energy distribution of the detected photons, as well as their spatial distribution on the detector plane (the signal “image”), can be compared with expectations in the case of a positive signal. The latter should correspond to the angular distribution of solar axion emission, spatially convoluted with the optics response or “point spread function.”

Because the background is measured and statistically subtracted from the ``signal data'', the signal-to-noise ratio in axion helioscopes goes with the background fluctuations  rather that the background itself ($\sqrt{n}$ versus $n$). In general the figure of merit of an axion helioscope $F_{\rm helio}$ can be defined as proportional to the signal to noise ratio for a given value of $\gagamma$, so that:

\begin{equation}
    F_{\rm helio} \propto B^2 L^2 {\cal A} \; \frac{\epsilon_d \epsilon_o}{\sqrt{ba}} \; \sqrt{\epsilon_t t}
    \label{eq:helioscope_fom}
\end{equation}

\noindent where $B$, $L$ and  $\cal A$ are the transverse magnetic field, length and cross-sectional area of the magnet respectively, $\epsilon_o$ is the throughput of the optics (or focalization efficiency), $a$ the signal spot size after focalization, $\epsilon_d$ the detection efficiency, $b$ the normalized (in area and time)
background of the detector, $\epsilon_t$ is the data-taking efficiency, i. e. the fraction of time the magnet tracks the Sun (a parameter that depends
on the extent of the platform movements) and $t$ the duration of the data taking campaign.

So far, we have assumed that the magnet bores are in vacuum. This is known as the baseline (or phase-I) configuration. In order to attain sensitivity to axion masses above the value at which ${\cal F}(qL)$ drops due to loss of coherence (i.e., $m_a \gtrsim 0.01$eV), the bores can be filled with a buffer gas\cite{vanBibber:1988ge}. This gas provides the photon with an effective mass and restores coherence for a narrow window of axion masses around the photon refractive mass.
In this gas phase (or phase-II) of the experiment, the pressure of the gas is varied in steps, and data taking follows a scanning procedure in which the experiment becomes sensitive to different small mass intervals at each step—similar to axion haloscopes, although in this case the relative width of each mass step is on the order of ${\cal O}(10^{-2})$.

The strategy described above has been followed by the CERN Axion Solar Telescope (CAST) experiment, using a decommissioned LHC test magnet that provides a 9~T field inside the two 10~m long, 5~cm diameter magnet bores. CAST has been active for about 20 years at CERN\footnote{The experiment definitely finished operation in 2021.}, going through several data taking campaigns, and represents the state-of-the-art in the search for solar axions. It has been the first axion helioscope using X-ray optics. The latest solar axion result~\cite{CAST:2024eil}
sets an upper bound on the axion-photon coupling of:

\begin{equation}
 \gagamma < 0.58 \times 10^{-10}{\rm~GeV}^{-1},
 \label{eq:cast}
\end{equation}

\noindent for $m_a\lesssim 0.01$~eV. Figure~\ref{fig:helioscopes} shows the full exclusion line. The wiggly extension at higher masses, up to about 1eV, is the result of scanning with a buffer gas in the bores~\cite{CAST:2008ixs,CAST:2011rjr,CAST:2013bqn}, which has allowed CAST to probe the QCD band in that mass range. The limit in Eq.(\ref{eq:cast}) competes with the strongest bound coming from astrophysics. Advancing beyond this bound to lower $\gagamma$ values is now highly motivated, as it would mean venturing into regions of parameter space allowed by astrophysical constraints~\cite{Irastorza:2011gs}.
CAST has also searched for solar axions produced via the axion-electron coupling~\cite{Barth:2013sma} (and axion-nucleon coupling in~\cite{CAST:2009klq,Andriamonje:2009dx}), although the very stringent astrophysical bound on this coupling remains, so far, unchallenged by experiments.

The successor of CAST is the International Axion Observatory (IAXO)~\cite{Armengaud:2014gea}, a new generation axion helioscope, aiming at the detection of solar axions with sensitivities to $\gagamma$ down to a few 10$^{-12}$~GeV$^{-1}$, a factor of 20 better than the current best limit from CAST (a factor of more than $10^4$ in signal-to-noise ratio).  This leap forward in sensitivity is achieved by the realization of a large-scale magnet, as well as by extensive use of X-ray focusing optics and low background detectors. 

The main element of IAXO is thus a new dedicated large superconducting magnet~\cite{Shilon:2013xma}, designed to maximize the helioscope figure of merit. The IAXO magnet will be a superconducting magnet following a large multi-bore toroidal configuration, to efficiently produce an intense magnetic field over a large volume. The design is inspired by the ATLAS barrel and end-cap toroids, the largest superconducting toroids ever built and presently in operation at CERN. Indeed the experience of CERN in the design, construction and operation of large superconducting magnets is a key aspect of the project.

As already mentioned, X-ray focalization relies on the fact that, at grazing incident angles, it is possible to realize X-ray mirrors with high reflectivity. IAXO envisions newly-built optics similar to those used onboard NASA's NuSTAR satellite mission, but optimized for the energies of the solar axion spectrum. Each of the eight $\sim$60~cm diameter magnet bores will be equipped with such optics. At the focal plane of each of the optics, IAXO will have low-background X-ray detectors. Several detection technologies are under consideration, but the most developed ones are small gaseous chambers read by pixelised microbulk Micromegas planes~\cite{Aznar:2015iia}. They involve low-background techniques typically developed in underground laboratories, like the use of radiopure detector components, appropriate shielding, and the use of offline discrimination algorithms. Alternative or additional X-ray detection technologies are also considered, like GridPix detectors, Magnetic Metallic Calorimeters, Transition Edge Sensors, or Silicon Drift Detectors. All of them show promising prospects to outperform the baseline Micromegas detectors in aspects like energy threshold or resolution, which are of interest, for example, to search for solar axions via the axion-electron coupling. This is because this process features both lower energies than the standard Primakoff ones, and monochromatic peaks in the spectrum.

An intermediate experimental stage called BabyIAXO~\cite{IAXO:2020wwp} is the near term goal of the collaboration. BabyIAXO will test magnet, optics and detectors at a technically representative scale for the full IAXO, and, at the same time, it will be operated and will take data as a fully-fledged helioscope experiment, with sensitivity beyond CAST (see Figure~\ref{fig:helioscopes}). It is currently under construction at DESY.

The expected sensitivity of BabyIAXO and IAXO in the $(\gagamma,m_a)$ plane is shown in Figure~\ref{fig:helioscopes}, both including also a phase II result at higher masses. The IAXO projection include two lines, one corresponding to nominal expectations and another one a more optimistic projection with a $\times10$ better $F_{\rm helio}$. The sensitivity of IAXO to $g_{ae}$ via the search of ABC axions (not shown in the plots) will be for the first time competitive with astrophysical bounds and in particular sufficient to probe a good part of the hinted range from the anomalous cooling of stars. We refer to \cite{Armengaud:2019uso} for more details on this and other the physics potential of BabyIAXO and IAXO. 

\subsection{Other solar axion detectors}

For completeness, let us mention that several alternative techniques to axion helioscopes have been explored. The AMELIE~\cite{Galan:2015msa} concept proposes a magnetized gaseous detector where the gas acts both as a conversion and detection medium. Coherence is lost, but a daily modulation of the signal remains. Although there is currently no experimental effort implementing the concept, competitive sensitivity could be achieved at higher masses, if high pressure or high-$Z$ gases are used.
Bragg-enhanced conversion in crystals~\cite{Buchmuller:1989rb,Paschos:1993yf,Creswick:1997pg} produces energy- and time-dependent patterns~\cite{Avignone:1997th,Morales:2001we,Bernabei:2001ny,Ahmed:2009ht,Armengaud:2013rta,Li:2015tsa,Xu:2016tap}, although do not compete with standard axion helioscopes~\cite{Cebrian:1998mu,Avignone:2010zn}. Ionization detectors can exploit the axioelectric effect~\cite{Ljubicic:2004gt,Derbin:2011gg,Derbin:2011zz,Derbin:2012yk,Bellini:2012kz,Abe:2012ut,Aprile:2014eoa,PandaX:2017ock,Akerib:2017uem}, while nuclear resonant absorption has also been studied~\cite{Moriyama:1995bz,Krcmar:1998xn,Krcmar:2001si,Derbin:2009jw}, though both remain less sensitive than astrophysical constraints.


\section{From data to physics}
\label{sec:analysis}

\subsection{Haloscope analyses}

To appreciate the difficulty in the endeavour of searching the axion dark matter signal, it may be useful to compare to some quantities from everyday life.
Today’s long-term evolution (LTE) standard for wireless broadband communication, used, among else, for mobile devices emits at the GHz
range, not far from what is probed at in typical axion experiments. 
To receive a good signal $10^{-6}$ Watt are needed, a weak signal can still
be received at, say 7 orders of magnitude below that level. 
Plugging typical values in Eq.~\ref{eq:haloscope_signal} it means that
your typical axion experiment needs to receive signals {\it at least a billion times} weaker than the weakest signal than your mobile phone would need for you to receive a call.
This endeavour clearly needs not only good hardware, but also a careful analysis technique.

\begin{figure}[h]
	\centering
	\includegraphics[width=0.7\textwidth]{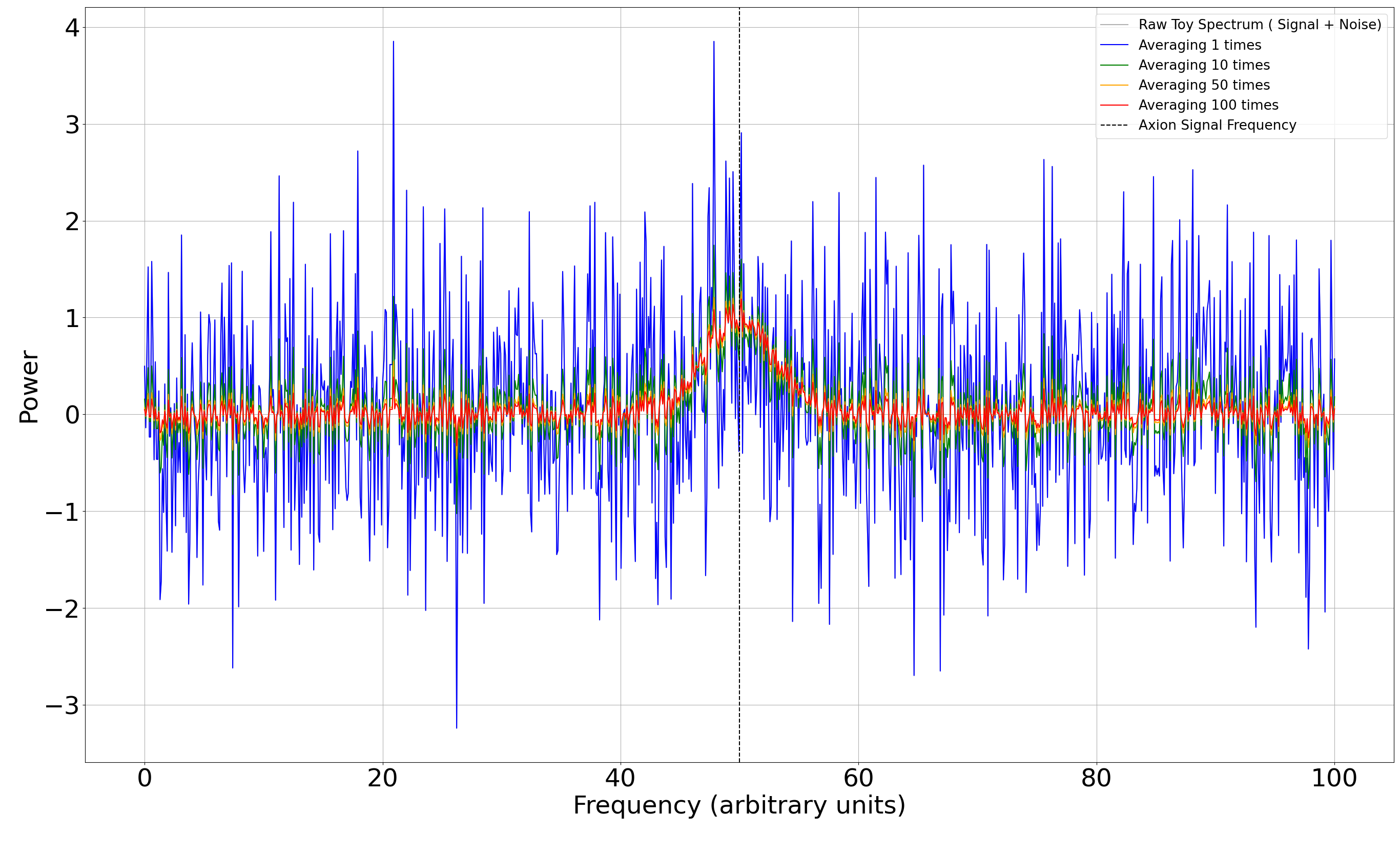}
	\caption{Simplified illustration of an `axion signal' emerging when averaging the signal+ electronic noise background components a number of times, (colored lines). The random background fluctuations average out while the axion signal persists (the signal is here modelled as Gaussian for simplicity).
 }
	\label{fig:signal}
\end{figure}

The basic idea of how the data in a haloscope search should look like is very simple: If the data-acquisition system records `just noise', random upwards and downwards fluctuations of the signal power should average out. If, instead, an axion is to be found at a fixed frequency, it should contribute to the read-out power only `positively', and a `signal over background' can emerge, cf. Figure \ref{fig:signal}.

With this simple insight, in principle `only' the expected axion lineshape needs to be known in order to proceed. Most analyses are fit to find an axion in the `standard halo model, i.e., given the assumptions that lead to equation \ref{eq:axion_veldistri}.
Typically, one wants that the bin-width of the data-acquisition is such that an axion can spread over several bins. This enables to fit an expected line-shape of the axion, while maintaining a reasonable scan time.

We want to highlight, however, that in order to best exploit the data, other possibilities can and should be studied. For example, the axion could also be part of cold dark matter flows.
In this case, the search procedure needs a better frequency resolution (and thus longer integration times than in the `standard' case, see, e.g. for an analysis \cite{ADMX:2023ctd} and \cite{OHare:2023rtm} a related theory model of this kind.

In the following we want to outline the typical analysis steps that are performed to reach the idealized state of affairs shown in the illustrative Figure \ref{fig:signal} without reference to a particular model.

In practice, data-acquisition systems are imperfect and data-taking is impacted by disturbances of often very sensitive equipment.
Thus, analysis procedures are set up to extract meaningful information from the acquired data.

The analysis procedures within the various experimental groups differ in their details. A seminal paper that many groups follow  (see e.g. \cite{Schneemann:2023bqc,QUAX:2024fut,CAST:2020rlf}) is the one of  HAYSTAC \cite{Brubaker:2017rna}, with an update \cite{Palken:2020wgs}.

Rather than diving into the details of any specific analysis, let us try to summarize the steps that typically are necessary for everyone:

\begin{enumerate}
        \item {\bf removal of `obvious' outliers:}
          recorded spectra in which the axion sensitivity cannot be reliably established can be quickly discarded. Examples could be problems in the monitoring of relevant parameters (frequency drifts in the cavity peak for example caused by vibrations, `noisy bins' from electronic faults, or other issues...)
      \item  {\bf averaging:} as outlined above, this is at the heart of the procedure (fig.~\ref{fig:signal}) 
    \item  {\bf normalization:} Signals appear relative to the baseline in a given spectrum. This baseline can vary. It makes therefore sense to normalize the raw spectra to an average baseline for further processing
    \item {\bf removal of the Lorentzian cavity shape:} For the detection algorithm we want to typically remove the Lorentzian shape. The assumption of the axion signal position on top of the Lorentzian is only needed when assessing the signal strength
    \item  {\bf Removal of structures which are different in scale compared to an axion signal:} A filter commonly applied in axion search analysis is the one established by A. Savitzky and M. J. E. Golay (SG) \cite{Savitzky:2002vxy}: It fits successive sets of subsequent data points with a polynomial through the least squares method. If the window width is much wider than that of the axion signal, dividing by this filter can distort and attenuate the axion signal, but should not remove it.
    The parameters of this filter depend very much on the experimental setup and a careful optimisation of this filter's parameter and an accounting of its effects are needed for a credible analysis. Note that sometimes the SG filter is applied multiple times.
\end{enumerate}

\vspace{0.2cm}
In summary, a procedure that analyses the obtained cavity power spectrum is needed that partially follows generic steps but often needs to be adapted to the specifics of the experiment.
In particular, thermal noise, amplifier noise, environmental interference as well as systematic effects must be understood. If any power excess is found, statistical tests determine if it is significant.
In case a possible signal is found, a re-scan should check the frequency range for reproducibility and correct behaviour (e.g. correct scaling with the magnetic field).
If no axions are found, exclusion limits on axion-photon coupling are set.
If a signal would be confirmed, it would be a revolutionary discovery.

\subsection{Helioscope analyses} %

As previously discussed in section~\ref{sec:solar}, solar axions possess energies in the range of a few keV (see Fig.~\ref{fig:axion_flux} and, for Primakoff axions, Eq~(\ref{eq:primakoffflux})). When these axions convert within the magnetic field of the helioscope, the emitted photons retain the same energy (and momentum) as the incident axions. Consequently, the signal anticipated in a helioscope detector should manifest as an excess of x-ray counts at corresponding energies above the background levels. The actual energy distribution of these signal counts may deviate from that predicted by Eq.~\ref{eq:primakoffflux} due to energy-dependent efficiency functions arising from various factors, including optical focusing efficiency (e.g., reflectivity of optical shells), quantum efficiency of the detector, and transmission characteristics of any intervening windows.

In the early versions of helioscopes, which lacked x-ray optics, the analysis of potential signals was conducted entirely in the spectral dimension. The detector background had to be determined experimentally through dedicated runs where the magnet was not oriented towards the Sun. This background was then subtracted from the "axion-sensitive" data collected during sun-tracking runs, and the results were fitted to the signal spectrum to identify (or exclude) a possible signal. In modern, optics-enhanced versions of the experiment, the detector is positioned at the focal plane of x-ray optics, with expected signal counts concentrated in a specific (small) area of the detector centered around this focal point—representing an "axion image" of the Sun as focalized by these optics. Besides improving the signal-to-noise ratio (as indicated by Eq.~\ref{eq:helioscope_fom}), this arrangement also allows for spatial information to be utilized in the statistical treatment of the data. Specifically, background can be experimentally determined using events outside of this spot (assuming that background remains sufficiently constant across the detector surface). Naturally, it is essential that the detector possesses adequate spatial resolution (pixelization) in the ($x$-$y$) plane perpendicular to the optical axis. The actual signal distribution in the ($x$-$y$) plane can be computed through detailed ray-tracing simulations utilizing an accurate model of the optics. These simulations are validated via long-exposure calibration runs with a distant x-ray source that generates quasi-parallel x-rays at the entrance of optics.

\begin{figure}
    \centering
    \includegraphics[width=\linewidth]{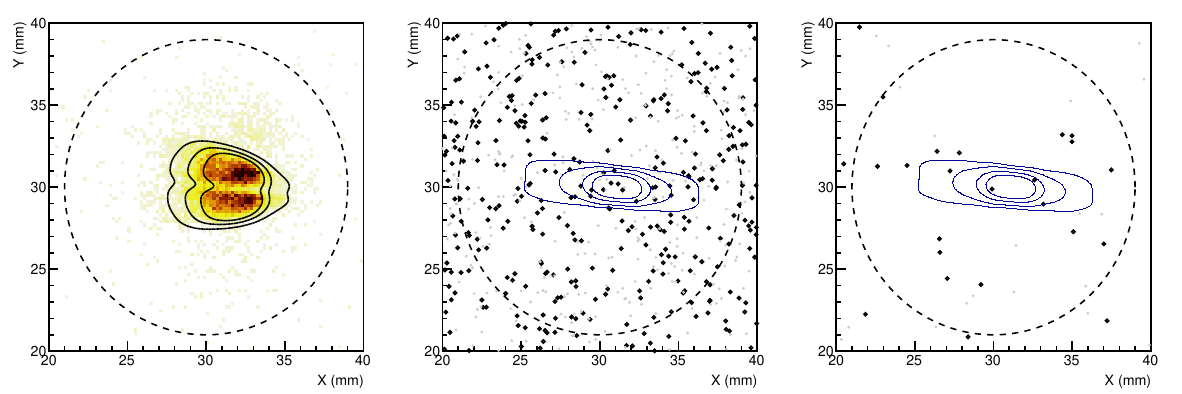}
    \caption{Illustrative example of data analysis from the 2017 CAST result~\cite{CAST:2017uph}. 2D hitmap of events detected in the ($x$-$y$) plane of the focal plane detector, in a typical in-situ calibration run (left), as well as in the background (middle) and tracking (right) data. The calibration is performed with an x-ray source placed $\sim$12~m away (at the opposite side the magnet). The contours in the calibration run represent the 95\%, 85\% and 68\% signal-encircling regions from ray-trace simulations, taking into account the source size and distance. In the tracking and background plots, grey full circles represent events that pass all detector cuts but that are in coincidence with the muon vetoes, and therefore rejected. Black open circles represent final counts. Closed contours in the background and tracking runs indicate the 99\%, 95\%, 85\% and 68\% signal-encircling regions out of detailed ray-trace simulations of the XRT plus spatial resolution of the detector. The large circle represents the region of detector exposed to daily energy calibration. Plots adapted from~\cite{CAST:2017uph}.}
    \label{fig:castresult}
\end{figure}

Fig~\ref{fig:castresult}, taken from one of the key CAST results~\cite{CAST:2017uph}, illustrates the aforementioned analysis process. The plots depict events recorded in the ($x$-$y$) plane of the CAST focal-plane Micromegas detector during, from left to right, a distant-source calibration run, all integrated background runs (i.e., with the magnet not oriented towards the sun), and all integrated ``signal'' runs (i.e., ``tracking'' the sun). Further details can be found in the figure caption. The difference in signal contours between the calibration plot and the signal plot arises because, in the former case, the x-ray source is at a finite distance (12 m), whereas it is considered to be at infinity in the latter. The calibration events align reasonably well with predicted contours, thereby validating the simulation.

From the ``tracking'' dataset (plot on the right), one extracts the main statistics to estimate the presence of a significant excess in the data. Due to the very low counting statistics (in this example only 3 counts were measured in the signal region), an unbinned likelihood function is used. The function includes both the spatial and spectral distributions of the signal to appropriately weigh each of the measured counts.  We refer to~\cite{CAST:2017uph} for further details.

\section{Physics topics addressed by the experiment}\label{physics}

Due to fixed mass-coupling relation of the `vanilla version' of the QCD axion with very weak effective couplings, typically only setups in which all experimental parameters are optimal become sensitive to these axions.
By contrast, novel parameter space of axion-like particles can often easily be tested \cite{Arias:2012az}  by haloscopes and helioscscopes  whose performance yet has to be optimized for reaching near-DSZF or KSVZ coupling values.
This makes especially haloscopes experiments attractive, since even `newcomers' can quickly achieve meaningful results relatively quickly.
But even beyond axions and ALPs, a number of physics scenarios can be probed with haloscopes, sometimes with little to no modifications in the setup.
In the following we present the arguably most popular such models that axion DM setups and helioscopes can probe.

\subsection{Dark Photons}
An axion haloscope can be sensitive to new physics beyond vanilla axions or axion-like particles. The simplest example for this is that
a by-product of axion searches is the sensitivity to dark matter Dark Photon \cite{Arias:2012az}.
In this case, besides the known EM-$U(1)$ gauge group, a second one is introduced, which comprises the `Dark Photon'.
Its existence can be motivated from BSM scenarios such as string theory \cite{Abel:2008ai}.

The interaction with the SM is via a kinetic mixing term:

\begin{equation} \label{eq:kinetic_mixing_lagrangian}
\mathcal{L}_{kin} = -\frac{1}{4} F_{\mu\nu} F^{\mu\nu} - \frac{1}{4} X_{\mu\nu} X^{\mu\nu} - \frac{\chi}{2} F_{\mu\nu} X^{\mu\nu},
\end{equation}

where $F_{\mu\nu}$ and $X_{\mu\nu}$ are the field strength tensors of the SM EM-$U(1)$ and the Dark Photon respectively. The parameter $\chi$ quantifies the strength of their mixing.
The Dark Photon has to have a even-so-tiny mass $m$, otherwise it would be indistinguishable from the SM photon.
As the Dark Photon has a small probability to oscillate into a photon, it can be detected in an axion Haloscope or helioscope even if $B=0$.

We should note that Dark Photon dark matter searches are not only a by-product of axion haloscopes, but are and have been searched also in dedicated experiments, using, for example an open resonator (a.k.a. a mirror) \cite{Horns:2012fx}, where the signal of a DM-induced photon would not appear in the focal point of the mirror but its centre point, see  e.g.\cite{Suzuki:2015sza, FUNKExperiment:2020ofv}.

However, there is an important difference in ALP/axion w.r.t. Dark Photon searches beyond the fact that one one requires an external magnetic field and the other one does not.
As the Dark Photon carries polarization, a Dark Photon dark matter search intrinsically makes assumptions about the polarization of the Dark Photons.
Most commonly, it is assumed that the Dark Photons are randomly polarized, but this need not be the case.
In fact, exploiting the knowledge of the movement of the earth and the solar system with respect to the dark matter background, the sensitivity for polarized Dark Photons can be improved with an appropriate choice of measurement intervals \cite{Caputo:2021eaa}.

For helioscopes, a dedicated Dark Photon search was performed by SHIPS \cite{Schwarz:2015lqa}, but it struggles to be competitive with astrophysical limits.

\subsection{Cosmic Axion Background}

The exact study of the CMB is a key factor in determining the amount of dark matter \cite{Cirelli:2024ssz}.
The axion, while studied and searched mostly as a dark matter candidate, could manifest itself in the CMB also as `Dark radiation' thus a {\it relativistic} component of particles: If axions exist, they could be produced in the early Universe also thermally and thus becoming relativistic. This could happen through the
decay of topological defects or  from the decay of another dark matter candidate during later epochs. In this case, the relativistic abundance of such axions would form a Cosmic axion Background
(CaB) \cite{PhysRevD.103.115004}, similar to the photons in the cosmic microwave background.

While the spatial coherence length of axion dark matter is on the order of hundreds of meters, the CaB would have a coherence length of the order of the cavity size and could induce a much broader signal in frequency then dark matter would.
A search strategy here is to look for daily modulations of the power excess, which requires mitigating and/or controlling gain fluctuations in the readout and amplifiers. An analysis that set first limits on a concrete variant of a CaB can be found in  \cite{ADMX:2023rsk}.

\subsection{High Frequency Gravitational Waves} %

An additional physics case for axion haloscopes that has received increased attention during the past few years \cite{Aggarwal:2020olq} is the possibility of detecting different implications of high-frequency gravitational waves  (HFGWs) passing through a haloscope.
While the details of the computations very much differ between the spin-0 and spin-2 case, intuitively and phenomenologically the mixing of the photon with axions and gravitons, respectively, is very similar  \cite{Raffelt:1987im}.
We know that gravitational waves exist. However, known sources of such waves at frequencies that a haloscope is sensitive to, are much weaker than what is currently detectable (the dominant local source being our sun \cite{Garcia-Cely:2024ujr} ). On the other hand, this means any detection of a HFGW could be a hint for new physics.

Axion haloscopes can contribute to HFGW searches via the  inverse Gerstenshtein effect: gravitons can convert to photons in the background of a magnetic field, upper limits have been obtained via this meachanism for helioscopes and LSW experiments \cite{Ejlli:2019bqj}, but also for haloscopes  \cite{Aggarwal:2020olq}.

For a more expensive recent overview of this emerging field, we refer to \cite{Gatti:2024mde,Aggarwal:2025noe}. Note also that for the detection of such signal a network of detectors that allows to check the correlation between signals from multiple, geographically separated cavities can be beneficial, see \cite{Schmieden:2023fzn}.
Lastly, is interesting to note that the magnets that are part of axion experiments themselves can act as `gravitational wave antennas' and and give access to probing somewhat lower GW frequencies \cite{Domcke:2024mfu}.

\subsection{Axion Quark Nuggets}

Axion Quark Nuggets (AQNs) are a theoretical form of dark matter, different from conventional axions but still connected to axion physics. The AQN model suggests that dark matter is composed of compact but macroscopic objects made of quarks and antiquarks surrounded by a `shell' of axions. These nuggets, if realised in nature,  are predicted to have significant mass (ranging from kilograms to planet-sized). The model from which they arise was developed to explain the asymmetry between matter and antimatter in Universe:
a fraction of matter is bound into heavy nuggets of quark matter in a colour superconducting phase \cite{Zhitnitsky:2004da}.
 Axion haloscopes could also potentially detect signals from AQNs despite the fact that AQNs would be macroscopic. However, analysis strategies have to be adapted for this scenario as AQNs would create a more `transient' or burst-like signal. If an AQN were to pass through a haloscope, it would produce a short-lived but strong burst of microwave photons via the concentration of axions surrounding the nugget.

The detection process then would focus on looking for brief spikes in photon counts, corresponding to the passage of AQNs.  Haloscopes, especially those with broadband detection capabilities, can be employed to look for transient signals instead of continuous narrowband ones, see this search performed with CAST-CAPP as an example \cite{Caspers:2024kjp}.

\subsection{Chameleons}

Chameleons are scalar particles introduced in certain models of dark energy to explain the accelerating expansion of the universe.
A new scalar field would naturally couple to Matter, and to be consistent with local tests of gravity the effects of the scalars must be suppressed or `screened' locally (for limitations of such constructions, see \cite{PhysRevLett.109.241301}.
To do so, they can be constructed to interact strongly with their environment
 in regions of high matter density. Effectively, they become heavy, while in low-density environments, they are much lighter and can effectively be `trapped' in the resonator. This makes chameleons detectable when coupling to electromagnetic fields in low-density or vacuum environments. Chameleons, similar to axions, are expected to couple weakly with photons. This coupling can be exploited to detect chameleons using similar methods developed for axion detection \cite{Ahlers:2007st}.  The first such searches where performed in \cite{Rybka_2010} for haloscopes.
In the helioscope case most recent searches were performed using the CAST magnet \cite{ArguedasCuendis:2019fxj}, a performance perspective for a large cavity in the FLASH experiment can be found in \cite{Alesini:2023qed}.

\subsection{Axions from other astrophysical sources}

Axion helioscopes can, in principle, be directed toward astrophysical objects other than the Sun. However, among these, only galactic core-collapse supernovae (SN) appear capable of producing detectable axion fluxes competitive with solar axions. Recent studies within the IAXO collaboration~\cite{Carenza:2025uib} have shown that a nearby SN event, occurring within a few hundred parsecs, could generate a transient axion flux observable by helioscopes such as IAXO or BabyIAXO—provided these are equipped with dedicated MeV-scale photon detectors. 

Such a detection would constitute a major advance in both axion physics and astrophysics. It would open a new channel for multi-messenger observations of supernovae and provide unique information about the SN core, including the role of pionic processes and the equation of state of dense nuclear matter—quantities that are otherwise inaccessible to laboratory experiments. Importantly, the axion signal from supernovae would probe regions of the axion parameter space, particularly at higher masses, that lie beyond the sensitivity reach of solar axion searches. This implies that supernova axions could be detected even in the absence of a solar axion signal.

To realize this potential, several conditions must be met. A dedicated MeV photon detector would need to be installed—ideally at the end of the magnet opposite to the X-ray optics~\cite{Ge:2020zww}—and the experiment would require an early alert of a supernova event, potentially provided by neutrino-based early warning systems such as SNEWS. The detection of axions from a core-collapse supernova would thus offer a transformative opportunity to probe fundamental particle physics and extreme astrophysical environments simultaneously.

\subsection{Post-discovery physics program}

A confirmed axion detection in a helioscope would open a rich post-discovery research program. First, helioscopes could determine the axion’s \emph{intrinsic parity} by analising the polarization of photons resulting from axion-photon conversion. These photons are expected to be either $p$-polarized (for pseudo-scalars like the axion) or $s$-polarized (for scalars), relative to the magnetic field orientation. The angular distribution of photoelectrons produced in gaseous detectors—following a $\cos^2 \phi$ pattern—can be exploited to reconstruct the photon polarization, something being studied in IAXO.

Second, helioscopes offer a unique opportunity to \emph{decode the axion model} by simultaneously probing the couplings to photons ($g_{a\gamma}$), electrons ($g_{ae}$), and potentially nucleons ($g_{aN}$). Each coupling contributes differently to the solar axion spectrum~\cite{IAXO:2019mpc}, enabling their disentanglement. Lowering the detection threshold below 1\,keV—down to 300\,eV in planned upgrades—significantly enhances sensitivity to $g_{ae}$~\cite{Jaeckel:2019mbt}. Specific nuclear processes, like the de-excitation of $^{57}$Fe~\cite{DiLuzio:2021ysg}, would allow probing $g_{aN}$ and potentially disentangling its proton and neutron components. Under suitable conditions, the axion mass could also be inferred from spectral features related to coherence loss~\cite{Dafni:2018tvj}.

Finally, axions can be powerful \emph{solar probes}. Their flux is sensitive to the internal solar magnetic field~\cite{OHare:2020wah}, temperature profile~\cite{Hoof:2023hyl}, and metal abundances~\cite{Jaeckel:2019bnh}, providing independent insight into the solar composition and potentially addressing the solar abundance problem~\cite{Hoof:2021wpx}. These capabilities render helioscopes uniquely suited for multidisciplinary studies in particle physics, astrophysics, and solar physics beyond axion discovery.

In its turn, a confirmed axion detection in an axion haloscope would provide a precise measurement of the axion mass and demonstrate that axions constitute (at least part of) the dark matter in the Universe. With the axion mass known, detectors could be rapidly optimized to accumulate high-statistics data, enabling precision studies of the signal. In particular, the detailed structure of the axion spectral line —arising from the velocity distribution of axions in the Galactic halo— would become accessible. This would open a novel observational window on the phase-space distribution of dark matter, providing information on Galactic dynamics, halo substructure, and even the merger history of the Milky Way.


\section{Complementarity with other experiments in the field}\label{sec:complementarity}

Beyond the two `pillars' in axion search described so far: haloscopes and helioscopes, and before describing more `indirect' search strategies, let us highlight once more the importance of the light-shining-through-walls approach.
In brief, the concept is to keep a large number of near-optical-wavelength photons stored in an optical cavity before a ‘wall’, i.e., an environment which blocks light of the frequency in use. In the absence of backgrounds, measuring photons beyond that wall would indicate BSM physics: If the photons are stored in a magnetic field in vacuum, the BSM process could be due to photons oscillating into axion-like particles.
This is the principle behind the ALPS-experiment at DESY and its successor ALPS-II \cite{Bahre:2013ywa}.
The importance in this approach lies in the fact that it does not need to make assumptions about, for example, the Dark Matter density. While currently, LSW/ALPS-II cannot compete in its parameter reach with haloscopes and helioscopes (cf. Fig.~\ref{fig:helioscopes}), given a suspected or known axion mass, such an experiment could be optimized in their reach to a certain mass range \cite{Hoof:2024gfk}.

\subsection{Astrophysical and cosmological probes}

Astrophysical searches for axions are powerful because they provide a natural laboratory to test axion properties under extreme conditions that cannot be reached in a laboratory setting or even in our own sun. Especially for this reason, these searches complement the ones described so far. 

Beyond our sun, also other stars and neutron stars can produce axions through well-understood processes discussed partially in the previous section: They can be be created in the dense, high-temperature cores of stars via the Primakoff effect or axion-electron interactions. Strong magnetic and gravitational fields make neutron stars a possible source of X-rays that are created from axion-photon conversion. 
If axions exist, they could also alter the energy loss rates of stars, affecting their lifetimes and observable properties. Prime examples are Red Giants and Horizontal Branch Stars: Excessive axion cooling would shorten their lifetimes. This is constrained by observations of globular clusters.
We also want to highlight efforts to connect axion physics with the gravitational waves physics: For example, using multi-messenger signals of black hole-neutron star binary inspirals to test the axion DM hypothesis, see \cite{PhysRevLett.124.161101}.
Strong constraints are also set by Supernovae (e.g., SN 1987A): If axions were too strongly coupled, they would have drained too much energy from the explosion. This is in conflict with observed neutrino signals.
An exhaustive recent review on the processes mentioned above (including extensive references) can be found in \cite{Caputo:2024oqc}.

In terms of cosmological observables most prominently, axion dark matter can leave imprints on the CMB anisotropies  and polarization (via interaction with CMB photons). Also, ultra-light axions would affect the formation of galaxies and cosmic structures. Ultra-light axions are also relevant in terms of  Black Hole Superradiance \cite{Brito:2015oca}: Axions in a specific mass range  can extract rotational energy from black holes via ``superradiance'', affecting black hole spin distributions. 
A more exhaustive recent review on the axion's impact on cosmology (including extensive references) is provided in \cite{OHare:2024nmr}.

In summary: If axions exist, their effects should be visible across multiple astrophysical sources and cosmological impacts, and provide an important complementary probe to direct axion searches.

\subsection{Non-axion laboratory experiments}

Compared to other experiments searching for popular Dark Matter candidates, axion searches are a relatively recent field and the technique to search them is different and complementary to other `strategic' dark matter searches.
Most prominently,
WIMP (Weakly Interacting Massive Particle) searches have been ongoing since the late 1980s, following the rise of the CDM paradigm and the realization that particles with weak-scale interactions could explain the observed abundance of dark matter through thermal freeze-out in a natural way.
WIMPs are a generic prediction of extensions to the Standard Model, such as supersymmetry, and are normally assumed to be much heavier (GeV to TeV range) than generic axions, see e.g.~\cite{Cirelli:2024ssz}. Direct searches for them  primarily involve nuclear (and more recently electron) recoil detection, which requires a set-up different from the ones used to probe the the axion-photon conversion process. Indirect detection can proceed via their annihilation products, or searches with particle beams (collider and fixed-target).

It is interesting to note that searches with particle beams can tell us something about the possible existence of heavy axions:
Recently, interest is sparking for the search of heavy mediators that enable interactions between the SM and the DM sector and also potentially explain various observations that the SM does not account for.
Collider searches are most sensitive to ALPs above the GeV scale, see, e.g. \cite{Antel:2023hkf}.

Also fixed-target experiments present a promising strategy to search for production and decays of heavy axions  with  masses up to the order of 1 GeV. Their advantage is that they can operate at comparably high intensities in a low-background environment. The mass and coupling range that can be probed directly by fixed-target experiments is of particular interest in models that describe hypothetical mediators between dark matter and SM particles, collectively referred to as Dark Sector portals.  A popular classification of Dark Sector portal benchmark models has been put forward in \cite{Beacham:2019nyx}. The portal complementary to the low-mass axion searches described so far is that of an axion-like particle which couples to the SM fermions and gauge bosons.
For references to theory and phenomenology of such axions, the interested reader is referred to \cite{Jerhot:2022chi}.

Let us close this section by mentioning highlighting a complementary axion probe in the flavour sector:
 While the `vanilla' version of the axion commonly is assumed to be flavour-diagonal, it can have flavour-violating couplings \cite{Wilczek:1982rv,Davidson:1981zd}.
Among the most sensitive searches for such axions in terms of absolute scale of flavour violation is the decay  $K\to \pi a$ \cite{NA62:2020xlg}, an overview of the current constraints in such models can be found in \cite{MartinCamalich:2020dfe}.

\begin{figure}[ht]
	\centering
	\includegraphics[width=13cm,height=6cm]{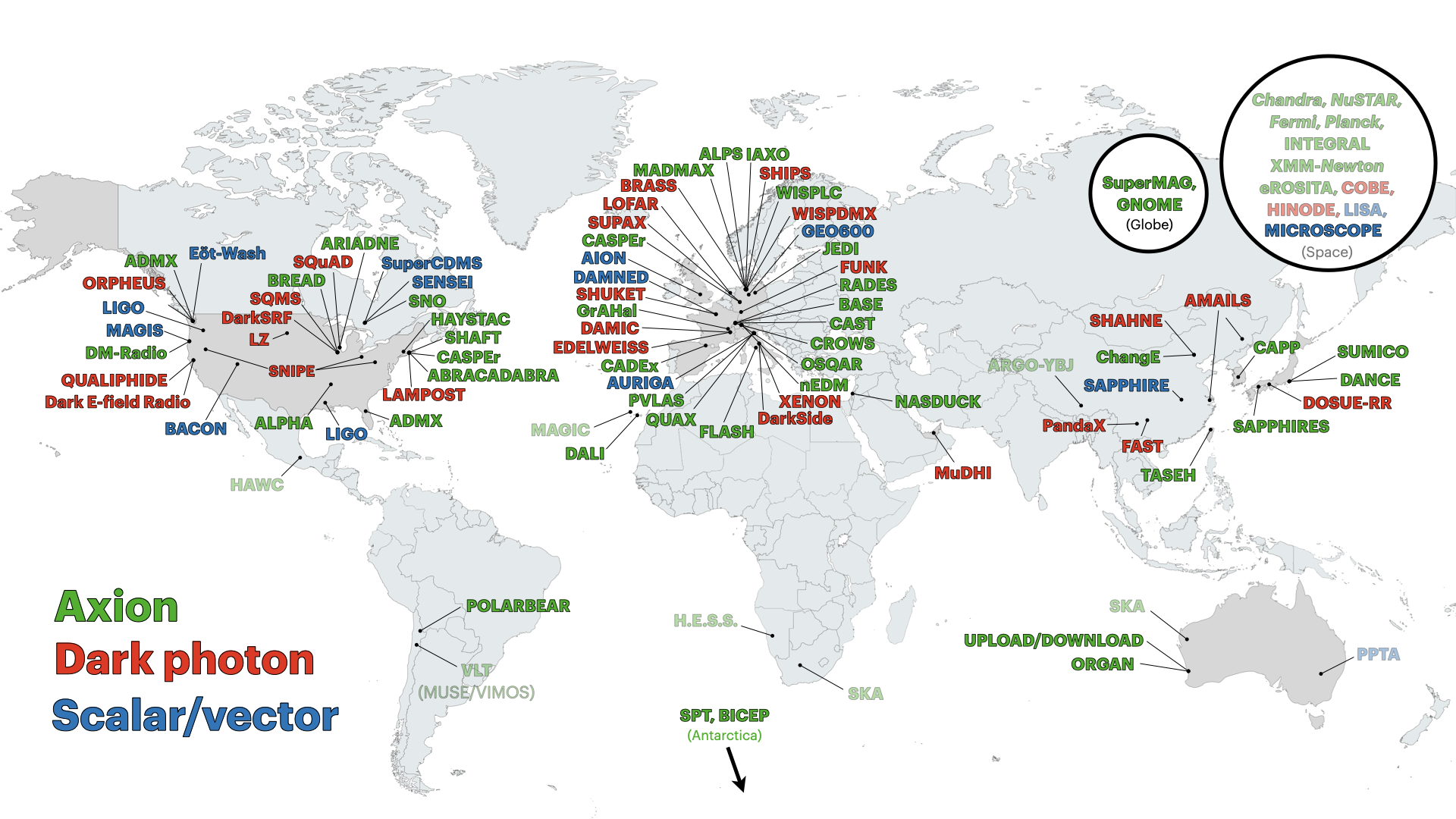}
	\caption{ Map of experiments with sensitivity to axions, Dark Photons or light BSM scalars.
The figure is taken taken from \cite{AxionLimits}, where also references to the various experiments are provided.}
	\label{fig:mapohare}
\end{figure}

In summary, it is interesting to keep in mind that experiments not built to detect axions can say at least something about non-standard variants
Figure \ref{fig:mapohare} shows a world-map with experiments sensitive to axions in the broader sense:

\begin{itemize}
    \item a number of experiments built for a different physics case have parasitic or direct sensitivity to axions (such as XENON)
    \item note that the map is constructed in an inclusive fashion and lists all efforts on a rather equal footing with no ranking on sensitivity or time-scales
    \item the experiments in the map often probe different couplings (photon, electron, nucleon...) and different mass ranges
\end{itemize}

The map is thus meant to illustrate the complementarity of approaches, and the fact that experiments sensitive to axions are running on 5 different continents. This map is thus rather dynamic. The message to take away from is that researchers with different experimental specialization often can contribute in a meaningful way to the search for axions and a diverse set of experiments could be the key to uncover the nature of Dark matter.
It should also not be forgotten that it is easily possible Dark matter consists of a mixture of different BSM particles, for example a mixture of WIMPs or primordial black holes and axions.

\section{Future development}\label{future}

Looking into the glass ball for axion searches, the most certain statement that can be made is probably that efforts are ramping up in all possible directions of `untouched' parameter space: Going to larger masses, Sect.~\ref{sec:highmass} (motivated by computations of the postinflationary axion mass, see references maintained at \cite{AxionLimits}), going to lower masses, Sect.~\ref{sec:lowmass} (motivated by vanilla axion models but also models of ultra-light ALPs), and going down to smaller couplings by improving experimental parameters such as the cavity $Q$ (Sect.~\ref{sec:superconductors}), or all-together the detection method (Sect~\ref{sec_quantum}))
In the following we present a selection of technologies and concepts that are likely relevant in the near- and mid-term future.

\subsection{Extending the parameter reach: larger masses \label{sec:highmass}}

In the `classical' haloscope setup, a limitation is a priory, that reaching higher masses (resonance frequencies) means  shrinking the cavity which negatively effects the volume and quality factor. Several developments are ongoing to overcome this issue and often requires altogether different detection concepts. 

Sticking to the `calssical haloscope concept' one attempt can be to go to multi-cavity structures, i.e. cavity structures where the cavity diameters do not set the frequency resonance scale \cite{Melcon:2018dba,Melcon:2020xvj,Jeong:2017hqs,Jeong:2020cwz}, but more complicated cavity structures can lead also to complications in designing a tuning mechanism. A viable option is also the `conducting rod resonator' (as employed by ORGAN \cite{Quiskamp:2024oet}).

Most relevant to higher masses is thus the concept of the  \textit{magnetized dish antenna} and its evolution, the \textit{dielectric haloscope}. A dielectric interface (e.g. a mirror, or the surface of a dielectric slab) immersed in a magnetic field parallel to the surface should emit electromagnetic radiation perpendicular to its surface, due to the presence of the dark matter axion field~\cite{Horns:2012jf}. This tiny signal can be made detectable if the emission of a large surface is made to concentrate in a small point, like e.g. in the case of the surface having a spherical shape. This technique has the advantage of being broad-band, with sensitivity to all axion masses at once~\footnote{In practice this is limited by the bandwidth of the photon sensor being used.}. This technique is being followed by the BRASS~\cite{BRASSweb} collaboration at U. of Hamburg, as well as G-LEAD at CEA/Saclay~\cite{PC_brun}. Many of the existing reflector geometries are difficult to
combine with a strong magnetic field (keeping field lines parallel to surface everywhere, a  dish-antenna  that tries to overcome this problem and is adapted to solenoidal geometry is pursued by BREAD with first ALP limits reported in \cite{Hoshino:2025fiz}.

Given that no resonance is involved in this scheme, very large areas are needed to obtain competitive sensitivities. Dielectric haloscopes are an evolution of this concept, in which several dielectric slabs are stacked together inside a magnetic field and placed in front of a metallic mirror. This increases the number of emitting surfaces and, in addition, constructive interference among the different emitted (and reflected) waves can be achieved for a frequency band if the disks are adjusted at precise positions. This effectively amplifies the resulting signal. The MADMAX collaboration~\cite{TheMADMAXWorkingGroup:2016hpc} plans to implement such a concept
in a 10 T B-field,  leading to a boost in power emitted by the system of a $>10^{4}$ with respect to a single metallic mirror in a relatively broad frequency band of 50 MHz. By adjusting the spacing between the discs the frequency range in which the boost occurs can be adjusted, with the goal of scanning an axion mass range between 40 and 400 $\mu$eV. 
The experiment is planned to be sited at DESY. A first smaller-scale demonstrating prototype \cite{Garcia:2024xzc} has been operated in the MORPURGO magnet at CERN, and a larger one will be built, before jumping to the full size experiment. 

Finally, let us mention that an implementation of the dielectric haloscope concept but at even higher frequencies (LAMPPOST) has been discussed in the literature, with potential sensitivity to 0.2 eV axions and above~\cite{Baryakhtar:2018doz}. 
First results searching for Dark Photons using this concept in combination with superconducting nano-wires \cite{Chiles:2021gxk} outline the promising capability of this idea. 

Another recent proposal to detect axion DM at high  mass values involves the use of certain antiferromagnetic topological insulators~\cite{Marsh:2018dlj,Schutte-Engel:2021bqm}. Such materials contain axion quasiparticles (AQs), that are longitudinal antiferromagnetic spin fluctuations. These AQs have similar dynamics to the axion field, including a mass mixing with the electric field in the presence of magnetic fields. The dispersion relation and boundary conditions permit resonant conversion of axion DM into THz photons in a way that is independent of the resonant frequency. An advantage of this method is the tunability of the resonance with applied magnetic field. The technique could be competitive in the search for DM axions of masses in the 1 to 10 meV range. 

A promising strategy are also the ``plasma haloscopes'', in which the resonant conversion is achieved by matching the axion mass to a plasma frequency. The advantage of this approach is that the plasma frequency is unrelated to the physical size of the device, allowing large conversion volumes. A concrete proposal using wire metamaterials as the plasma, with the plasma frequency tuned by varying the interwire spacing, points to potentially competitive sensitivity for axion masses at $35-400$~eV~\cite{Lawson:2019brd}.
The ALPHA collaboration \cite{ALPHA:2022rxj} proposes to build a plasma haloscope from wire metamaterials in which the full-scale experiment could discover QCD axions over almost a decade of parameter space.

\subsection{Extending the parameter reach: developments towards lower masses \label{sec:lowmass}}

Conversely to the high-mass situation in `classical' haloscope setup, the limitation to reach low masses (resonance frequencies) means expanding the cavity which becomes impractical at a certain scale. 
At the few hundred MHz scale, when a huge magnetic volume is available, such searches are still viable. Proposals for haloscope efforts in this frequency range exist for example by using the babyIAXO magnet as a haloscope \cite{Ahyoune:2023gfw} with RADES or re-purposing the FINUDA magnet \cite{Alesini:2023qed} with FLASH.
 This parameter space is also targeted by efforts in ADMX, see \cite{Chakrabarty:2023rha}, as well as the hybrid magnet system in Grenoble \cite{Pugnat:2024sxb}.

At a certain scale however, the size of magnets saturates and developments are ongoing to overcome this issue using novel detection concepts. 

At the very low masses, DM axions can produce an oscillation of the optical linear polarization of a laser beam in a bow-tie cavity. The DANCE experiment has already provided proof-of-concept results~\cite{Oshima:2021irp} with a table-top setup, while large potential for improvement exists in scale-up projections.

The techniques mentioned above are all based on the axion-photon coupling. If the axion has relevant fermionic couplings, the axion DM field would couple with nuclear spins like a fictitious magnetic field and produce the precession of nuclear spins. Moreover, by virtue of the same Peccei-Quinn term that solves the strong CP problem, the DM axion field should induce oscillating electric-dipole-moments (EDM) in the nuclei. Both effects can be searched for by nuclear magnetic resonance (NMR) methods. The CASPEr project~\cite{Budker:2013hfa,JacksonKimball:2017elr} is exploring several NMR-based implementations to search for axion DM along these directions. The prospects of the technique may reach relevant QCD models for very low axion masses ($\lesssim 10^{-8}$eV). A conceptually similar concept is done by the QUAX experiment, but invoking the electron coupling using magnetic materials~\cite{Barbieri:2016vwg}. In this case, the sample is inserted in a resonant cavity and the spin-precession resonance hybridises with the electromagnetic mode of the cavity. The experiment focuses on a particular axion mass $m_a\sim200~\mu$eV, but sensitivity to QCD models will require lowering the detection noise below the quantum limit. The recent experiment NASDUCK~\cite{Bloch:2021vnn} has reported competitive limits on $g_{ap}$ and $g_{an}$ from ALP DM interacting with atomic spins, using a quantum detector based on spin-polarized xenon gas. Another technique recently proposed is to search for the axion/ALP induced EDM in the future proton storage ring develop to measure the static proton EDM~\cite{Chang:2017ruk}.

DM axions can produce atomic excitations in a target material to levels with an energy difference equal to the axion mass. This can again happen via the axion interactions to the nuclezi or electron spins. The use of the Zeeman effect has been proposed~\cite{Sikivie:2014lha} to split the ground state of atoms to effectively create atomic transition of energy levels that are tunable to the axion mass, by changing the external magnetic field. 

The AXIOMA~\cite{1367-2630-17-11-113025,Braggio:2017oyt} project has started feasibility studies to experimentally implement this detection concept. Sensitivity to axion models (with fermion couplings) in the ballpark of $10^{-4}-10^{-3}$~eV could eventually be achieved if target materials of $\sim$kg mass are instrumented and cooled down to mK temperatures. 
For a more thorough review of the possibilities that atomic physics offer to axion physics we refer to section 1.4 of Ref.~\cite{Agrawal:2021dbo}.

For very low axion masses (well below $\mu$eV), it may be more effective to attempt the detection of the tiny oscillating $B$-field associated with the axion dark matter field in an external constant magnetic field, by means of a carefully placed pick-up coil inside a large magnet~\cite{Sikivie:2013laa,Chaudhuri:2014dla,Kahn:2016aff}. Resonance amplification can be achieved externally by an $LC$-circuit, which makes tuning in principle easier than in conventional haloscopes. A broad-band non-resonant mode of operation is also possible ~\cite{Kahn:2016aff}.  Several teams are studying implementations of this concept~\cite{Silva-Feaver:2016qhh,Kahn:2016aff}. Two of them, the ABRACADABRA~\cite{Ouellet:2018beu,Salemi:2021gck} and SHAFT~\cite{Gramolin:2020ict} experiments, have recently released results with small table-top demonstrators, reaching sensitivities similar to the CAST bound for masses in the 10$^{-11}-$10$^{-8}$~eV range. Another similar implementation, that of BEAST~\cite{McAllister:2018ndu}, has obtained better sensitivities in a narrower mass range around 10$^{-11}$~eV, although its principle has been doubted by the community~\cite{Ouellet:2018nfr,Beutter:2018xfx}. Similarly, the more recent result from the ADMX SLIC pilot experiment has probed a few narrow regions around $2\times10^{-7}$~eV and down to $\sim 10^{-12}$~GeV$^{-1}$\cite{Crisosto:2019fcj}. Finally, the BASE experiment, whose main goal is the study of antimatter at CERN, has recently released a result adapting its setup to the search of axions following this concept~\cite{Devlin:2021fpq}. In general, this technique could reach sensitivity down to the QCD axion for masses $m_a \lesssim 10^{-6}$~eV, if implemented in magnet volumes of few~m$^3$ volumes and a few~T fields.

\subsection{Quantum sensing for axion haloscopes \label{sec_quantum}}

Eq. (\ref{eq:haloscope_fom}) illustrates the improvement in sensitivity for lower system noise temperature $T_{\rm sys}$. Typically,  $T_{\rm sys}$ is  the sum of the physical temperature of the cavity, which sets the thermal photon background, and the sensor’s intrinsic noise: $T_{\rm sys} = T_{\rm phys} + T_{\rm sen}$. Simply cooling the cavity to lower $T_{\rm phys}$ enhances sensitivity only if the sensor’s noise $T_{\rm sen}$ is also sufficiently suppressed. This can be achieved by employing quantum-limited amplifiers based on quantum circuits. Prominent examples include Josephson Parametric Amplifiers (JPAs), Superconducting Quantum Interference Devices (SQUIDs), and Traveling Wave Parametric Amplifiers (TWPAs), although other solid-state circuit designs may also be applicable. 

However, $T_{\rm phys}$ itself is fundamentally constrained by the Standard Quantum Limit (SQL)~\cite{PhysRevD.26.1817,PhysRev.128.2407} defined as $T_{\rm SQL} \sim hf/k_{\rm B}$ (for a frequency $f$, being $h$ and $k_{\rm B}$ the Planck and  Boltzmann constants). As a reference, for $f \sim 10$~GHz, $T_{\rm SQL} \sim 500$~mK. When operating below $T_{\rm SQL}$, the cavity is depleted of photons (its photon occupation number is below unity), and, by virtue of the Heisenberg uncertainty principle, any conventional heterodyne detection scheme will measure a minimum effective power equal to one photon due to the vacuum fluctuations~\cite{Lamoreaux:2013koa}. Leading axion dark matter experiments like ADMX and CAPP are already routinely operating close to $T_{\rm SQL}$, thanks to quantum-limited amplifiers like JPAs or TWPAs.

Even the SQL can be surpassed by novel quantum metrology techniques. One option is preparing the vacuum state of the cavity is a squeezed state, which is a quantum state of light in which one of the quadratures features reduced variance (while the other has a larger-than-normal variance, in order to respect Heisenberg principle). By injecting squeezed states into the detection chain using a Josephson Parametric Amplifier (JPA), one can effectively suppress quantum noise in the measured quadrature. The HAYSTAC experiment has pioneered the use of a squeezed-vacuum state receiver to enhance sensitivity in axion searches. This led to an improvement in scan rate by a factor of approximately 2 compared to operation at the SQL, without increasing integration time~\cite{Malnou:2018dxn,HAYSTAC:2024squeezed}. This quantum enhancement has allowed HAYSTAC to achieve near-DFSZ sensitivity in the 16.96--17.28~$\mu$eV mass range.

\begin{figure}
    \centering
    \includegraphics[width=0.5\linewidth]{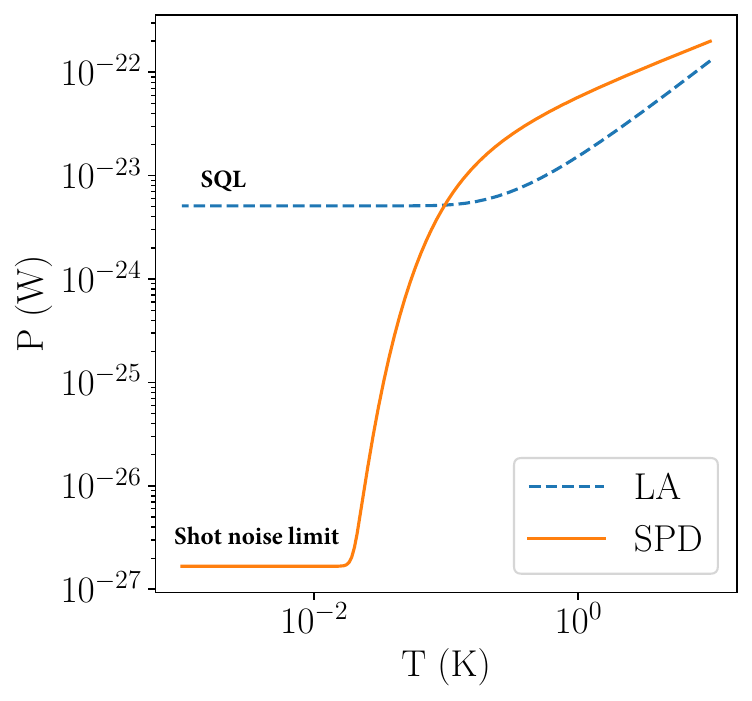}
    \caption{Minimum detectable power $P$ versus temperature $T$ by single-photon detection (SPD, solid orange line) or linear amplification (LA, dashed blue line), assuming the ideal situation in which only thermal emission contributes to the noise. LA-based detection is limited by the SQL for $T<T_{\rm SQL}$, while SPD is not. 
 The plot has been computed for parameters similar to RADES \cite{Ahyoune:2024klt}, with $Q=10^{5}$ and an integration time of 10{$^4$} s.}
    \label{fig:LA_vs_SQL}
\end{figure}

A more radical approach to surpassing the SQL involves the use of single microwave photon detectors (SMPD) to directly measure axion-induced photons in the cavity. In this scheme, sensitivity to photon number is maximized by accepting complete uncertainty in the phase—allowing measurement precision to be limited only by Poissonian shot noise, rather than quantum fluctuations. In the ideal case of negligible intrinsic dark counts, such detectors could lower the effective noise temperature $T_{\rm phys}$ by several orders of magnitude compared to the SQL, as illustrated in Fig.~\ref{fig:LA_vs_SQL}. Since $T_{\rm SQL}$ increases with frequency, the relative advantage of single-photon counters becomes more pronounced at higher masses. It has been shown that, assuming typical haloscope parameters, single-photon detection becomes competitive—and ultimately preferred—over quantum-limited linear amplification above $\sim$10~GHz~\cite{Lamoreaux:2013koa}.

In practice, counting microwave photons with sufficiently low dark count rate, and in the high-$B$-field environment of an axion haloscope is a considerable challenge.  Key research directions currently focus on developing viable experimental implementations of these technologies. Pioneering efforts such as the CARRACK experiment explored the use of Rydberg atoms for single-photon detection~\cite{Yamamoto:2000si}, though the R\&D has since been discontinued. Recently, significant progress has been made in developing quantum devices with single photon sensitivity at (or close to) microwave energies. Some examples are transition-edge sensors (TES)~\cite{pepe2024TES,Paolucci:2021kle} and kinetic inductance detectors (KIDs)~\cite{smith2024scaling}, graphene-based calorimeters~\cite{huang2024graphene}, superconducting nanowire single-photon detectors (SNSPDs)~\cite{natarajan2012superconducting} or different implementations using superconducting qubits (transmons)~\cite{Flurin:2021smpd,Dixit:2020ymh}. Not all these works have achieved single-photon sensitivity to sufficiently low frequencies, or with sufficiently low dark count rate, but continuos efforts are ongoing. The SMPD technique in \cite{Flurin:2021smpd} has been aso far the only one successfully integrated in an operating axion haloscope~\cite{Braggio:2024xed}, reaching a dark count rate a factor $\sim$20 below SQL.

\subsection{Superconductors for cavities \label{sec:superconductors}}

The figure of merit in haloscopes increases by the power of four with the strength of the magnetic field, and linearly with the quality factor, see Eqn. \ref{eq:haloscope_fom}. Therefore we aim at magnets with fields as high as possible, as well as large quality factors. To profit from both these factors simultaneously, one can deduct  requirements for the ideal coating of a haloscope cavity: we needed a type II superconductor with a critical magnetic field $B_{c2}$ well above several Tesla at few Kelvin temperatures and below. To profit, coatings should also possess a RF surface resistance {R$_s$} lower than copper at such conditions.

 The pathway to exploit such coatings in haloscopes has already started, but future work and further improvements are likely:

 \begin{itemize}
     \item high-temperature superconductors like REBCO (Rare-earth barium copper oxide), can be commercially purchased in the form of tapes. A hastelloy substrate is stripped off, such that the REBCO layer is exposed to the radiofrequency fields. Depending on the cavity frequency and field strength, cavity quality factor improvements of factors up to 10 compared to copper are obtained with cavities of flat sections \cite{Ahn:2021fgb}, or less with rounded surfaces \cite{Golm:2021ooj,Ahyoune:2024klt}.
     Such HTS-based superconductors are also synergetic with developments/research for future accelerators \cite{Romanov:2020epk}.
     \item  Using Niobium-based superconductors in haloscopes, like Nb$_3$Sn \cite{Golm:2021ooj} and Nb \cite{Schmieden:2024wqp}, NbTi \cite{Alesini:2019ajt} offer very high quality factor improvements and can be sputtered on rounded surfaces. However here the quality factor quickly degrades with increasing magnetic fields and thus is only worth up to below $\sim$1 Tesla
 \end{itemize}

\section{Conclusions}
\label{sec:conclusions}

In recent years new physics searches (be it direct or indirect), have excluded large portions of prime dark matter candidate parameter space. While it can be frustrating to `open a signal box' with no or too few events inside, the take-away message from this could instead be positive: There is clear evidence for dark matter but the low-hanging fruits are not its Nature! We have to put even more effort:
This is an exciting time for axion physics because of major theoretical, experimental, and technological advancements that are rapidly expanding our ability to detect axions and explore their role in fundamental physics.

From Figures \ref{fig:haloscopes} and ~\ref{fig:helioscopes}  which show experimental and observational bounds on the $\gagamma$-$m_a$ plane, it can be seen that probing the `simplest incarnations' of axion Dark Matter and the most interesting target region for solar axions is within reach.

Young researchers and students in axion physics can profit from working with
\begin{itemize}
    \item a diverse set set of novel, partially cutting-edge technologies
    \item in comparably small teams/collaborations where it is common to understand most details of the experimental setup and the analysis methods in depth
    \item in an environment where experimental and theoretical cross-talk is strong and it is not uncommon for a researcher to publish in both domains
\end{itemize}

If interest in axion physics (Fig~\ref{fig:titlepage} !) and technological progress keep their pace, the coming years will bring us much closer to discover or excluding axions.

\begin{ack}[Acknowledgments]%
 We thank M. Jiménez for help updating plots~\ref{fig:helioscopes} and ~\ref{fig:haloscopes}.
BD acknowledges funding through the European Research Council under grant ERC-2018-StG-802836 (AxScale) and by Deutsche Forschungsgemeinschaft (DFG) through Grant No. 532766533..
This research was also supported under Horizon Europe programme (ERC-2023-SyG DarkQuantum, grant agreement No. 101118911).
IGI acknowledges funding from the European Research Council (ERC) under the European Union’s Horizon 2020 research and innovation programme (ERC-2017-AdG IAXO+, grant agreement No. 788781). IGI also acknowledges funding from the Agencia Estatal de Investigación (AEI) under the grant agreement PID2022-137268NB-C51 funded by MCIN/AEI/10.13039/501100011033/FEDER, as well as funds from “European Union NextGenerationEU/PRTR” (Planes complementarios, Programa de Astrofísica y Física de Altas Energías).
\end{ack}


\bibliographystyle{Numbered-Style} 
\bibliography{axionbib}

\begin{thebibliography*}{100}
\providecommand{\bibtype}[1]{}
\providecommand{\url}[1]{{\tt #1}}
\providecommand{\urlprefix}{URL }
\expandafter\ifx\csname urlstyle\endcsname\relax
  \providecommand{\doi}[1]{doi:\discretionary{}{}{}#1}\else
  \providecommand{\doi}{doi:\discretionary{}{}{}\begingroup
  \urlstyle{rm}\Url}\fi
\providecommand{\bibinfo}[2]{#2}
\providecommand{\eprint}[2][]{\url{#2}}
\makeatletter\def\@biblabel#1{\bibinfo{label}{[#1]}}\makeatother

\bibtype{Article}%
\bibitem{Maximilien:2767149}
\bibinfo{author}{Brice Maximilien}, \bibinfo{title}{{The RADES detector at the
  CAST experiment}}, \bibinfo{journal}{CDS database}  (\bibinfo{year}{2021}),
  \bibinfo{note}{general Photo},
  \bibinfo{url}{\urlprefix\url{https://cds.cern.ch/record/2767149}}.

\bibtype{Article}%
\bibitem{Cirelli:2024ssz}
\bibinfo{author}{Marco Cirelli}, \bibinfo{author}{Alessandro Strumia},
  \bibinfo{author}{Jure Zupan}, \bibinfo{title}{{Dark Matter}},
  \bibinfo{journal}{to appear}  (\bibinfo{year}{2024}), \eprint{2406.01705}.

\bibtype{Article}%
\bibitem{Balazs:2024uyj}
\bibinfo{author}{Csaba Balazs}, \bibinfo{author}{Torsten Bringmann},
  \bibinfo{author}{Felix Kahlhoefer}, \bibinfo{author}{Martin White},
  \bibinfo{title}{{A Primer on Dark Matter}}, \bibinfo{journal}{to appear}
  (\bibinfo{year}{2024}), \eprint{2411.05062}.

\bibtype{Article}%
\bibitem{OHare:2024nmr}
\bibinfo{author}{Ciaran A.~J. O'Hare}, \bibinfo{title}{{Cosmology of axion dark
  matter}}, \bibinfo{journal}{PoS} \bibinfo{volume}{COSMICWISPers}
  (\bibinfo{year}{2024}) \bibinfo{pages}{040},
  \bibinfo{doi}{\doi{10.22323/1.454.0040}}, \eprint{2403.17697}.

\bibtype{Article}%
\bibitem{DiLuzio:2020wdo}
\bibinfo{author}{Luca Di~Luzio}, \bibinfo{author}{Maurizio Giannotti},
  \bibinfo{author}{Enrico Nardi}, \bibinfo{author}{Luca Visinelli},
  \bibinfo{title}{{The landscape of QCD axion models}}, \bibinfo{journal}{Phys.
  Rept.} \bibinfo{volume}{870} (\bibinfo{year}{2020}) \bibinfo{pages}{1--117},
  \bibinfo{doi}{\doi{10.1016/j.physrep.2020.06.002}}, \eprint{2003.01100}.

\bibtype{Article}%
\bibitem{Sikivie:2024isv}
\bibinfo{author}{Pierre Sikivie}, \bibinfo{title}{{Axion dark matter}},
  \bibinfo{journal}{Nucl. Phys. B} \bibinfo{volume}{1003}
  (\bibinfo{year}{2024}) \bibinfo{pages}{116500},
  \bibinfo{doi}{\doi{10.1016/j.nuclphysb.2024.116500}}.

\bibtype{Article}%
\bibitem{Yu:2023gdq}
\bibinfo{author}{Felix Yu}, \bibinfo{title}{{Primer on Axion Physics}},
  \bibinfo{journal}{Annalen Phys.} \bibinfo{volume}{536} (\bibinfo{number}{1})
  (\bibinfo{year}{2024}) \bibinfo{pages}{2300106},
  \bibinfo{doi}{\doi{10.1002/andp.202300106}}, \eprint{2308.08612}.

\bibtype{Article}%
\bibitem{Irastorza:2021tdu}
\bibinfo{author}{Igor~Garcia Irastorza}, \bibinfo{title}{{An introduction to
  axions and their detection}}, \bibinfo{journal}{SciPost Phys. Lect. Notes}
  \bibinfo{volume}{45} (\bibinfo{year}{2022}) \bibinfo{pages}{1},
  \bibinfo{doi}{\doi{10.21468/SciPostPhysLectNotes.45}}, \eprint{2109.07376}.

\bibtype{Article}%
\bibitem{Irastorza:2018dyq}
\bibinfo{author}{Igor~G. Irastorza}, \bibinfo{author}{Javier Redondo},
  \bibinfo{title}{{New experimental approaches in the search for axion-like
  particles}}, \bibinfo{journal}{Prog. Part. Nucl. Phys.} \bibinfo{volume}{102}
  (\bibinfo{year}{2018}) \bibinfo{pages}{89--159},
  \bibinfo{doi}{\doi{10.1016/j.ppnp.2018.05.003}}, \eprint{1801.08127}.

\bibtype{Inproceedings}%
\bibitem{Adams:2022pbo}
\bibinfo{author}{C.~B. Adams}, et al., \bibinfo{title}{{Axion Dark Matter}},
  in: \bibinfo{booktitle}{{Snowmass 2021}} \bibinfo{year}{2022}, p.
  \bibinfo{pages}{all}, \eprint{2203.14923}.

\bibtype{Article}%
\bibitem{Baryakhtar:2025jwh}
\bibinfo{author}{Masha Baryakhtar}, \bibinfo{author}{Leslie Rosenberg},
  \bibinfo{author}{Gray Rybka}, \bibinfo{title}{{Searching for the QCD Dark
  Matter Axion}}, \bibinfo{journal}{to appear}  (\bibinfo{year}{2025}),
  \eprint{2504.10607}.

\bibtype{Article}%
\bibitem{Peccei:1977hh}
\bibinfo{author}{R.~D. Peccei}, \bibinfo{author}{Helen~R. Quinn},
  \bibinfo{title}{{CP conservation in the Presence of Instantons}},
  \bibinfo{journal}{Phys. Rev. Lett.} \bibinfo{volume}{38}
  (\bibinfo{year}{1977}) \bibinfo{pages}{1440--1443},
  \bibinfo{doi}{\doi{10.1103/PhysRevLett.38.1440}}.

\bibtype{Article}%
\bibitem{Peccei:1977ur}
\bibinfo{author}{R.~D. Peccei}, \bibinfo{author}{Helen~R. Quinn},
  \bibinfo{title}{Constraints imposed by {CP} conservation in the presence of
  instantons}, \bibinfo{journal}{Phys. Rev.} \bibinfo{volume}{D16}
  (\bibinfo{year}{1977}) \bibinfo{pages}{1791--1797}.

\bibtype{Article}%
\bibitem{Weinberg:1977ma}
\bibinfo{author}{Steven Weinberg}, \bibinfo{title}{{A New Light Boson?}},
  \bibinfo{journal}{Phys. Rev. Lett.} \bibinfo{volume}{40}
  (\bibinfo{year}{1978}) \bibinfo{pages}{223--226},
  \bibinfo{doi}{\doi{10.1103/PhysRevLett.40.223}}.

\bibtype{Article}%
\bibitem{Wilczek:1977pj}
\bibinfo{author}{Frank Wilczek}, \bibinfo{title}{{Problem of Strong P and T
  Invariance in the Presence of Instantons}}, \bibinfo{journal}{Phys. Rev.
  Lett.} \bibinfo{volume}{40} (\bibinfo{year}{1978}) \bibinfo{pages}{279--282},
  \bibinfo{doi}{\doi{10.1103/PhysRevLett.40.279}}.

\bibtype{Article}%
\bibitem{Preskill:1982cy}
\bibinfo{author}{John Preskill}, \bibinfo{author}{Mark~B. Wise},
  \bibinfo{author}{Frank Wilczek}, \bibinfo{title}{{Cosmology of the Invisible
  Axion}}, \bibinfo{journal}{Phys. Lett. B} \bibinfo{volume}{120}
  (\bibinfo{year}{1983}) \bibinfo{pages}{127--132},
  \bibinfo{doi}{\doi{10.1016/0370-2693(83)90637-8}}.

\bibtype{Article}%
\bibitem{Abbott:1982af}
\bibinfo{author}{L.~F. Abbott}, \bibinfo{author}{P. Sikivie},
  \bibinfo{title}{{A Cosmological Bound on the Invisible Axion}},
  \bibinfo{journal}{Phys. Lett. B} \bibinfo{volume}{120} (\bibinfo{year}{1983})
  \bibinfo{pages}{133--136}, \bibinfo{doi}{\doi{10.1016/0370-2693(83)90638-X}}.

\bibtype{Article}%
\bibitem{Dine:1982ah}
\bibinfo{author}{Michael Dine}, \bibinfo{author}{Willy Fischler},
  \bibinfo{title}{{The Not So Harmless Axion}}, \bibinfo{journal}{Phys. Lett.
  B} \bibinfo{volume}{120} (\bibinfo{year}{1983}) \bibinfo{pages}{137--141},
  \bibinfo{doi}{\doi{10.1016/0370-2693(83)90639-1}}.

\bibtype{Article}%
\bibitem{Sikivie:1983ip}
\bibinfo{author}{P. Sikivie}, \bibinfo{title}{{Experimental tests of the
  invisible axion}}, \bibinfo{journal}{Phys. Rev. Lett.} \bibinfo{volume}{51}
  (\bibinfo{year}{1983}) \bibinfo{pages}{1415},
  \bibinfo{doi}{\doi{10.1103/PhysRevLett.51.1415}}.

\bibtype{Article}%
\bibitem{Wuensch:1989sa}
\bibinfo{author}{Walter Wuensch}, \bibinfo{author}{S. De~Panfilis-Wuensch},
  \bibinfo{author}{Y.~K. Semertzidis}, \bibinfo{author}{J.~T. Rogers},
  \bibinfo{author}{A.~C. Melissinos}, \bibinfo{author}{H.~J. Halama},
  \bibinfo{author}{B.~E. Moskowitz}, \bibinfo{author}{A.~G. Prodell},
  \bibinfo{author}{W.~B. Fowler}, \bibinfo{author}{F.~A. Nezrick},
  \bibinfo{title}{{Results of a Laboratory Search for Cosmic Axions and Other
  Weakly Coupled Light Particles}}, \bibinfo{journal}{Phys. Rev. D}
  \bibinfo{volume}{40} (\bibinfo{year}{1989}) \bibinfo{pages}{3153},
  \bibinfo{doi}{\doi{10.1103/PhysRevD.40.3153}}.

\bibtype{Article}%
\bibitem{Lazarus:1992ry}
\bibinfo{author}{D.~M. Lazarus}, et al., \bibinfo{title}{{A Search for solar
  axions}}, \bibinfo{journal}{Phys. Rev. Lett.} \bibinfo{volume}{69}
  (\bibinfo{year}{1992}) \bibinfo{pages}{2333--2336},
  \bibinfo{doi}{\doi{10.1103/PhysRevLett.69.2333}}.

\bibtype{Article}%
\bibitem{Moriyama:1998kd}
\bibinfo{author}{Shigetaka Moriyama}, et al., \bibinfo{title}{{Direct search
  for solar axions by using strong magnetic field and X-ray detectors}},
  \bibinfo{journal}{Phys. Lett.} \bibinfo{volume}{B434} (\bibinfo{year}{1998})
  \bibinfo{pages}{147}, \bibinfo{doi}{\doi{10.1016/S0370-2693(98)00766-7}},
  \eprint{hep-ex/9805026}.

\bibtype{Article}%
\bibitem{Inoue:2002qy}
\bibinfo{author}{Yoshizumi Inoue}, et al., \bibinfo{title}{{Search for
  sub-electronvolt solar axions using coherent conversion of axions into
  photons in magnetic field and gas helium}}, \bibinfo{journal}{Phys. Lett.}
  \bibinfo{volume}{B536} (\bibinfo{year}{2002}) \bibinfo{pages}{18--23},
  \bibinfo{doi}{\doi{10.1016/S0370-2693(02)01822-1}},
  \eprint{astro-ph/0204388}.

\bibtype{Article}%
\bibitem{deSalas:2020hbh}
\bibinfo{author}{Pablo~F. de Salas}, \bibinfo{author}{Axel Widmark},
  \bibinfo{title}{{Dark matter local density determination: recent observations
  and future prospects}}, \bibinfo{journal}{Rept. Prog. Phys.}
  \bibinfo{volume}{84} (\bibinfo{number}{10}) (\bibinfo{year}{2021})
  \bibinfo{pages}{104901}, \bibinfo{doi}{\doi{10.1088/1361-6633/ac24e7}},
  \eprint{2012.11477}.

\bibtype{Article}%
\bibitem{Lentz:2017aay}
\bibinfo{author}{Erik~W. Lentz}, \bibinfo{author}{Thomas~R. Quinn},
  \bibinfo{author}{Leslie~J. Rosenberg}, \bibinfo{author}{Michael~J. Tremmel},
  \bibinfo{title}{{A New Signal Model for Axion Cavity Searches from N-Body
  Simulations}}, \bibinfo{journal}{Astrophys. J.} \bibinfo{volume}{845}
  (\bibinfo{number}{2}) (\bibinfo{year}{2017}) \bibinfo{pages}{121},
  \bibinfo{doi}{\doi{10.3847/1538-4357/aa80dd}}, \eprint{1703.06937}.

\bibtype{Article}%
\bibitem{Sikivie:1995dp}
\bibinfo{author}{Pierre Sikivie}, \bibinfo{author}{I.~I. Tkachev},
  \bibinfo{author}{Yun Wang}, \bibinfo{title}{{The Velocity peaks in the cold
  dark matter spectrum on earth}}, \bibinfo{journal}{Phys. Rev. Lett.}
  \bibinfo{volume}{75} (\bibinfo{year}{1995}) \bibinfo{pages}{2911--2915},
  \bibinfo{doi}{\doi{10.1103/PhysRevLett.75.2911}}, \eprint{astro-ph/9504052}.

\bibtype{Article}%
\bibitem{Chakrabarty:2020qgm}
\bibinfo{author}{Sankha~S. Chakrabarty}, \bibinfo{author}{Yaqi Han},
  \bibinfo{author}{Anthony~H. Gonzalez}, \bibinfo{author}{Pierre Sikivie},
  \bibinfo{title}{{Implications of triangular features in the Gaia skymap for
  the Caustic Ring Model of the Milky Way halo}}, \bibinfo{journal}{Phys. Dark
  Univ.} \bibinfo{volume}{33} (\bibinfo{year}{2021}) \bibinfo{pages}{100838},
  \bibinfo{doi}{\doi{10.1016/j.dark.2021.100838}}, \eprint{2007.10509}.

\bibtype{Article}%
\bibitem{OHare:2023rtm}
\bibinfo{author}{Ciaran A.~J. O'Hare}, \bibinfo{author}{Giovanni Pierobon},
  \bibinfo{author}{Javier Redondo}, \bibinfo{title}{{Axion Minicluster Streams
  in the Solar Neighborhood}}, \bibinfo{journal}{Phys. Rev. Lett.}
  \bibinfo{volume}{133} (\bibinfo{number}{8}) (\bibinfo{year}{2024})
  \bibinfo{pages}{081001}, \bibinfo{doi}{\doi{10.1103/PhysRevLett.133.081001}},
  \eprint{2311.17367}.

\bibtype{Article}%
\bibitem{Kim:2020kfo}
\bibinfo{author}{Dongok Kim}, \bibinfo{author}{Junu Jeong},
  \bibinfo{author}{SungWoo Youn}, \bibinfo{author}{Younggeun Kim},
  \bibinfo{author}{Yannis~K. Semertzidis}, \bibinfo{title}{{Revisiting the
  detection rate for axion haloscopes}}, \bibinfo{journal}{JCAP}
  \bibinfo{volume}{03} (\bibinfo{year}{2020}) \bibinfo{pages}{066},
  \bibinfo{doi}{\doi{10.1088/1475-7516/2020/03/066}}, \eprint{2001.05605}.

\bibtype{Article}%
\bibitem{Du:2018uak}
\bibinfo{author}{N. Du}, et al. (\bibinfo{collaboration}{ADMX}),
  \bibinfo{title}{{A Search for Invisible Axion Dark Matter with the Axion Dark
  Matter Experiment}}, \bibinfo{journal}{Phys. Rev. Lett.}
  \bibinfo{volume}{120} (\bibinfo{number}{15}) (\bibinfo{year}{2018})
  \bibinfo{pages}{151301}, \bibinfo{doi}{\doi{10.1103/PhysRevLett.120.151301}},
  \eprint{1804.05750}.

\bibtype{Article}%
\bibitem{Braine:2019fqb}
\bibinfo{author}{T. Braine}, et al. (\bibinfo{collaboration}{ADMX}),
  \bibinfo{title}{{Extended Search for the Invisible Axion with the Axion Dark
  Matter Experiment}}, \bibinfo{journal}{Phys. Rev. Lett.}
  \bibinfo{volume}{124} (\bibinfo{number}{10}) (\bibinfo{year}{2020})
  \bibinfo{pages}{101303}, \bibinfo{doi}{\doi{10.1103/PhysRevLett.124.101303}},
  \eprint{1910.08638}.

\bibtype{Article}%
\bibitem{ADMX:2021nhd}
\bibinfo{author}{C. Bartram}, et al. (\bibinfo{collaboration}{ADMX}),
  \bibinfo{title}{{Search for ''Invisible'' Axion Dark Matter in the
  $3.3\text{-}4.2~{\mu}$eV Mass Range}}, \bibinfo{journal}{to appear}
  (\bibinfo{year}{2021}), \eprint{2110.06096}.

\bibtype{Article}%
\bibitem{ADMX:2025vom}
\bibinfo{author}{G. Carosi}, et al. (\bibinfo{collaboration}{ADMX}),
  \bibinfo{title}{{Search for Axion Dark Matter from 1.1 to 1.3 GHz with
  ADMX}}, \bibinfo{journal}{to appear}  (\bibinfo{year}{2025}),
  \eprint{2504.07279}.

\bibtype{Article}%
\bibitem{CAPP:2023pace}
\bibinfo{author}{J. Kim}, \bibinfo{author}{O. Kwon}, \bibinfo{author}{C.
  Kutlu}, \bibinfo{author}{W. Chung}, \bibinfo{author}{A. Matlashov},
  \bibinfo{author}{Y.~K. Semertzidis}, \bibinfo{title}{{Near-Quantum-Noise
  Axion Dark Matter Search at CAPP around 5.3 $\mu$eV}},
  \bibinfo{journal}{Phys. Rev. Lett.} \bibinfo{volume}{130}
  (\bibinfo{year}{2023}) \bibinfo{pages}{091602},
  \bibinfo{doi}{\doi{10.1103/PhysRevLett.130.091602}}, \eprint{2211.07929}.

\bibtype{Article}%
\bibitem{CAPP:2023capp18t}
\bibinfo{author}{B. Yang}, \bibinfo{author}{H. Yoon}, \bibinfo{author}{M. Ahn},
  \bibinfo{author}{Y. Lee}, \bibinfo{author}{J. Yoo}, \bibinfo{title}{{Extended
  axion dark matter search using the CAPP18T haloscope}},
  \bibinfo{journal}{Phys. Rev. Lett.} \bibinfo{volume}{131}
  (\bibinfo{year}{2023}) \bibinfo{pages}{081801},
  \bibinfo{doi}{\doi{10.1103/PhysRevLett.131.081801}}, \eprint{2306.08565}.

\bibtype{Article}%
\bibitem{CAPP:2024max}
\bibinfo{author}{S. Ahn}, \bibinfo{author}{J.~M. Kim}, \bibinfo{author}{B.~I.
  Ivanov}, \bibinfo{author}{O. Kwon}, \bibinfo{author}{H.~S. Byun},
  \bibinfo{author}{Y.~K. Semertzidis}, \bibinfo{title}{{Extensive search for
  axion dark matter over 1 GHz with CAPP's Main Axion eXperiment (CAPP-MAX)}},
  \bibinfo{journal}{Phys. Rev. X} \bibinfo{volume}{14} (\bibinfo{year}{2024})
  \bibinfo{pages}{031023}, \bibinfo{doi}{\doi{10.1103/PhysRevX.14.031023}},
  \eprint{2402.07941}.

\bibtype{Article}%
\bibitem{Alesini:2019ajt}
\bibinfo{author}{D. Alesini}, et al., \bibinfo{title}{{Galactic axions search
  with a superconducting resonant cavity}}, \bibinfo{journal}{Phys. Rev. D}
  \bibinfo{volume}{99} (\bibinfo{number}{10}) (\bibinfo{year}{2019})
  \bibinfo{pages}{101101}, \bibinfo{doi}{\doi{10.1103/PhysRevD.99.101101}},
  \eprint{1903.06547}.

\bibtype{Article}%
\bibitem{Ahyoune:2024klt}
\bibinfo{author}{S. Ahyoune}, et al., \bibinfo{title}{{RADES axion search
  results with a high-temperature superconducting cavity in an 11.7 T magnet}},
  \bibinfo{journal}{JHEP} \bibinfo{volume}{04} (\bibinfo{year}{2025})
  \bibinfo{pages}{113}, \bibinfo{doi}{\doi{10.1007/JHEP04(2025)113}},
  \eprint{2403.07790}.

\bibtype{Article}%
\bibitem{Boutan:2018uoc}
\bibinfo{author}{C. Boutan}, et al. (\bibinfo{collaboration}{ADMX}),
  \bibinfo{title}{{Piezoelectrically Tuned Multimode Cavity Search for Axion
  Dark Matter}}, \bibinfo{journal}{Phys. Rev. Lett.} \bibinfo{volume}{121}
  (\bibinfo{number}{26}) (\bibinfo{year}{2018}) \bibinfo{pages}{261302},
  \bibinfo{doi}{\doi{10.1103/PhysRevLett.121.261302}}, \eprint{1901.00920}.

\bibtype{Article}%
\bibitem{Melcon:2020xvj}
\bibinfo{author}{A. \'Alvarez~Melc\'on}, et al., \bibinfo{title}{{Scalable
  haloscopes for axion dark matter detection in the 30$\mu$eV range with
  RADES}}, \bibinfo{journal}{JHEP} \bibinfo{volume}{07} (\bibinfo{year}{2020})
  \bibinfo{pages}{084}, \bibinfo{doi}{\doi{10.1007/JHEP07(2020)084}},
  \eprint{2002.07639}.

\bibtype{Article}%
\bibitem{CAST:2020rlf}
\bibinfo{author}{A.~{\'A}lvarez Melc{\'o}n}, et al.
  (\bibinfo{collaboration}{CAST}), \bibinfo{title}{{First results of the
  CAST-RADES haloscope search for axions at 34.67 $\mu$eV}},
  \bibinfo{journal}{JHEP} \bibinfo{volume}{10} (\bibinfo{year}{2020})
  \bibinfo{pages}{075}, \bibinfo{doi}{\doi{10.1007/JHEP10(2021)075}},
  \eprint{2104.13798}.

\bibtype{Article}%
\bibitem{Jeong:2020cwz}
\bibinfo{author}{Junu Jeong}, \bibinfo{author}{SungWoo Youn},
  \bibinfo{author}{Sungjae Bae}, \bibinfo{author}{Jihngeun Kim},
  \bibinfo{author}{Taehyeon Seong}, \bibinfo{author}{Jihn~E. Kim},
  \bibinfo{author}{Yannis~K. Semertzidis}, \bibinfo{title}{{Search for
  Invisible Axion Dark Matter with a Multiple-Cell Haloscope}},
  \bibinfo{journal}{Phys. Rev. Lett.} \bibinfo{volume}{125}
  (\bibinfo{number}{22}) (\bibinfo{year}{2020}) \bibinfo{pages}{221302},
  \bibinfo{doi}{\doi{10.1103/PhysRevLett.125.221302}}, \eprint{2008.10141}.

\bibtype{Article}%
\bibitem{Andriamonje:2007ew}
\bibinfo{author}{S. Andriamonje}, et al. (\bibinfo{collaboration}{CAST}),
  \bibinfo{title}{{An improved limit on the axion-photon coupling from the CAST
  experiment}}, \bibinfo{journal}{JCAP} \bibinfo{volume}{0704}
  (\bibinfo{year}{2007}) \bibinfo{pages}{010}, \eprint{hep-ex/0702006}.

\bibtype{Article}%
\bibitem{Hoof:2021mld}
\bibinfo{author}{Sebastian Hoof}, \bibinfo{author}{Joerg Jaeckel},
  \bibinfo{author}{Lennert~J. Thormaehlen}, \bibinfo{title}{{Quantifying
  uncertainties in the solar axion flux and their impact on determining axion
  model parameters}}, \bibinfo{journal}{JCAP} \bibinfo{volume}{09}
  (\bibinfo{year}{2021}) \bibinfo{pages}{006},
  \bibinfo{doi}{\doi{10.1088/1475-7516/2021/09/006}}, \eprint{2101.08789}.

\bibtype{Article}%
\bibitem{Caputo:2020quz}
\bibinfo{author}{Andrea Caputo}, \bibinfo{author}{Alexander~J. Millar},
  \bibinfo{author}{Edoardo Vitagliano}, \bibinfo{title}{{Revisiting
  longitudinal plasmon-axion conversion in external magnetic fields}},
  \bibinfo{journal}{Phys. Rev. D} \bibinfo{volume}{101} (\bibinfo{number}{12})
  (\bibinfo{year}{2020}) \bibinfo{pages}{123004},
  \bibinfo{doi}{\doi{10.1103/PhysRevD.101.123004}}, \eprint{2005.00078}.

\bibtype{Article}%
\bibitem{Guarini:2020hps}
\bibinfo{author}{Ersilia Guarini}, \bibinfo{author}{Pierluca Carenza},
  \bibinfo{author}{Javier Galan}, \bibinfo{author}{Maurizio Giannotti},
  \bibinfo{author}{Alessandro Mirizzi}, \bibinfo{title}{{Production of
  axionlike particles from photon conversions in large-scale solar magnetic
  fields}}, \bibinfo{journal}{Phys. Rev. D} \bibinfo{volume}{102}
  (\bibinfo{number}{12}) (\bibinfo{year}{2020}) \bibinfo{pages}{123024},
  \bibinfo{doi}{\doi{10.1103/PhysRevD.102.123024}}, \eprint{2010.06601}.

\bibtype{Article}%
\bibitem{OHare:2020wum}
\bibinfo{author}{Ciaran A.~J. O'Hare}, \bibinfo{author}{Andrea Caputo},
  \bibinfo{author}{Alexander~J. Millar}, \bibinfo{author}{Edoardo Vitagliano},
  \bibinfo{title}{{Axion helioscopes as solar magnetometers}},
  \bibinfo{journal}{Phys. Rev. D} \bibinfo{volume}{102} (\bibinfo{number}{4})
  (\bibinfo{year}{2020}) \bibinfo{pages}{043019},
  \bibinfo{doi}{\doi{10.1103/PhysRevD.102.043019}}, \eprint{2006.10415}.

\bibtype{Article}%
\bibitem{Redondo:2013wwa}
\bibinfo{author}{Javier Redondo}, \bibinfo{title}{{Solar axion flux from the
  axion-electron coupling}}, \bibinfo{journal}{JCAP} \bibinfo{volume}{1312}
  (\bibinfo{year}{2013}) \bibinfo{pages}{008},
  \bibinfo{doi}{\doi{10.1088/1475-7516/2013/12/008}}, \eprint{1310.0823}.

\bibtype{Article}%
\bibitem{Barth:2013sma}
\bibinfo{author}{K. Barth}, \bibinfo{author}{A. Belov}, \bibinfo{author}{B.
  Beltran}, \bibinfo{author}{H. Brauninger}, \bibinfo{author}{J.M. Carmona}, et
  al., \bibinfo{title}{{CAST constraints on the axion-electron coupling}},
  \bibinfo{journal}{JCAP} \bibinfo{volume}{1305} (\bibinfo{year}{2013})
  \bibinfo{pages}{010}, \bibinfo{doi}{\doi{10.1088/1475-7516/2013/05/010}},
  \eprint{1302.6283}.

\bibtype{Techreport}%
\bibitem{Irastorza:1567109}
\bibinfo{author}{Igor~G Irastorza} (\bibinfo{collaboration}{IAXO}),
  \bibinfo{title}{{The International Axion Observatory IAXO. Letter of Intent
  to the CERN SPS committee}}, \bibinfo{type}{\bibinfo{comment}{Tech. Rep.}}
  \bibinfo{number}{CERN-SPSC-2013-022. SPSC-I-242},
  \bibinfo{institution}{CERN}, \bibinfo{address}{Geneva} \bibinfo{year}{2013}.

\bibtype{Article}%
\bibitem{DiLuzio:2021qct}
\bibinfo{author}{Luca Di~Luzio}, et al., \bibinfo{title}{{Probing the
  axion\textendash{}nucleon coupling with the next generation of~axion
  helioscopes}}, \bibinfo{journal}{Eur. Phys. J. C} \bibinfo{volume}{82}
  (\bibinfo{number}{2}) (\bibinfo{year}{2022}) \bibinfo{pages}{120},
  \bibinfo{doi}{\doi{10.1140/epjc/s10052-022-10061-1}}, \eprint{2111.06407}.

\bibtype{Article}%
\bibitem{Zioutas:2004hi}
\bibinfo{author}{K. Zioutas}, et al. (\bibinfo{collaboration}{CAST}),
  \bibinfo{title}{{First results from the CERN Axion Solar Telescope (CAST)}},
  \bibinfo{journal}{Phys. Rev. Lett.} \bibinfo{volume}{94}
  (\bibinfo{year}{2005}) \bibinfo{pages}{121301},
  \bibinfo{doi}{\doi{10.1103/PhysRevLett.94.121301}}, \eprint{hep-ex/0411033}.

\bibtype{Article}%
\bibitem{vanBibber:1988ge}
\bibinfo{author}{K. van Bibber}, \bibinfo{author}{P.~M. McIntyre},
  \bibinfo{author}{D.~E. Morris}, \bibinfo{author}{G.~G. Raffelt},
  \bibinfo{title}{{Design for a practical laboratory detector for solar
  axions}}, \bibinfo{journal}{Phys. Rev.} \bibinfo{volume}{D39}
  (\bibinfo{year}{1989}) \bibinfo{pages}{2089},
  \bibinfo{doi}{\doi{10.1103/PhysRevD.39.2089}}.

\bibtype{Article}%
\bibitem{CAST:2024eil}
\bibinfo{author}{K. Altenm\"uller}, et al. (\bibinfo{collaboration}{CAST}),
  \bibinfo{title}{{New Upper Limit on the Axion-Photon Coupling with an
  Extended CAST Run with a Xe-Based Micromegas Detector}},
  \bibinfo{journal}{Phys. Rev. Lett.} \bibinfo{volume}{133}
  (\bibinfo{number}{22}) (\bibinfo{year}{2024}) \bibinfo{pages}{221005},
  \bibinfo{doi}{\doi{10.1103/PhysRevLett.133.221005}}, \eprint{2406.16840}.

\bibtype{Article}%
\bibitem{CAST:2008ixs}
\bibinfo{author}{E. Arik}, et al. (\bibinfo{collaboration}{CAST}),
  \bibinfo{title}{{Probing eV-scale axions with CAST}}, \bibinfo{journal}{JCAP}
  \bibinfo{volume}{02} (\bibinfo{year}{2009}) \bibinfo{pages}{008},
  \bibinfo{doi}{\doi{10.1088/1475-7516/2009/02/008}}, \eprint{0810.4482}.

\bibtype{Article}%
\bibitem{CAST:2011rjr}
\bibinfo{author}{S. Aune}, et al. (\bibinfo{collaboration}{CAST}),
  \bibinfo{title}{{CAST search for sub-eV mass solar axions with 3He buffer
  gas}}, \bibinfo{journal}{Phys. Rev. Lett.} \bibinfo{volume}{107}
  (\bibinfo{year}{2011}) \bibinfo{pages}{261302},
  \bibinfo{doi}{\doi{10.1103/PhysRevLett.107.261302}}, \eprint{1106.3919}.

\bibtype{Article}%
\bibitem{CAST:2013bqn}
\bibinfo{author}{M. Arik}, et al. (\bibinfo{collaboration}{CAST}),
  \bibinfo{title}{{Search for Solar Axions by the CERN Axion Solar Telescope
  with $^3$He Buffer Gas: Closing the Hot Dark Matter Gap}},
  \bibinfo{journal}{Phys. Rev. Lett.} \bibinfo{volume}{112}
  (\bibinfo{number}{9}) (\bibinfo{year}{2014}) \bibinfo{pages}{091302},
  \bibinfo{doi}{\doi{10.1103/PhysRevLett.112.091302}}, \eprint{1307.1985}.

\bibtype{Article}%
\bibitem{Irastorza:2011gs}
\bibinfo{author}{Igor~G. Irastorza}, \bibinfo{author}{F.T. Avignone},
  \bibinfo{author}{S. Caspi}, \bibinfo{author}{J.M. Carmona},
  \bibinfo{author}{T. Dafni}, et al., \bibinfo{title}{{Towards a new generation
  axion helioscope}}, \bibinfo{journal}{JCAP} \bibinfo{volume}{1106}
  (\bibinfo{year}{2011}) \bibinfo{pages}{013},
  \bibinfo{doi}{\doi{10.1088/1475-7516/2011/06/013}}, \eprint{1103.5334}.

\bibtype{Article}%
\bibitem{CAST:2009klq}
\bibinfo{author}{S. Andriamonje}, et al. (\bibinfo{collaboration}{CAST}),
  \bibinfo{title}{{Search for solar axion emission from $^7Li$ and
  $D(p,\gamma)^3He$ nuclear decays with the CAST $\gamma$-ray calorimeter}},
  \bibinfo{journal}{JCAP} \bibinfo{volume}{03} (\bibinfo{year}{2010})
  \bibinfo{pages}{032}, \bibinfo{doi}{\doi{10.1088/1475-7516/2010/03/032}},
  \eprint{0904.2103}.

\bibtype{Article}%
\bibitem{Andriamonje:2009dx}
\bibinfo{author}{S. Andriamonje}, et al. (\bibinfo{collaboration}{CAST}),
  \bibinfo{title}{{Search for 14.4-keV solar axions emitted in the
  M1-transition of Fe-57 nuclei with CAST}}, \bibinfo{journal}{JCAP}
  \bibinfo{volume}{0912} (\bibinfo{year}{2009}) \bibinfo{pages}{002},
  \bibinfo{doi}{\doi{10.1088/1475-7516/2009/12/002}}, \eprint{0906.4488}.

\bibtype{Article}%
\bibitem{Armengaud:2014gea}
\bibinfo{author}{E. Armengaud}, \bibinfo{author}{F.T. Avignone},
  \bibinfo{author}{M. Betz}, \bibinfo{author}{P. Brax}, \bibinfo{author}{P.
  Brun}, et al., \bibinfo{title}{{Conceptual Design of the International Axion
  Observatory (IAXO)}}, \bibinfo{journal}{JINST} \bibinfo{volume}{9}
  (\bibinfo{year}{2014}) \bibinfo{pages}{T05002},
  \bibinfo{doi}{\doi{10.1088/1748-0221/9/05/T05002}}, \eprint{1401.3233}.

\bibtype{Article}%
\bibitem{Shilon:2013xma}
\bibinfo{author}{I. Shilon}, \bibinfo{author}{A. Dudarev}, \bibinfo{author}{H.
  Silva}, \bibinfo{author}{U. Wagner}, \bibinfo{author}{H.~H.~J. ten Kate},
  \bibinfo{title}{{The Superconducting Toroid for the New International AXion
  Observatory (IAXO)}}, \bibinfo{journal}{IEEE Trans. Appl. Supercond.}
  \bibinfo{volume}{24} (\bibinfo{number}{3}) (\bibinfo{year}{2014})
  \bibinfo{pages}{4500104}, \bibinfo{doi}{\doi{10.1109/TASC.2013.2280654}},
  \eprint{1309.2117}.

\bibtype{Article}%
\bibitem{Aznar:2015iia}
\bibinfo{author}{F. Aznar}, et al., \bibinfo{title}{{A Micromegas-based
  low-background x-ray detector coupled to a slumped-glass telescope for axion
  research}}, \bibinfo{journal}{JCAP} \bibinfo{volume}{1512}
  (\bibinfo{year}{2015}) \bibinfo{pages}{008}, \eprint{1509.06190}.

\bibtype{Article}%
\bibitem{IAXO:2020wwp}
\bibinfo{author}{A. Abeln}, et al. (\bibinfo{collaboration}{IAXO}),
  \bibinfo{title}{{Conceptual design of BabyIAXO, the intermediate stage
  towards the International Axion Observatory}}, \bibinfo{journal}{JHEP}
  \bibinfo{volume}{05} (\bibinfo{year}{2021}) \bibinfo{pages}{137},
  \bibinfo{doi}{\doi{10.1007/JHEP05(2021)137}}, \eprint{2010.12076}.

\bibtype{Article}%
\bibitem{Armengaud:2019uso}
\bibinfo{author}{E. Armengaud}, et al. (\bibinfo{collaboration}{IAXO}),
  \bibinfo{title}{{Physics potential of the International Axion Observatory
  (IAXO)}}, \bibinfo{journal}{JCAP} \bibinfo{volume}{1906}
  (\bibinfo{number}{06}) (\bibinfo{year}{2019}) \bibinfo{pages}{047},
  \bibinfo{doi}{\doi{10.1088/1475-7516/2019/06/047}}, \eprint{1904.09155}.

\bibtype{Article}%
\bibitem{Galan:2015msa}
\bibinfo{author}{J. Gal\'{a}n}, et al., \bibinfo{title}{{Exploring 0.1-10$\,$eV
  axions with a new helioscope concept}}, \bibinfo{journal}{JCAP}
  \bibinfo{volume}{1512} (\bibinfo{year}{2015}) \bibinfo{pages}{012},
  \eprint{1508.03006}.

\bibtype{Article}%
\bibitem{Buchmuller:1989rb}
\bibinfo{author}{W. Buchm\"uller}, \bibinfo{author}{F. Hoogeveen},
  \bibinfo{title}{{Coherent production of light scalar particles in Bragg
  scattering}}, \bibinfo{journal}{Phys.Lett.} \bibinfo{volume}{B237}
  (\bibinfo{year}{1990}) \bibinfo{pages}{278},
  \bibinfo{doi}{\doi{10.1016/0370-2693(90)91444-G}}.

\bibtype{Article}%
\bibitem{Paschos:1993yf}
\bibinfo{author}{E.~A. Paschos}, \bibinfo{author}{K. Zioutas},
  \bibinfo{title}{{A Proposal for solar axion detection via Bragg scattering}},
  \bibinfo{journal}{Phys. Lett.} \bibinfo{volume}{B323} (\bibinfo{year}{1994})
  \bibinfo{pages}{367--372}, \bibinfo{doi}{\doi{10.1016/0370-2693(94)91233-5}}.

\bibtype{Article}%
\bibitem{Creswick:1997pg}
\bibinfo{author}{R.~J. Creswick}, et al., \bibinfo{title}{{Theory for the
  direct detection of solar axions by coherent Primakoff conversion in
  germanium detectors}}, \bibinfo{journal}{Phys. Lett.} \bibinfo{volume}{B427}
  (\bibinfo{year}{1998}) \bibinfo{pages}{235--240},
  \bibinfo{doi}{\doi{10.1016/S0370-2693(98)00183-X}}, \eprint{hep-ph/9708210}.

\bibtype{Article}%
\bibitem{Avignone:1997th}
\bibinfo{author}{III Avignone, F.~T.}, et al. (\bibinfo{collaboration}{SOLAX}),
  \bibinfo{title}{{Experimental Search for Solar Axions via Coherent Primakoff
  Conversion in a Germanium Spectrometer}}, \bibinfo{journal}{Phys. Rev. Lett.}
  \bibinfo{volume}{81} (\bibinfo{year}{1998}) \bibinfo{pages}{5068--5071},
  \bibinfo{doi}{\doi{10.1103/PhysRevLett.81.5068}}, \eprint{astro-ph/9708008}.

\bibtype{Article}%
\bibitem{Morales:2001we}
\bibinfo{author}{A. Morales}, et al. (\bibinfo{collaboration}{COSME}),
  \bibinfo{title}{{Particle Dark Matter and Solar Axion Searches with a small
  germanium detector at the Canfranc Underground Laboratory}},
  \bibinfo{journal}{Astropart. Phys.} \bibinfo{volume}{16}
  (\bibinfo{year}{2002}) \bibinfo{pages}{325--332},
  \bibinfo{doi}{\doi{10.1016/S0927-6505(01)00117-7}}, \eprint{hep-ex/0101037}.

\bibtype{Article}%
\bibitem{Bernabei:2001ny}
\bibinfo{author}{R. Bernabei}, et al., \bibinfo{title}{{Search for solar axions
  by Primakoff effect in NaI crystals}}, \bibinfo{journal}{Phys. Lett.}
  \bibinfo{volume}{B515} (\bibinfo{year}{2001}) \bibinfo{pages}{6--12},
  \bibinfo{doi}{\doi{10.1016/S0370-2693(01)00840-1}}.

\bibtype{Article}%
\bibitem{Ahmed:2009ht}
\bibinfo{author}{Z. Ahmed}, et al. (\bibinfo{collaboration}{CDMS}),
  \bibinfo{title}{{Search for Axions with the CDMS Experiment}},
  \bibinfo{journal}{Phys. Rev. Lett.} \bibinfo{volume}{103}
  (\bibinfo{year}{2009}) \bibinfo{pages}{141802},
  \bibinfo{doi}{\doi{10.1103/PhysRevLett.103.141802}}, \eprint{0902.4693}.

\bibtype{Article}%
\bibitem{Armengaud:2013rta}
\bibinfo{author}{E. Armengaud}, et al., \bibinfo{title}{{Axion searches with
  the EDELWEISS-II experiment}}, \bibinfo{journal}{JCAP} \bibinfo{volume}{11}
  (\bibinfo{year}{2013}) \bibinfo{pages}{067},
  \bibinfo{doi}{\doi{10.1088/1475-7516/2013/11/067}}, \eprint{1307.1488}.

\bibtype{Article}%
\bibitem{Li:2015tsa}
\bibinfo{author}{Dawei. Li}, \bibinfo{author}{R.~J. Creswick},
  \bibinfo{author}{F.~T. Avignone}, \bibinfo{author}{Yuanxu. Wang},
  \bibinfo{title}{{Theoretical Estimate of the Sensitivity of the CUORE
  Detector to Solar Axions}}, \bibinfo{journal}{JCAP} \bibinfo{volume}{1510}
  (\bibinfo{year}{2015}) \bibinfo{pages}{065},
  \bibinfo{doi}{\doi{10.1088/1475-7516/2015/10/065}}, \eprint{1507.00603}.

\bibtype{Article}%
\bibitem{Xu:2016tap}
\bibinfo{author}{Wenqin Xu}, \bibinfo{author}{Steven~R. Elliott},
  \bibinfo{title}{{Solar Axion Search Technique with Correlated Signals from
  Multiple Detectors}}, \bibinfo{journal}{Astropart. Phys.}
  \bibinfo{volume}{89} (\bibinfo{year}{2017}) \bibinfo{pages}{39--50},
  \bibinfo{doi}{\doi{10.1016/j.astropartphys.2017.01.008}},
  \eprint{1610.03886}.

\bibtype{Article}%
\bibitem{Cebrian:1998mu}
\bibinfo{author}{S. Cebri{\'a}n}, et al., \bibinfo{title}{{Prospects of solar
  axion searches with crystal detectors}}, \bibinfo{journal}{Astropart. Phys.}
  \bibinfo{volume}{10} (\bibinfo{year}{1999}) \bibinfo{pages}{397--404},
  \bibinfo{doi}{\doi{10.1016/S0927-6505(98)00069-3}},
  \eprint{astro-ph/9811359}.

\bibtype{Article}%
\bibitem{Avignone:2010zn}
\bibinfo{author}{F.~T. Avignone}, \bibinfo{author}{R.~J. Creswick},
  \bibinfo{author}{S. Nussinov}, \bibinfo{title}{{The experimental challenge of
  detecting solar axion-like particles to test cosmological ALP-photon
  oscillation hypothesis}}, \bibinfo{journal}{Astropart. Phys.}
  \bibinfo{volume}{34} (\bibinfo{year}{2011}) \bibinfo{pages}{640--642},
  \bibinfo{doi}{\doi{10.1016/j.astropartphys.2010.12.012}}, \eprint{1002.2718}.

\bibtype{Article}%
\bibitem{Ljubicic:2004gt}
\bibinfo{author}{A. Ljubicic}, \bibinfo{author}{D. Kekez}, \bibinfo{author}{Z.
  Krecak}, \bibinfo{author}{T. Ljubicic}, \bibinfo{title}{{Search for hadronic
  axions using axioelectric effect}}, \bibinfo{journal}{Phys.Lett.}
  \bibinfo{volume}{B599} (\bibinfo{year}{2004}) \bibinfo{pages}{143--147},
  \bibinfo{doi}{\doi{10.1016/j.physletb.2004.08.038}}, \eprint{hep-ex/0403045}.

\bibtype{Article}%
\bibitem{Derbin:2011gg}
\bibinfo{author}{A.V. Derbin}, \bibinfo{author}{A.S. Kayunov},
  \bibinfo{author}{V.N. Muratova}, \bibinfo{author}{D.A. Semenov},
  \bibinfo{author}{E.V. Unzhakov}, \bibinfo{title}{{Constraints on the
  axion-electron coupling for solar axions produced by Compton process and
  bremsstrahlung}}, \bibinfo{journal}{Phys.Rev.} \bibinfo{volume}{D83}
  (\bibinfo{year}{2011}) \bibinfo{pages}{023505},
  \bibinfo{doi}{\doi{10.1103/PhysRevD.83.023505}}, \eprint{1101.2290}.

\bibtype{Article}%
\bibitem{Derbin:2011zz}
\bibinfo{author}{A.V. Derbin}, \bibinfo{author}{V.N. Muratova},
  \bibinfo{author}{D.A. Semenov}, \bibinfo{author}{E.V. Unzhakov},
  \bibinfo{title}{{New limit on the mass of 14.4-keV solar axions emitted in an
  M1 transition in Fe-57 nuclei}}, \bibinfo{journal}{Phys.Atom.Nucl.}
  \bibinfo{volume}{74} (\bibinfo{year}{2011}) \bibinfo{pages}{596--602},
  \bibinfo{doi}{\doi{10.1134/S1063778811040041}}.

\bibtype{Article}%
\bibitem{Derbin:2012yk}
\bibinfo{author}{A.V. Derbin}, \bibinfo{author}{I.S. Drachnev},
  \bibinfo{author}{A.S. Kayunov}, \bibinfo{author}{V.N. Muratova},
  \bibinfo{title}{{Search for solar axions produced by Compton process and
  bremsstrahlung using axioelectric effect}}, \bibinfo{journal}{JETP Lett.}
  \bibinfo{volume}{95} (\bibinfo{year}{2012}) \bibinfo{pages}{379},
  \eprint{1206.4142}.

\bibtype{Article}%
\bibitem{Bellini:2012kz}
\bibinfo{author}{G. Bellini}, et al. (\bibinfo{collaboration}{Borexino}),
  \bibinfo{title}{{Search for Solar Axions Produced in $p(d,\rm{^3He})A$
  Reaction with Borexino Detector}}, \bibinfo{journal}{Phys.Rev.}
  \bibinfo{volume}{D85} (\bibinfo{year}{2012}) \bibinfo{pages}{092003},
  \bibinfo{doi}{\doi{10.1103/PhysRevD.85.092003}}, \eprint{1203.6258}.

\bibtype{Article}%
\bibitem{Abe:2012ut}
\bibinfo{author}{K. Abe}, et al., \bibinfo{title}{{Search for solar axions in
  XMASS, a large liquid-xenon detector}}, \bibinfo{journal}{Phys. Lett.}
  \bibinfo{volume}{B724} (\bibinfo{year}{2013}) \bibinfo{pages}{46--50},
  \bibinfo{doi}{\doi{10.1016/j.physletb.2013.05.060}}, \eprint{1212.6153}.

\bibtype{Article}%
\bibitem{Aprile:2014eoa}
\bibinfo{author}{E. Aprile}, et al. (\bibinfo{collaboration}{XENON100}),
  \bibinfo{title}{{First Axion Results from the XENON100 Experiment}},
  \bibinfo{journal}{Phys. Rev.} \bibinfo{volume}{D90} (\bibinfo{number}{6})
  (\bibinfo{year}{2014}) \bibinfo{pages}{062009},
  \bibinfo{doi}{\doi{10.1103/PhysRevD.90.062009, 10.1103/PhysRevD.95.029904}},
  \bibinfo{note}{[Erratum: Phys. Rev.D95,no.2,029904(2017)]},
  \eprint{1404.1455}.

\bibtype{Article}%
\bibitem{PandaX:2017ock}
\bibinfo{author}{Changbo Fu}, et al. (\bibinfo{collaboration}{PandaX}),
  \bibinfo{title}{{Limits on Axion Couplings from the First 80 Days of Data of
  the PandaX-II Experiment}}, \bibinfo{journal}{Phys. Rev. Lett.}
  \bibinfo{volume}{119} (\bibinfo{number}{18}) (\bibinfo{year}{2017})
  \bibinfo{pages}{181806}, \bibinfo{doi}{\doi{10.1103/PhysRevLett.119.181806}},
  \eprint{1707.07921}.

\bibtype{Article}%
\bibitem{Akerib:2017uem}
\bibinfo{author}{D.~S. Akerib}, et al. (\bibinfo{collaboration}{LUX}),
  \bibinfo{title}{{First Searches for Axions and Axionlike Particles with the
  LUX Experiment}}, \bibinfo{journal}{Phys. Rev. Lett.} \bibinfo{volume}{118}
  (\bibinfo{number}{26}) (\bibinfo{year}{2017}) \bibinfo{pages}{261301},
  \bibinfo{doi}{\doi{10.1103/PhysRevLett.118.261301}}, \eprint{1704.02297}.

\bibtype{Article}%
\bibitem{Moriyama:1995bz}
\bibinfo{author}{Shigetaka Moriyama}, \bibinfo{title}{{A Proposal to search for
  a monochromatic component of solar axions using Fe-57}},
  \bibinfo{journal}{Phys.Rev.Lett.} \bibinfo{volume}{75} (\bibinfo{year}{1995})
  \bibinfo{pages}{3222--3225},
  \bibinfo{doi}{\doi{10.1103/PhysRevLett.75.3222}}, \eprint{hep-ph/9504318}.

\bibtype{Article}%
\bibitem{Krcmar:1998xn}
\bibinfo{author}{M. Krcmar}, \bibinfo{author}{Z. Krecak}, \bibinfo{author}{M.
  Stipcevic}, \bibinfo{author}{A. Ljubicic}, \bibinfo{author}{D.A. Bradley},
  \bibinfo{title}{{Search for invisible axions using Fe-57}},
  \bibinfo{journal}{Phys.Lett.} \bibinfo{volume}{B442} (\bibinfo{year}{1998})
  \bibinfo{pages}{38}, \bibinfo{doi}{\doi{10.1016/S0370-2693(98)01231-3}},
  \eprint{nucl-ex/9801005}.

\bibtype{Article}%
\bibitem{Krcmar:2001si}
\bibinfo{author}{M. Krcmar}, \bibinfo{author}{Z. Krecak}, \bibinfo{author}{A.
  Ljubicic}, \bibinfo{author}{M. Stipcevic}, \bibinfo{author}{D.A. Bradley},
  \bibinfo{title}{{Search for solar axions using Li-7}},
  \bibinfo{journal}{Phys.Rev.} \bibinfo{volume}{D64} (\bibinfo{year}{2001})
  \bibinfo{pages}{115016}, \bibinfo{doi}{\doi{10.1103/PhysRevD.64.115016}},
  \eprint{hep-ex/0104035}.

\bibtype{Article}%
\bibitem{Derbin:2009jw}
\bibinfo{author}{A.V. Derbin}, \bibinfo{author}{S.V. Bakhlanov},
  \bibinfo{author}{A.I. Egorov}, \bibinfo{author}{I.A. Mitropolsky},
  \bibinfo{author}{V.N. Muratova}, et al., \bibinfo{title}{{Search for Solar
  Axions Produced by Primakoff Conversion Using Resonant Absorption by Tm-169
  Nuclei}}, \bibinfo{journal}{Phys.Lett.} \bibinfo{volume}{B678}
  (\bibinfo{year}{2009}) \bibinfo{pages}{181--185},
  \bibinfo{doi}{\doi{10.1016/j.physletb.2009.06.016}}, \eprint{0904.3443}.

\bibtype{Article}%
\bibitem{ADMX:2023ctd}
\bibinfo{author}{C. Bartram}, et al. (\bibinfo{collaboration}{ADMX}),
  \bibinfo{title}{{Nonvirialized axion search sensitive to Doppler effects in
  the Milky~Way halo}}, \bibinfo{journal}{Phys. Rev. D} \bibinfo{volume}{109}
  (\bibinfo{number}{8}) (\bibinfo{year}{2024}) \bibinfo{pages}{083014},
  \bibinfo{doi}{\doi{10.1103/PhysRevD.109.083014}}, \eprint{2311.07748}.

\bibtype{Article}%
\bibitem{Schneemann:2023bqc}
\bibinfo{author}{Tim Schneemann}, \bibinfo{author}{Kristof Schmieden},
  \bibinfo{author}{Matthias Schott}, \bibinfo{title}{{First results of the
  SUPAX Experiment: Probing Dark Photons}}, \bibinfo{journal}{to appear}
  (\bibinfo{year}{2023}), \eprint{2308.08337}.

\bibtype{Article}%
\bibitem{QUAX:2024fut}
\bibinfo{author}{A. Rettaroli}, et al. (\bibinfo{collaboration}{QUAX}),
  \bibinfo{title}{{Search for axion dark matter with the QUAX\textendash{}LNF
  tunable haloscope}}, \bibinfo{journal}{Phys. Rev. D} \bibinfo{volume}{110}
  (\bibinfo{number}{2}) (\bibinfo{year}{2024}) \bibinfo{pages}{022008},
  \bibinfo{doi}{\doi{10.1103/PhysRevD.110.022008}}, \eprint{2402.19063}.

\bibtype{Article}%
\bibitem{Brubaker:2017rna}
\bibinfo{author}{B.M. Brubaker}, \bibinfo{author}{L. Zhong},
  \bibinfo{author}{S.K. Lamoreaux}, \bibinfo{author}{K.W. Lehnert},
  \bibinfo{author}{K.A. van Bibber}, \bibinfo{title}{{HAYSTAC axion search
  analysis procedure}}, \bibinfo{journal}{Phys. Rev. D} \bibinfo{volume}{96}
  (\bibinfo{number}{12}) (\bibinfo{year}{2017}) \bibinfo{pages}{123008},
  \bibinfo{doi}{\doi{10.1103/PhysRevD.96.123008}}, \eprint{1706.08388}.

\bibtype{Article}%
\bibitem{Palken:2020wgs}
\bibinfo{author}{D.A. Palken}, et al., \bibinfo{title}{{Improved analysis
  framework for axion dark matter searches}}, \bibinfo{journal}{Phys. Rev. D}
  \bibinfo{volume}{101} (\bibinfo{number}{12}) (\bibinfo{year}{2020})
  \bibinfo{pages}{123011}, \bibinfo{doi}{\doi{10.1103/PhysRevD.101.123011}},
  \eprint{2003.08510}.

\bibtype{Article}%
\bibitem{Savitzky:2002vxy}
\bibinfo{author}{Abraham. Savitzky}, \bibinfo{author}{M.~J.~E. Golay},
  \bibinfo{title}{{Smoothing and Differentiation of Data by Simplified Least
  Squares Procedures.}}, \bibinfo{journal}{Anal. Chem.} \bibinfo{volume}{36}
  (\bibinfo{number}{8}) (\bibinfo{year}{1964}) \bibinfo{pages}{1627--1639},
  \bibinfo{doi}{\doi{10.1021/ac60214a047}}.

\bibtype{Article}%
\bibitem{CAST:2017uph}
\bibinfo{author}{V. Anastassopoulos}, et al. (\bibinfo{collaboration}{CAST}),
  \bibinfo{title}{{New CAST Limit on the Axion-Photon Interaction}},
  \bibinfo{journal}{Nature Phys.} \bibinfo{volume}{13} (\bibinfo{year}{2017})
  \bibinfo{pages}{584--590}, \bibinfo{doi}{\doi{10.1038/nphys4109}},
  \eprint{1705.02290}.

\bibtype{Article}%
\bibitem{Arias:2012az}
\bibinfo{author}{Paola Arias}, \bibinfo{author}{Davide Cadamuro},
  \bibinfo{author}{Mark Goodsell}, \bibinfo{author}{Joerg Jaeckel},
  \bibinfo{author}{Javier Redondo}, \bibinfo{author}{Andreas Ringwald},
  \bibinfo{title}{{WISPy Cold Dark Matter}}, \bibinfo{journal}{JCAP}
  \bibinfo{volume}{1206} (\bibinfo{year}{2012}) \bibinfo{pages}{013},
  \bibinfo{doi}{\doi{10.1088/1475-7516/2012/06/013}}, \eprint{1201.5902}.

\bibtype{Article}%
\bibitem{Abel:2008ai}
\bibinfo{author}{S.~A. Abel}, \bibinfo{author}{M.~D. Goodsell},
  \bibinfo{author}{J. Jaeckel}, \bibinfo{author}{V.~V. Khoze},
  \bibinfo{author}{A. Ringwald}, \bibinfo{title}{{Kinetic Mixing of the Photon
  with Hidden U(1)s in String Phenomenology}}, \bibinfo{journal}{JHEP}
  \bibinfo{volume}{07} (\bibinfo{year}{2008}) \bibinfo{pages}{124},
  \bibinfo{doi}{\doi{10.1088/1126-6708/2008/07/124}}, \eprint{0803.1449}.

\bibtype{Article}%
\bibitem{Horns:2012fx}
\bibinfo{author}{D. Horns}, \bibinfo{author}{M. Meyer},
  \bibinfo{title}{{Indications for a pair-production anomaly from the
  propagation of VHE gamma-rays}}, \bibinfo{journal}{JCAP}
  \bibinfo{volume}{1202} (\bibinfo{year}{2012}) \bibinfo{pages}{033},
  \bibinfo{doi}{\doi{10.1088/1475-7516/2012/02/033}}, \eprint{1201.4711}.

\bibtype{Article}%
\bibitem{Suzuki:2015sza}
\bibinfo{author}{J. Suzuki}, \bibinfo{author}{T. Horie}, \bibinfo{author}{Y.
  Inoue}, \bibinfo{author}{M. Minowa}, \bibinfo{title}{{Experimental Search for
  Hidden Photon CDM in the eV mass range with a Dish Antenna}},
  \bibinfo{journal}{JCAP} \bibinfo{volume}{09} (\bibinfo{year}{2015})
  \bibinfo{pages}{042}, \bibinfo{doi}{\doi{10.1088/1475-7516/2015/09/042}},
  \eprint{1504.00118}.

\bibtype{Article}%
\bibitem{FUNKExperiment:2020ofv}
\bibinfo{author}{A. Andrianavalomahefa}, et al. (\bibinfo{collaboration}{FUNK
  Experiment}), \bibinfo{title}{{Limits from the Funk Experiment on the Mixing
  Strength of Hidden-Photon Dark Matter in the Visible and Near-Ultraviolet
  Wavelength Range}}, \bibinfo{journal}{Phys. Rev. D} \bibinfo{volume}{102}
  (\bibinfo{number}{4}) (\bibinfo{year}{2020}) \bibinfo{pages}{042001},
  \bibinfo{doi}{\doi{10.1103/PhysRevD.102.042001}}, \eprint{2003.13144}.

\bibtype{Article}%
\bibitem{Caputo:2021eaa}
\bibinfo{author}{Andrea Caputo}, \bibinfo{author}{Alexander~J. Millar},
  \bibinfo{author}{Ciaran A.~J. O'Hare}, \bibinfo{author}{Edoardo Vitagliano},
  \bibinfo{title}{{Dark photon limits: A handbook}}, \bibinfo{journal}{Phys.
  Rev. D} \bibinfo{volume}{104} (\bibinfo{number}{9}) (\bibinfo{year}{2021})
  \bibinfo{pages}{095029}, \bibinfo{doi}{\doi{10.1103/PhysRevD.104.095029}},
  \eprint{2105.04565}.

\bibtype{Article}%
\bibitem{Schwarz:2015lqa}
\bibinfo{author}{Matthias Schwarz}, \bibinfo{author}{Ernst-Axel Knabbe},
  \bibinfo{author}{Axel Lindner}, \bibinfo{author}{Javier Redondo},
  \bibinfo{author}{Andreas Ringwald}, \bibinfo{author}{Magnus Schneide},
  \bibinfo{author}{Jaroslaw Susol}, \bibinfo{author}{G\"unter Wiedemann},
  \bibinfo{title}{{Results from the Solar Hidden Photon Search (SHIPS)}},
  \bibinfo{journal}{JCAP} \bibinfo{volume}{08} (\bibinfo{year}{2015})
  \bibinfo{pages}{011}, \bibinfo{doi}{\doi{10.1088/1475-7516/2015/08/011}},
  \eprint{1502.04490}.

\bibtype{Article}%
\bibitem{PhysRevD.103.115004}
\bibinfo{author}{Jeff~A. Dror}, \bibinfo{author}{Hitoshi Murayama},
  \bibinfo{author}{Nicholas~L. Rodd}, \bibinfo{title}{Cosmic axion background},
  \bibinfo{journal}{Phys. Rev. D} \bibinfo{volume}{103} (\bibinfo{year}{2021})
  \bibinfo{pages}{115004}, \bibinfo{doi}{\doi{10.1103/PhysRevD.103.115004}},
  \bibinfo{url}{\urlprefix\url{https://link.aps.org/doi/10.1103/PhysRevD.103.115004}}.

\bibtype{Article}%
\bibitem{ADMX:2023rsk}
\bibinfo{author}{T. Nitta}, et al. (\bibinfo{collaboration}{ADMX}),
  \bibinfo{title}{{Search for a Dark-Matter-Induced Cosmic Axion Background
  with ADMX}}, \bibinfo{journal}{Phys. Rev. Lett.} \bibinfo{volume}{131}
  (\bibinfo{number}{10}) (\bibinfo{year}{2023}) \bibinfo{pages}{101002},
  \bibinfo{doi}{\doi{10.1103/PhysRevLett.131.101002}}, \eprint{2303.06282}.

\bibtype{Article}%
\bibitem{Aggarwal:2020olq}
\bibinfo{author}{Nancy Aggarwal}, et al., \bibinfo{title}{{Challenges and
  opportunities of gravitational-wave searches at MHz to GHz frequencies}},
  \bibinfo{journal}{Living Rev. Rel.} \bibinfo{volume}{24}
  (\bibinfo{number}{1}) (\bibinfo{year}{2021}) \bibinfo{pages}{4},
  \bibinfo{doi}{\doi{10.1007/s41114-021-00032-5}}, \eprint{2011.12414}.

\bibtype{Article}%
\bibitem{Raffelt:1987im}
\bibinfo{author}{Georg Raffelt}, \bibinfo{author}{Leo Stodolsky},
  \bibinfo{title}{{Mixing of the Photon with Low Mass Particles}},
  \bibinfo{journal}{Phys.Rev.} \bibinfo{volume}{D37} (\bibinfo{year}{1988})
  \bibinfo{pages}{1237}, \bibinfo{doi}{\doi{10.1103/PhysRevD.37.1237}}.

\bibtype{Article}%
\bibitem{Garcia-Cely:2024ujr}
\bibinfo{author}{Camilo Garc\'\i{}a-Cely}, \bibinfo{author}{Andreas Ringwald},
  \bibinfo{title}{{Complete Gravitational-Wave Spectrum of the Sun}},
  \bibinfo{journal}{to appear}  (\bibinfo{year}{2024}), \eprint{2407.18297}.

\bibtype{Article}%
\bibitem{Ejlli:2019bqj}
\bibinfo{author}{Aldo Ejlli}, \bibinfo{author}{Damian Ejlli},
  \bibinfo{author}{Adrian~Mike Cruise}, \bibinfo{author}{Giampaolo Pisano},
  \bibinfo{author}{Hartmut Grote}, \bibinfo{title}{{Upper limits on the
  amplitude of ultra-high-frequency gravitational waves from graviton to photon
  conversion}}, \bibinfo{journal}{Eur. Phys. J. C} \bibinfo{volume}{79}
  (\bibinfo{number}{12}) (\bibinfo{year}{2019}) \bibinfo{pages}{1032},
  \bibinfo{doi}{\doi{10.1140/epjc/s10052-019-7542-5}}, \eprint{1908.00232}.

\bibtype{Article}%
\bibitem{Gatti:2024mde}
\bibinfo{author}{Claudio Gatti}, \bibinfo{author}{Luca Visinelli},
  \bibinfo{author}{Michael Zantedeschi}, \bibinfo{title}{{Cavity detection of
  gravitational waves: Where do we stand?}}, \bibinfo{journal}{Phys. Rev. D}
  \bibinfo{volume}{110} (\bibinfo{number}{2}) (\bibinfo{year}{2024})
  \bibinfo{pages}{023018}, \bibinfo{doi}{\doi{10.1103/PhysRevD.110.023018}},
  \eprint{2403.18610}.

\bibtype{Article}%
\bibitem{Aggarwal:2025noe}
\bibinfo{author}{Nancy Aggarwal}, et al., \bibinfo{title}{{Challenges and
  Opportunities of Gravitational Wave Searches above 10 kHz}},
  \bibinfo{journal}{to appear}  (\bibinfo{year}{2025}), \eprint{2501.11723}.

\bibtype{Article}%
\bibitem{Schmieden:2023fzn}
\bibinfo{author}{Kristof Schmieden}, \bibinfo{author}{Matthias Schott},
  \bibinfo{title}{{A Global Network of Cavities to Search for Gravitational
  Waves (GravNet): A novel scheme to hunt gravitational waves signatures from
  the early universe}}, \bibinfo{journal}{PoS} \bibinfo{volume}{EPS-HEP2023}
  (\bibinfo{year}{2024}) \bibinfo{pages}{102},
  \bibinfo{doi}{\doi{10.22323/1.449.0102}}, \eprint{2308.11497}.

\bibtype{Article}%
\bibitem{Domcke:2024mfu}
\bibinfo{author}{Valerie Domcke}, \bibinfo{author}{Sebastian A.~R. Ellis},
  \bibinfo{author}{Nicholas~L. Rodd}, \bibinfo{title}{{Magnets are Weber Bar
  Gravitational Wave Detectors}}, \bibinfo{journal}{Phys. Rev. Lett.}
  \bibinfo{volume}{134} (\bibinfo{number}{23}) (\bibinfo{year}{2025})
  \bibinfo{pages}{231401}, \bibinfo{doi}{\doi{10.1103/966v-r5fm}},
  \eprint{2408.01483}.

\bibtype{Article}%
\bibitem{Zhitnitsky:2004da}
\bibinfo{author}{A. Zhitnitsky}, \bibinfo{title}{{'Nonbaryonic' dark matter as
  baryonic colour superconductor}}, \bibinfo{journal}{J. Phys. G}
  \bibinfo{volume}{30} (\bibinfo{year}{2004}) \bibinfo{pages}{S513--S517},
  \bibinfo{doi}{\doi{10.1088/0954-3899/30/1/061}}.

\bibtype{Article}%
\bibitem{Caspers:2024kjp}
\bibinfo{author}{F. Caspers}, et al., \bibinfo{title}{{Daily modulations and
  broadband strategy in axion searches: An application with the CAST-CAPP
  detector}}, \bibinfo{journal}{Phys. Rev. D} \bibinfo{volume}{111}
  (\bibinfo{number}{8}) (\bibinfo{year}{2025}) \bibinfo{pages}{082009},
  \bibinfo{doi}{\doi{10.1103/PhysRevD.111.082009}}, \eprint{2405.10972}.

\bibtype{Article}%
\bibitem{PhysRevLett.109.241301}
\bibinfo{author}{Junpu Wang}, \bibinfo{author}{Lam Hui},
  \bibinfo{author}{Justin Khoury}, \bibinfo{title}{No-Go Theorems for
  Generalized Chameleon Field Theories}, \bibinfo{journal}{Phys. Rev. Lett.}
  \bibinfo{volume}{109} (\bibinfo{year}{2012}) \bibinfo{pages}{241301},
  \bibinfo{doi}{\doi{10.1103/PhysRevLett.109.241301}},
  \bibinfo{url}{\urlprefix\url{https://link.aps.org/doi/10.1103/PhysRevLett.109.241301}}.

\bibtype{Article}%
\bibitem{Ahlers:2007st}
\bibinfo{author}{M. Ahlers}, \bibinfo{author}{A. Lindner}, \bibinfo{author}{A.
  Ringwald}, \bibinfo{author}{L. Schrempp}, \bibinfo{author}{C. Weniger},
  \bibinfo{title}{{Alpenglow - A Signature for Chameleons in Axion-Like
  Particle Search Experiments}}, \bibinfo{journal}{Phys. Rev. D}
  \bibinfo{volume}{77} (\bibinfo{year}{2008}) \bibinfo{pages}{015018},
  \bibinfo{doi}{\doi{10.1103/PhysRevD.77.015018}}, \eprint{0710.1555}.

\bibtype{Article}%
\bibitem{Rybka_2010}
\bibinfo{author}{G. Rybka}, \bibinfo{author}{M. Hotz}, \bibinfo{author}{L.~J
  Rosenberg}, \bibinfo{author}{S.~J. Asztalos}, \bibinfo{author}{G. Carosi},
  \bibinfo{author}{C. Hagmann}, \bibinfo{author}{D. Kinion},
  \bibinfo{author}{K. van Bibber}, \bibinfo{author}{J. Hoskins},
  \bibinfo{author}{C. Martin}, \bibinfo{author}{P. Sikivie},
  \bibinfo{author}{D.~B. Tanner}, \bibinfo{author}{R. Bradley},
  \bibinfo{author}{J. Clarke}, \bibinfo{title}{Search for Chameleon Scalar
  Fields with the Axion Dark Matter Experiment}, \bibinfo{journal}{Physical
  Review Letters} \bibinfo{volume}{105} (\bibinfo{number}{5})
  (\bibinfo{year}{2010}), ISSN \bibinfo{issn}{1079-7114},
  \bibinfo{doi}{\doi{10.1103/physrevlett.105.051801}},
  \bibinfo{url}{\urlprefix\url{http://dx.doi.org/10.1103/PhysRevLett.105.051801}}.

\bibtype{Article}%
\bibitem{ArguedasCuendis:2019fxj}
\bibinfo{author}{S. Arguedas~Cuendis}, et al., \bibinfo{title}{{First Results
  on the Search for Chameleons with the KWISP Detector at CAST}},
  \bibinfo{journal}{Phys. Dark Univ.} \bibinfo{volume}{26}
  (\bibinfo{year}{2019}) \bibinfo{pages}{100367},
  \bibinfo{doi}{\doi{10.1016/j.dark.2019.100367}}, \eprint{1906.01084}.

\bibtype{Article}%
\bibitem{Alesini:2023qed}
\bibinfo{author}{David Alesini}, et al., \bibinfo{title}{{The future search for
  low-frequency axions and new physics with the FLASH resonant cavity
  experiment at Frascati National Laboratories}}, \bibinfo{journal}{Phys. Dark
  Univ.} \bibinfo{volume}{42} (\bibinfo{year}{2023}) \bibinfo{pages}{101370},
  \bibinfo{doi}{\doi{10.1016/j.dark.2023.101370}}, \eprint{2309.00351}.

\bibtype{Article}%
\bibitem{Carenza:2025uib}
\bibinfo{author}{P. Carenza}, \bibinfo{author}{J.~A. Garc\'\i{}a~Pascual},
  \bibinfo{author}{M. Giannotti}, \bibinfo{author}{I.~G. Irastorza},
  \bibinfo{author}{M. Kaltschmidt}, \bibinfo{author}{A. Lella},
  \bibinfo{author}{A. Lindner}, \bibinfo{author}{G. Lucente},
  \bibinfo{author}{A. Mirizzi}, \bibinfo{author}{M.~J. Puyuelo},
  \bibinfo{title}{{Detecting Supernova Axions with IAXO}}, \bibinfo{journal}{to
  appear}  (\bibinfo{year}{2025}), \eprint{2502.19476}.

\bibtype{Article}%
\bibitem{Ge:2020zww}
\bibinfo{author}{Shao-Feng Ge}, \bibinfo{author}{Koichi Hamaguchi},
  \bibinfo{author}{Koichi Ichimura}, \bibinfo{author}{Koji Ishidoshiro},
  \bibinfo{author}{Yoshiki Kanazawa}, \bibinfo{author}{Yasuhiro Kishimoto},
  \bibinfo{author}{Natsumi Nagata}, \bibinfo{author}{Jiaming Zheng},
  \bibinfo{title}{{Supernova-scope for the Direct Search of Supernova Axions}},
  \bibinfo{journal}{JCAP} \bibinfo{volume}{11} (\bibinfo{year}{2020})
  \bibinfo{pages}{059}, \bibinfo{doi}{\doi{10.1088/1475-7516/2020/11/059}},
  \eprint{2008.03924}.

\bibtype{Article}%
\bibitem{IAXO:2019mpc}
\bibinfo{author}{E.~et~al. Armengaud}, \bibinfo{title}{Physics potential of the
  International Axion Observatory (IAXO)}, \bibinfo{journal}{JCAP}
  \bibinfo{volume}{2019} (\bibinfo{number}{06}) (\bibinfo{year}{2019})
  \bibinfo{pages}{047}, \eprint{1904.09155}.

\bibtype{Article}%
\bibitem{Jaeckel:2019mbt}
\bibinfo{author}{J. Jaeckel}, \bibinfo{author}{L.~J. Thormaehlen},
  \bibinfo{title}{Distinguishing Axion Models with IAXO},
  \bibinfo{journal}{JCAP} \bibinfo{volume}{2019} (\bibinfo{number}{03})
  (\bibinfo{year}{2019}) \bibinfo{pages}{039}, \eprint{1811.09278}.

\bibtype{Article}%
\bibitem{DiLuzio:2021ysg}
\bibinfo{author}{L.~et~al. Di~Luzio}, \bibinfo{title}{Probing the
  axion--nucleon coupling with the next generation of axion helioscopes},
  \bibinfo{journal}{Eur. Phys. J. C} \bibinfo{volume}{82}
  (\bibinfo{year}{2022}) \bibinfo{pages}{120}, \eprint{2111.06407}.

\bibtype{Article}%
\bibitem{Dafni:2018tvj}
\bibinfo{author}{T.~et~al. Dafni}, \bibinfo{title}{Weighing the solar axion},
  \bibinfo{journal}{Phys. Rev. D} \bibinfo{volume}{99} (\bibinfo{year}{2019})
  \bibinfo{pages}{035037}, \eprint{1811.09290}.

\bibtype{Article}%
\bibitem{OHare:2020wah}
\bibinfo{author}{Ciaran A.~J. O'Hare}, \bibinfo{author}{Edoardo Vitagliano},
  \bibinfo{title}{{Cornering the axion with $CP$-violating interactions}},
  \bibinfo{journal}{Phys. Rev. D} \bibinfo{volume}{102} (\bibinfo{number}{11})
  (\bibinfo{year}{2020}) \bibinfo{pages}{115026},
  \bibinfo{doi}{\doi{10.1103/PhysRevD.102.115026}}, \eprint{2010.03889}.

\bibtype{Article}%
\bibitem{Hoof:2023hyl}
\bibinfo{author}{S.~et~al. Hoof}, \bibinfo{title}{Axion helioscopes as solar
  thermometers}, \bibinfo{journal}{JCAP} \bibinfo{volume}{2023}
  (\bibinfo{number}{10}) (\bibinfo{year}{2023}) \bibinfo{pages}{024},
  \eprint{2306.00077}.

\bibtype{Article}%
\bibitem{Jaeckel:2019bnh}
\bibinfo{author}{J. Jaeckel}, \bibinfo{author}{L.~J. Thormaehlen},
  \bibinfo{title}{Axions as a probe of solar metals}, \bibinfo{journal}{Phys.
  Rev. D} \bibinfo{volume}{100} (\bibinfo{year}{2019}) \bibinfo{pages}{123020},
  \eprint{1908.10878}.

\bibtype{Article}%
\bibitem{Hoof:2021wpx}
\bibinfo{author}{S.~et~al. Hoof}, \bibinfo{title}{Quantifying uncertainties in
  the solar axion flux and their impact on determining axion model parameters},
  \bibinfo{journal}{JCAP} \bibinfo{volume}{2021} (\bibinfo{number}{09})
  (\bibinfo{year}{2021}) \bibinfo{pages}{006}, \eprint{2101.08789}.

\bibtype{Article}%
\bibitem{Bahre:2013ywa}
\bibinfo{author}{Robin B\"ahre}, et al., \bibinfo{title}{{Any light particle
  search II \textemdash{}Technical Design Report}}, \bibinfo{journal}{JINST}
  \bibinfo{volume}{8} (\bibinfo{year}{2013}) \bibinfo{pages}{T09001},
  \bibinfo{doi}{\doi{10.1088/1748-0221/8/09/T09001}}, \eprint{1302.5647}.

\bibtype{Article}%
\bibitem{Hoof:2024gfk}
\bibinfo{author}{Sebastian Hoof}, \bibinfo{author}{Joerg Jaeckel},
  \bibinfo{author}{Giuseppe Lucente}, \bibinfo{title}{{Ultimate
  light-shining-through-a-wall experiments to establish QCD axions as the
  dominant form of dark matter}}, \bibinfo{journal}{Phys. Rev. D}
  \bibinfo{volume}{111} (\bibinfo{number}{1}) (\bibinfo{year}{2025})
  \bibinfo{pages}{015003}, \bibinfo{doi}{\doi{10.1103/PhysRevD.111.015003}},
  \eprint{2407.04772}.

\bibtype{Article}%
\bibitem{PhysRevLett.124.161101}
\bibinfo{author}{Thomas D.~P. Edwards}, \bibinfo{author}{Marco Chianese},
  \bibinfo{author}{Bradley~J. Kavanagh}, \bibinfo{author}{Samaya~M. Nissanke},
  \bibinfo{author}{Christoph Weniger}, \bibinfo{title}{Unique Multimessenger
  Signal of QCD Axion Dark Matter}, \bibinfo{journal}{Phys. Rev. Lett.}
  \bibinfo{volume}{124} (\bibinfo{year}{2020}) \bibinfo{pages}{161101},
  \bibinfo{doi}{\doi{10.1103/PhysRevLett.124.161101}},
  \bibinfo{url}{\urlprefix\url{https://link.aps.org/doi/10.1103/PhysRevLett.124.161101}}.

\bibtype{Article}%
\bibitem{Caputo:2024oqc}
\bibinfo{author}{Andrea Caputo}, \bibinfo{author}{Georg Raffelt},
  \bibinfo{title}{{Astrophysical Axion Bounds: The 2024 Edition}},
  \bibinfo{journal}{PoS} \bibinfo{volume}{COSMICWISPers} (\bibinfo{year}{2024})
  \bibinfo{pages}{041}, \bibinfo{doi}{\doi{10.22323/1.454.0041}},
  \eprint{2401.13728}.

\bibtype{Article}%
\bibitem{Brito:2015oca}
\bibinfo{author}{Richard Brito}, \bibinfo{author}{Vitor Cardoso},
  \bibinfo{author}{Paolo Pani}, \bibinfo{title}{{Superradiance}: {New Frontiers
  in Black Hole Physics}}, \bibinfo{journal}{Lect. Notes Phys.}
  \bibinfo{volume}{906} (\bibinfo{year}{2015}) \bibinfo{pages}{pp.1--237},
  \bibinfo{doi}{\doi{10.1007/978-3-319-19000-6}}, \eprint{1501.06570}.

\bibtype{Article}%
\bibitem{Antel:2023hkf}
\bibinfo{author}{C. Antel}, et al., \bibinfo{title}{{Feebly-interacting
  particles: FIPs 2022 Workshop Report}}, \bibinfo{journal}{Eur. Phys. J. C}
  \bibinfo{volume}{83} (\bibinfo{number}{12}) (\bibinfo{year}{2023})
  \bibinfo{pages}{1122}, \bibinfo{doi}{\doi{10.1140/epjc/s10052-023-12168-5}},
  \eprint{2305.01715}.

\bibtype{Article}%
\bibitem{Beacham:2019nyx}
\bibinfo{author}{J. Beacham}, et al., \bibinfo{title}{{Physics Beyond Colliders
  at CERN: Beyond the Standard Model Working Group Report}},
  \bibinfo{journal}{J. Phys. G} \bibinfo{volume}{47} (\bibinfo{number}{1})
  (\bibinfo{year}{2020}) \bibinfo{pages}{010501},
  \bibinfo{doi}{\doi{10.1088/1361-6471/ab4cd2}}, \eprint{1901.09966}.

\bibtype{Article}%
\bibitem{Jerhot:2022chi}
\bibinfo{author}{Jan Jerhot}, \bibinfo{author}{Babette D\"obrich},
  \bibinfo{author}{Fatih Ertas}, \bibinfo{author}{Felix Kahlhoefer},
  \bibinfo{author}{Tommaso Spadaro}, \bibinfo{title}{{ALPINIST: Axion-Like
  Particles In Numerous Interactions Simulated and Tabulated}},
  \bibinfo{journal}{JHEP} \bibinfo{volume}{07} (\bibinfo{year}{2022})
  \bibinfo{pages}{094}, \bibinfo{doi}{\doi{10.1007/JHEP07(2022)094}},
  \eprint{2201.05170}.

\bibtype{Article}%
\bibitem{Wilczek:1982rv}
\bibinfo{author}{Frank Wilczek}, \bibinfo{title}{{Axions and Family Symmetry
  Breaking}}, \bibinfo{journal}{Phys. Rev. Lett.} \bibinfo{volume}{49}
  (\bibinfo{year}{1982}) \bibinfo{pages}{1549--1552},
  \bibinfo{doi}{\doi{10.1103/PhysRevLett.49.1549}}.

\bibtype{Article}%
\bibitem{Davidson:1981zd}
\bibinfo{author}{Aharon Davidson}, \bibinfo{author}{Kameshwar~C. Wali},
  \bibinfo{title}{{Minimal flavor unification via multigenerational
  Peccei-Quinn Symmetry}}, \bibinfo{journal}{Phys. Rev. Lett.}
  \bibinfo{volume}{48} (\bibinfo{year}{1982}) \bibinfo{pages}{11},
  \bibinfo{doi}{\doi{10.1103/PhysRevLett.48.11}}.

\bibtype{Article}%
\bibitem{NA62:2020xlg}
\bibinfo{author}{Eduardo Cortina~Gil}, et al. (\bibinfo{collaboration}{NA62}),
  \bibinfo{title}{{Search for a feebly interacting particle $X$ in the decay
  $K^{+}\rightarrow\pi^{+}X$}}, \bibinfo{journal}{JHEP} \bibinfo{volume}{03}
  (\bibinfo{year}{2021}) \bibinfo{pages}{058},
  \bibinfo{doi}{\doi{10.1007/JHEP03(2021)058}}, \eprint{2011.11329}.

\bibtype{Article}%
\bibitem{MartinCamalich:2020dfe}
\bibinfo{author}{Jorge Martin~Camalich}, \bibinfo{author}{Maxim Pospelov},
  \bibinfo{author}{Pham Ngoc~Hoa Vuong}, \bibinfo{author}{Robert Ziegler},
  \bibinfo{author}{Jure Zupan}, \bibinfo{title}{{Quark Flavor Phenomenology of
  the QCD Axion}}, \bibinfo{journal}{Phys. Rev. D} \bibinfo{volume}{102}
  (\bibinfo{number}{1}) (\bibinfo{year}{2020}) \bibinfo{pages}{015023},
  \bibinfo{doi}{\doi{10.1103/PhysRevD.102.015023}}, \eprint{2002.04623}.

\bibtype{Misc}%
\bibitem{AxionLimits}
\bibinfo{author}{Ciaran O'Hare}, \bibinfo{title}{cajohare/AxionLimits:
  AxionLimits},
  \bibinfo{howpublished}{\url{https://cajohare.github.io/AxionLimits/}}
  \bibinfo{year}{2020}, \bibinfo{doi}{\doi{10.5281/zenodo.3932430}}.

\bibtype{Article}%
\bibitem{Melcon:2018dba}
\bibinfo{author}{Alejandro~\'Alvarez Melc\'on}, et al., \bibinfo{title}{{Axion
  Searches with Microwave Filters: the RADES project}}, \bibinfo{journal}{JCAP}
  \bibinfo{volume}{1805} (\bibinfo{number}{05}) (\bibinfo{year}{2018})
  \bibinfo{pages}{040}, \bibinfo{doi}{\doi{10.1088/1475-7516/2018/05/040}},
  \eprint{1803.01243}.

\bibtype{Article}%
\bibitem{Jeong:2017hqs}
\bibinfo{author}{Junu Jeong}, \bibinfo{author}{SungWoo Youn},
  \bibinfo{author}{Saebyeok Ahn}, \bibinfo{author}{Jihn~E. Kim},
  \bibinfo{author}{Yannis~K. Semertzidis}, \bibinfo{title}{{Concept of
  multiple-cell cavity for axion dark matter search}}, \bibinfo{journal}{Phys.
  Lett. B} \bibinfo{volume}{777} (\bibinfo{year}{2018})
  \bibinfo{pages}{412--419},
  \bibinfo{doi}{\doi{10.1016/j.physletb.2017.12.066}}, \eprint{1710.06969}.

\bibtype{Article}%
\bibitem{Quiskamp:2024oet}
\bibinfo{author}{Aaron~P. Quiskamp}, \bibinfo{author}{Graeme~R. Flower},
  \bibinfo{author}{Steven Samuels}, \bibinfo{author}{Ben~T. McAllister},
  \bibinfo{author}{Paul Altin}, \bibinfo{author}{Eugene~N. Ivanov},
  \bibinfo{author}{Maxim Goryachev}, \bibinfo{author}{Michael~E. Tobar},
  \bibinfo{title}{{Near-quantum-limited axion dark matter search with the ORGAN
  experiment around 26{\,}{\,}{\ensuremath{\mu}}eV}}, \bibinfo{journal}{Phys.
  Rev. D} \bibinfo{volume}{111} (\bibinfo{number}{9}) (\bibinfo{year}{2025})
  \bibinfo{pages}{095007}, \bibinfo{doi}{\doi{10.1103/PhysRevD.111.095007}},
  \eprint{2407.18586}.

\bibtype{Article}%
\bibitem{Horns:2012jf}
\bibinfo{author}{Dieter Horns}, \bibinfo{author}{Joerg Jaeckel},
  \bibinfo{author}{Axel Lindner}, \bibinfo{author}{Andrei Lobanov},
  \bibinfo{author}{Javier Redondo}, et al., \bibinfo{title}{{Searching for
  WISPy Cold Dark Matter with a Dish Antenna}}, \bibinfo{journal}{JCAP}
  \bibinfo{volume}{1304} (\bibinfo{year}{2013}) \bibinfo{pages}{016},
  \bibinfo{doi}{\doi{10.1088/1475-7516/2013/04/016}}, \eprint{1212.2970}.

\bibtype{Misc}%
\bibitem{BRASSweb}
\bibinfo{howpublished}{http://www.iexp.uni-hamburg.de/groups/astroparticle/brass/brassweb.htm}.

\bibtype{Misc}%
\bibitem{PC_brun}
\bibinfo{howpublished}{{P. Brun, private communication}}.

\bibtype{Article}%
\bibitem{Hoshino:2025fiz}
\bibinfo{author}{Gabe Hoshino}, et al., \bibinfo{title}{{First Axion-Like
  Particle Results from a Broadband Search for Wave-Like Dark Matter in the 44
  to 52 $\mu$eV Range with a Coaxial Dish Antenna}}, \bibinfo{journal}{to
  appear}  (\bibinfo{year}{2025}), \eprint{2501.17119}.

\bibtype{Article}%
\bibitem{TheMADMAXWorkingGroup:2016hpc}
\bibinfo{author}{Allen Caldwell}, \bibinfo{author}{Gia Dvali},
  \bibinfo{author}{B{\'e}la Majorovits}, \bibinfo{author}{Alexander Millar},
  \bibinfo{author}{Georg Raffelt}, \bibinfo{author}{Javier Redondo},
  \bibinfo{author}{Olaf Reimann}, \bibinfo{author}{Frank Simon},
  \bibinfo{author}{Frank Steffen} (\bibinfo{collaboration}{MADMAX Working
  Group}), \bibinfo{title}{{Dielectric Haloscopes: A New Way to Detect Axion
  Dark Matter}}, \bibinfo{journal}{Phys. Rev. Lett.} \bibinfo{volume}{118}
  (\bibinfo{number}{9}) (\bibinfo{year}{2017}) \bibinfo{pages}{091801},
  \bibinfo{doi}{\doi{10.1103/PhysRevLett.118.091801}}, \eprint{1611.05865}.

\bibtype{Article}%
\bibitem{Garcia:2024xzc}
\bibinfo{author}{B.~Ary dos~Santos Garcia}, et al., \bibinfo{title}{{First
  search for axion dark matter with a Madmax prototype}}, \bibinfo{journal}{to
  appear}  (\bibinfo{year}{2024}), \eprint{2409.11777}.

\bibtype{Article}%
\bibitem{Baryakhtar:2018doz}
\bibinfo{author}{Masha Baryakhtar}, \bibinfo{author}{Junwu Huang},
  \bibinfo{author}{Robert Lasenby}, \bibinfo{title}{{Axion and hidden photon
  dark matter detection with multilayer optical haloscopes}},
  \bibinfo{journal}{Phys. Rev. D} \bibinfo{volume}{98} (\bibinfo{number}{3})
  (\bibinfo{year}{2018}) \bibinfo{pages}{035006},
  \bibinfo{doi}{\doi{10.1103/PhysRevD.98.035006}}, \eprint{1803.11455}.

\bibtype{Article}%
\bibitem{Chiles:2021gxk}
\bibinfo{author}{Jeff Chiles}, et al., \bibinfo{title}{{New Constraints on Dark
  Photon Dark Matter with Superconducting Nanowire Detectors in an Optical
  Haloscope}}, \bibinfo{journal}{Phys. Rev. Lett.} \bibinfo{volume}{128}
  (\bibinfo{number}{23}) (\bibinfo{year}{2022}) \bibinfo{pages}{231802},
  \bibinfo{doi}{\doi{10.1103/PhysRevLett.128.231802}}, \eprint{2110.01582}.

\bibtype{Article}%
\bibitem{Marsh:2018dlj}
\bibinfo{author}{David J.~E. Marsh}, \bibinfo{author}{Kin-Chung Fong},
  \bibinfo{author}{Erik~W. Lentz}, \bibinfo{author}{Libo\v{r} Smejkal},
  \bibinfo{author}{Mazhar~N. Ali}, \bibinfo{title}{{Proposal to Detect Dark
  Matter using Axionic Topological Antiferromagnets}}, \bibinfo{journal}{Phys.
  Rev. Lett.} \bibinfo{volume}{123} (\bibinfo{number}{12})
  (\bibinfo{year}{2019}) \bibinfo{pages}{121601},
  \bibinfo{doi}{\doi{10.1103/PhysRevLett.123.121601}}, \eprint{1807.08810}.

\bibtype{Article}%
\bibitem{Schutte-Engel:2021bqm}
\bibinfo{author}{Jan Sch{\"u}tte-Engel}, \bibinfo{author}{David J.~E. Marsh},
  \bibinfo{author}{Alexander~J. Millar}, \bibinfo{author}{Akihiko Sekine},
  \bibinfo{author}{Francesca Chadha-Day}, \bibinfo{author}{Sebastian Hoof},
  \bibinfo{author}{Mazhar~N. Ali}, \bibinfo{author}{Kin-Chung Fong},
  \bibinfo{author}{Edward Hardy}, \bibinfo{author}{Libor {\v{S}}mejkal},
  \bibinfo{title}{{Axion quasiparticles for axion dark matter detection}},
  \bibinfo{journal}{JCAP} \bibinfo{volume}{08} (\bibinfo{year}{2021})
  \bibinfo{pages}{066}, \bibinfo{doi}{\doi{10.1088/1475-7516/2021/08/066}},
  \eprint{2102.05366}.

\bibtype{Article}%
\bibitem{Lawson:2019brd}
\bibinfo{author}{Matthew Lawson}, \bibinfo{author}{Alexander~J. Millar},
  \bibinfo{author}{Matteo Pancaldi}, \bibinfo{author}{Edoardo Vitagliano},
  \bibinfo{author}{Frank Wilczek}, \bibinfo{title}{{Tunable axion plasma
  haloscopes}}, \bibinfo{journal}{Phys. Rev. Lett.} \bibinfo{volume}{123}
  (\bibinfo{number}{14}) (\bibinfo{year}{2019}) \bibinfo{pages}{141802},
  \bibinfo{doi}{\doi{10.1103/PhysRevLett.123.141802}}, \eprint{1904.11872}.

\bibtype{Article}%
\bibitem{ALPHA:2022rxj}
\bibinfo{author}{Alexander~J. Millar}, et al. (\bibinfo{collaboration}{ALPHA}),
  \bibinfo{title}{{Searching for dark matter with plasma haloscopes}},
  \bibinfo{journal}{Phys. Rev. D} \bibinfo{volume}{107} (\bibinfo{number}{5})
  (\bibinfo{year}{2023}) \bibinfo{pages}{055013},
  \bibinfo{doi}{\doi{10.1103/PhysRevD.107.055013}}, \eprint{2210.00017}.

\bibtype{Article}%
\bibitem{Ahyoune:2023gfw}
\bibinfo{author}{Saiyd Ahyoune}, et al., \bibinfo{title}{{A Proposal for a
  Low-Frequency Axion Search in the 1\textendash{}2 $\mu$ eV Range and Below
  with the BabyIAXO Magnet}}, \bibinfo{journal}{Annalen Phys.}
  \bibinfo{volume}{535} (\bibinfo{number}{12}) (\bibinfo{year}{2023})
  \bibinfo{pages}{2300326}, \bibinfo{doi}{\doi{10.1002/andp.202300326}},
  \eprint{2306.17243}.

\bibtype{Article}%
\bibitem{Chakrabarty:2023rha}
\bibinfo{author}{S. Chakrabarty}, et al., \bibinfo{title}{{Low frequency,
  100\textendash{}600~MHz, searches with axion cavity haloscopes}},
  \bibinfo{journal}{Phys. Rev. D} \bibinfo{volume}{109} (\bibinfo{number}{4})
  (\bibinfo{year}{2024}) \bibinfo{pages}{042004},
  \bibinfo{doi}{\doi{10.1103/PhysRevD.109.042004}}, \eprint{2303.07116}.

\bibtype{Article}%
\bibitem{Pugnat:2024sxb}
\bibinfo{author}{Pierre Pugnat}, et al., \bibinfo{title}{{GrAHal-CAPP for axion
  dark matter search with unprecedented sensitivity in the 1\textendash{}3
  $\mu$eV mass range}}, \bibinfo{journal}{Front. in Phys.} \bibinfo{volume}{12}
  (\bibinfo{year}{2024}) \bibinfo{pages}{1358810},
  \bibinfo{doi}{\doi{10.3389/fphy.2024.1358810}}.

\bibtype{Inproceedings}%
\bibitem{Oshima:2021irp}
\bibinfo{author}{Yuka Oshima}, \bibinfo{author}{Hiroki Fujimoto},
  \bibinfo{author}{Masaki Ando}, \bibinfo{author}{Tomohiro Fujita},
  \bibinfo{author}{Yuta Michimura}, \bibinfo{author}{Koji Nagano},
  \bibinfo{author}{Ippei Obata}, \bibinfo{author}{Taihei Watanabe},
  \bibinfo{title}{{Dark matter Axion search with riNg Cavity Experiment DANCE:
  Current sensitivity}}, in: \bibinfo{booktitle}{{55th Rencontres de Moriond on
  Gravitation}} \bibinfo{year}{2021}, p. \bibinfo{pages}{x},
  \eprint{2105.06252}.

\bibtype{Article}%
\bibitem{Budker:2013hfa}
\bibinfo{author}{Dmitry Budker}, \bibinfo{author}{Peter~W. Graham},
  \bibinfo{author}{Micah Ledbetter}, \bibinfo{author}{Surjeet Rajendran},
  \bibinfo{author}{Alex Sushkov}, \bibinfo{title}{{Cosmic Axion Spin Precession
  Experiment (CASPEr)}}, \bibinfo{journal}{Phys.Rev.} \bibinfo{volume}{X4}
  (\bibinfo{year}{2014}) \bibinfo{pages}{021030},
  \bibinfo{doi}{\doi{10.1103/PhysRevX.4.021030}}, \eprint{1306.6089}.

\bibtype{Article}%
\bibitem{JacksonKimball:2017elr}
\bibinfo{author}{Derek~F. Jackson~Kimball}, et al., \bibinfo{title}{{Overview
  of the Cosmic Axion Spin Precession Experiment (CASPEr)}},
  \bibinfo{journal}{Springer Proc. Phys.} \bibinfo{volume}{245}
  (\bibinfo{year}{2020}) \bibinfo{pages}{105--121},
  \bibinfo{doi}{\doi{10.1007/978-3-030-43761-9_13}}, \eprint{1711.08999}.

\bibtype{Article}%
\bibitem{Barbieri:2016vwg}
\bibinfo{author}{R. Barbieri}, \bibinfo{author}{C. Braggio},
  \bibinfo{author}{G. Carugno}, \bibinfo{author}{C.~S. Gallo},
  \bibinfo{author}{A. Lombardi}, \bibinfo{author}{A. Ortolan},
  \bibinfo{author}{R. Pengo}, \bibinfo{author}{G. Ruoso},
  \bibinfo{author}{C.~C. Speake}, \bibinfo{title}{{Searching for galactic
  axions through magnetized media: the QUAX proposal}}, \bibinfo{journal}{Phys.
  Dark Univ.} \bibinfo{volume}{15} (\bibinfo{year}{2017})
  \bibinfo{pages}{135--141}, \bibinfo{doi}{\doi{10.1016/j.dark.2017.01.003}},
  \eprint{1606.02201}.

\bibtype{Article}%
\bibitem{Bloch:2021vnn}
\bibinfo{author}{Itay~M. Bloch}, \bibinfo{author}{Gil Ronen},
  \bibinfo{author}{Roy Shaham}, \bibinfo{author}{Ori Katz},
  \bibinfo{author}{Tomer Volansky}, \bibinfo{author}{Or Katz}
  (\bibinfo{collaboration}{NASDUCK}), \bibinfo{title}{{New constraints on
  axion-like dark matter using a Floquet quantum detector}},
  \bibinfo{journal}{Sci. Adv.} \bibinfo{volume}{8} (\bibinfo{number}{5})
  (\bibinfo{year}{2022}) \bibinfo{pages}{abl8919},
  \bibinfo{doi}{\doi{10.1126/sciadv.abl8919}}, \eprint{2105.04603}.

\bibtype{Article}%
\bibitem{Chang:2017ruk}
\bibinfo{author}{Seung~Pyo Chang}, \bibinfo{author}{Selcuk Haciomeroglu},
  \bibinfo{author}{On Kim}, \bibinfo{author}{Soohyung Lee},
  \bibinfo{author}{Seongtae Park}, \bibinfo{author}{Yannis~K. Semertzidis},
  \bibinfo{title}{{Axionlike dark matter search using the storage ring EDM
  method}}, \bibinfo{journal}{Phys. Rev. D} \bibinfo{volume}{99}
  (\bibinfo{number}{8}) (\bibinfo{year}{2019}) \bibinfo{pages}{083002},
  \bibinfo{doi}{\doi{10.1103/PhysRevD.99.083002}}, \eprint{1710.05271}.

\bibtype{Article}%
\bibitem{Sikivie:2014lha}
\bibinfo{author}{P. Sikivie}, \bibinfo{title}{{Axion Dark Matter Detection
  using Atomic Transitions}}, \bibinfo{journal}{Phys. Rev. Lett.}
  \bibinfo{volume}{113} (\bibinfo{number}{20}) (\bibinfo{year}{2014})
  \bibinfo{pages}{201301}, \bibinfo{doi}{\doi{10.1103/PhysRevLett.113.201301}},
  \bibinfo{note}{[Erratum: Phys.Rev.Lett. 125, 029901 (2020)]},
  \eprint{1409.2806}.

\bibtype{Article}%
\bibitem{1367-2630-17-11-113025}
\bibinfo{author}{L Santamaria}, \bibinfo{author}{C Braggio}, \bibinfo{author}{G
  Carugno}, \bibinfo{author}{V~Di Sarno}, \bibinfo{author}{P Maddaloni},
  \bibinfo{author}{G Ruoso}, \bibinfo{title}{Axion dark matter detection by
  laser spectroscopy of ultracold molecular oxygen: a proposal},
  \bibinfo{journal}{New Journal of Physics} \bibinfo{volume}{17}
  (\bibinfo{number}{11}) (\bibinfo{year}{2015}) \bibinfo{pages}{113025},
  \bibinfo{url}{\urlprefix\url{http://stacks.iop.org/1367-2630/17/i=11/a=113025}}.

\bibtype{Article}%
\bibitem{Braggio:2017oyt}
\bibinfo{author}{C. Braggio}, et al., \bibinfo{title}{{Axion dark matter
  detection by laser induced fluorescence in rare-earth doped materials}},
  \bibinfo{journal}{Sci. Rep.} \bibinfo{volume}{7} (\bibinfo{year}{2017})
  \bibinfo{pages}{15168}, \bibinfo{doi}{\doi{10.1038/s41598-017-15413-6}},
  \eprint{1707.06103}.

\bibtype{Article}%
\bibitem{Agrawal:2021dbo}
\bibinfo{author}{Prateek Agrawal}, et al., \bibinfo{title}{{Feebly-interacting
  particles: FIPs 2020 workshop report}}, \bibinfo{journal}{Eur. Phys. J. C}
  \bibinfo{volume}{81} (\bibinfo{number}{11}) (\bibinfo{year}{2021})
  \bibinfo{pages}{1015}, \bibinfo{doi}{\doi{10.1140/epjc/s10052-021-09703-7}},
  \eprint{2102.12143}.

\bibtype{Article}%
\bibitem{Sikivie:2013laa}
\bibinfo{author}{P. Sikivie}, \bibinfo{author}{N. Sullivan},
  \bibinfo{author}{D.B. Tanner}, \bibinfo{title}{{Proposal for Axion Dark
  Matter Detection Using an LC Circuit}}, \bibinfo{journal}{Phys. Rev. Lett.}
  \bibinfo{volume}{112} (\bibinfo{number}{13}) (\bibinfo{year}{2014})
  \bibinfo{pages}{131301}, \bibinfo{doi}{\doi{10.1103/PhysRevLett.112.131301}},
  \eprint{1310.8545}.

\bibtype{Article}%
\bibitem{Chaudhuri:2014dla}
\bibinfo{author}{Saptarshi Chaudhuri}, \bibinfo{author}{Peter~W. Graham},
  \bibinfo{author}{Kent Irwin}, \bibinfo{author}{Jeremy Mardon},
  \bibinfo{author}{Surjeet Rajendran}, \bibinfo{author}{Yue Zhao},
  \bibinfo{title}{{Radio for hidden-photon dark matter detection}},
  \bibinfo{journal}{Phys. Rev.} \bibinfo{volume}{D92} (\bibinfo{number}{7})
  (\bibinfo{year}{2015}) \bibinfo{pages}{075012},
  \bibinfo{doi}{\doi{10.1103/PhysRevD.92.075012}}, \eprint{1411.7382}.

\bibtype{Article}%
\bibitem{Kahn:2016aff}
\bibinfo{author}{Yonatan Kahn}, \bibinfo{author}{Benjamin~R. Safdi},
  \bibinfo{author}{Jesse Thaler}, \bibinfo{title}{{Broadband and Resonant
  Approaches to Axion Dark Matter Detection}}, \bibinfo{journal}{Phys. Rev.
  Lett.} \bibinfo{volume}{117} (\bibinfo{number}{14}) (\bibinfo{year}{2016})
  \bibinfo{pages}{141801}, \bibinfo{doi}{\doi{10.1103/PhysRevLett.117.141801}},
  \eprint{1602.01086}.

\bibtype{Article}%
\bibitem{Silva-Feaver:2016qhh}
\bibinfo{author}{Maximiliano Silva-Feaver}, et al., \bibinfo{title}{{Design
  Overview of DM Radio Pathfinder Experiment}}, \bibinfo{journal}{IEEE Trans.
  Appl. Supercond.} \bibinfo{volume}{27} (\bibinfo{number}{4})
  (\bibinfo{year}{2017}) \bibinfo{pages}{1400204},
  \bibinfo{doi}{\doi{10.1109/TASC.2016.2631425}}, \eprint{1610.09344}.

\bibtype{Article}%
\bibitem{Ouellet:2018beu}
\bibinfo{author}{Jonathan~L. Ouellet}, et al., \bibinfo{title}{{First Results
  from ABRACADABRA-10 cm: A Search for Sub-$\mu$eV Axion Dark Matter}},
  \bibinfo{journal}{Phys. Rev. Lett.} \bibinfo{volume}{122}
  (\bibinfo{number}{12}) (\bibinfo{year}{2019}) \bibinfo{pages}{121802},
  \bibinfo{doi}{\doi{10.1103/PhysRevLett.122.121802}}, \eprint{1810.12257}.

\bibtype{Article}%
\bibitem{Salemi:2021gck}
\bibinfo{author}{Chiara~P. Salemi}, et al., \bibinfo{title}{{Search for
  Low-Mass Axion Dark Matter with ABRACADABRA-10~cm}}, \bibinfo{journal}{Phys.
  Rev. Lett.} \bibinfo{volume}{127} (\bibinfo{number}{8})
  (\bibinfo{year}{2021}) \bibinfo{pages}{081801},
  \bibinfo{doi}{\doi{10.1103/PhysRevLett.127.081801}}, \eprint{2102.06722}.

\bibtype{Article}%
\bibitem{Gramolin:2020ict}
\bibinfo{author}{Alexander~V. Gramolin}, \bibinfo{author}{Deniz Aybas},
  \bibinfo{author}{Dorian Johnson}, \bibinfo{author}{Janos Adam},
  \bibinfo{author}{Alexander~O. Sushkov}, \bibinfo{title}{{Search for
  axion-like dark matter with ferromagnets}}, \bibinfo{journal}{Nature Phys.}
  \bibinfo{volume}{17} (\bibinfo{number}{1}) (\bibinfo{year}{2021})
  \bibinfo{pages}{79--84}, \bibinfo{doi}{\doi{10.1038/s41567-020-1006-6}},
  \eprint{2003.03348}.

\bibtype{Article}%
\bibitem{McAllister:2018ndu}
\bibinfo{author}{Ben~T. McAllister}, \bibinfo{author}{Maxim Goryachev},
  \bibinfo{author}{Jeremy Bourhill}, \bibinfo{author}{Eugene~N. Ivanov},
  \bibinfo{author}{Michael~E. Tobar}, \bibinfo{title}{{Broadband Axion Dark
  Matter Haloscopes via Electric Sensing}}, \bibinfo{journal}{to appear}
  (\bibinfo{year}{2018}), \eprint{1803.07755}.

\bibtype{Article}%
\bibitem{Ouellet:2018nfr}
\bibinfo{author}{Jonathan Ouellet}, \bibinfo{author}{Zachary Bogorad},
  \bibinfo{title}{{Solutions to Axion Electrodynamics in Various Geometries}},
  \bibinfo{journal}{Phys. Rev. D} \bibinfo{volume}{99} (\bibinfo{number}{5})
  (\bibinfo{year}{2019}) \bibinfo{pages}{055010},
  \bibinfo{doi}{\doi{10.1103/PhysRevD.99.055010}}, \eprint{1809.10709}.

\bibtype{Article}%
\bibitem{Beutter:2018xfx}
\bibinfo{author}{Marc Beutter}, \bibinfo{author}{Andreas Pargner},
  \bibinfo{author}{Thomas Schwetz}, \bibinfo{author}{Elisa Todarello},
  \bibinfo{title}{{Axion-electrodynamics: a quantum field calculation}},
  \bibinfo{journal}{JCAP} \bibinfo{volume}{02} (\bibinfo{year}{2019})
  \bibinfo{pages}{026}, \bibinfo{doi}{\doi{10.1088/1475-7516/2019/02/026}},
  \eprint{1812.05487}.

\bibtype{Article}%
\bibitem{Crisosto:2019fcj}
\bibinfo{author}{N. Crisosto}, \bibinfo{author}{P. Sikivie},
  \bibinfo{author}{N.S. Sullivan}, \bibinfo{author}{D.B. Tanner},
  \bibinfo{author}{J. Yang}, \bibinfo{author}{G. Rybka}, \bibinfo{title}{{ADMX
  SLIC: Results from a Superconducting $LC$ Circuit Investigating Cold
  Axions}}, \bibinfo{journal}{Phys. Rev. Lett.} \bibinfo{volume}{124}
  (\bibinfo{number}{24}) (\bibinfo{year}{2020}) \bibinfo{pages}{241101},
  \bibinfo{doi}{\doi{10.1103/PhysRevLett.124.241101}}, \eprint{1911.05772}.

\bibtype{Article}%
\bibitem{Devlin:2021fpq}
\bibinfo{author}{Jack~A. Devlin}, et al., \bibinfo{title}{{Constraints on the
  Coupling between Axionlike Dark Matter and Photons Using an Antiproton
  Superconducting Tuned Detection Circuit in a Cryogenic Penning Trap}},
  \bibinfo{journal}{Phys. Rev. Lett.} \bibinfo{volume}{126}
  (\bibinfo{number}{4}) (\bibinfo{year}{2021}) \bibinfo{pages}{041301},
  \bibinfo{doi}{\doi{10.1103/PhysRevLett.126.041301}}, \eprint{2101.11290}.

\bibtype{Article}%
\bibitem{PhysRevD.26.1817}
\bibinfo{author}{Carlton~M. Caves}, \bibinfo{title}{Quantum limits on noise in
  linear amplifiers}, \bibinfo{journal}{Phys. Rev. D} \bibinfo{volume}{26}
  (\bibinfo{year}{1982}) \bibinfo{pages}{1817--1839},
  \bibinfo{doi}{\doi{10.1103/PhysRevD.26.1817}},
  \bibinfo{url}{\urlprefix\url{https://link.aps.org/doi/10.1103/PhysRevD.26.1817}}.

\bibtype{Article}%
\bibitem{PhysRev.128.2407}
\bibinfo{author}{H.~A. Haus}, \bibinfo{author}{J.~A. Mullen},
  \bibinfo{title}{Quantum Noise in Linear Amplifiers}, \bibinfo{journal}{Phys.
  Rev.} \bibinfo{volume}{128} (\bibinfo{year}{1962})
  \bibinfo{pages}{2407--2413}, \bibinfo{doi}{\doi{10.1103/PhysRev.128.2407}},
  \bibinfo{url}{\urlprefix\url{https://link.aps.org/doi/10.1103/PhysRev.128.2407}}.

\bibtype{Article}%
\bibitem{Lamoreaux:2013koa}
\bibinfo{author}{S.K. Lamoreaux}, \bibinfo{author}{K.A. van Bibber},
  \bibinfo{author}{K.W. Lehnert}, \bibinfo{author}{G. Carosi},
  \bibinfo{title}{{Analysis of single-photon and linear amplifier detectors for
  microwave cavity dark matter axion searches}}, \bibinfo{journal}{Phys.Rev.}
  \bibinfo{volume}{D88} (\bibinfo{number}{3}) (\bibinfo{year}{2013})
  \bibinfo{pages}{035020}, \bibinfo{doi}{\doi{10.1103/PhysRevD.88.035020}},
  \eprint{1306.3591}.

\bibtype{Article}%
\bibitem{Malnou:2018dxn}
\bibinfo{author}{M. Malnou}, \bibinfo{author}{D.A. Palken},
  \bibinfo{author}{B.M. Brubaker}, \bibinfo{author}{Leila~R. Vale},
  \bibinfo{author}{Gene~C. Hilton}, \bibinfo{author}{K.W. Lehnert},
  \bibinfo{title}{{Squeezed vacuum used to accelerate the search for a weak
  classical signal}}, \bibinfo{journal}{Phys. Rev. X} \bibinfo{volume}{9}
  (\bibinfo{number}{2}) (\bibinfo{year}{2019}) \bibinfo{pages}{021023},
  \bibinfo{doi}{\doi{10.1103/PhysRevX.9.021023}}, \bibinfo{note}{[Erratum:
  Phys.Rev.X 10, 039902 (2020)]}, \eprint{1809.06470}.

\bibtype{Article}%
\bibitem{HAYSTAC:2024squeezed}
\bibinfo{author}{Y.~Zhu et~al. (HAYSTAC~Collaboration)}, \bibinfo{title}{Dark
  Matter Axion Search with HAYSTAC Phase II}, \bibinfo{journal}{arXiv preprint}
   (\bibinfo{year}{2024}), \eprint{2409.08998},
  \bibinfo{url}{\urlprefix\url{https://arxiv.org/abs/2409.08998}}.

\bibtype{Inproceedings}%
\bibitem{Yamamoto:2000si}
\bibinfo{author}{K. Yamamoto}, et al., \bibinfo{title}{{The Rydberg atom cavity
  axion search}}, in: \bibinfo{booktitle}{{3rd International Heidelberg
  Conference on Dark Matter in Astro and Particle Physics}}
  \bibinfo{year}{2000}, pp. \bibinfo{pages}{638--645}, \eprint{hep-ph/0101200}.

\bibtype{Phdthesis}%
\bibitem{pepe2024TES}
\bibinfo{author}{Carlo Pepe}, \bibinfo{title}{Development of Superconducting
  Single-Particle Detector Transition-Edge Sensor}, \bibinfo{type}{Ph.d.
  thesis}, \bibinfo{school}{Politecnico di Torino}, \bibinfo{address}{Turin,
  Italy} \bibinfo{year}{2024}, \bibinfo{note}{doctoral Program in Metrology,
  Supervisor: Dr. M. Rajteri},
  \bibinfo{url}{\urlprefix\url{https://tesidottorato.depositolegale.it/bitstream/20.500.14242/170753/1/Phd_thesis_c_pepe_development_of_superconducting_single_particle_detector_transition_edge_sensor.pdf}}.

\bibtype{Article}%
\bibitem{Paolucci:2021kle}
\bibinfo{author}{Federico Paolucci}, \bibinfo{author}{Francesco Giazotto},
  \bibinfo{title}{{GHz Superconducting Single-Photon Detectors for Dark Matter
  Search}}, \bibinfo{journal}{Instruments} \bibinfo{volume}{5}
  (\bibinfo{number}{2}) (\bibinfo{year}{2021}) \bibinfo{pages}{14},
  \bibinfo{doi}{\doi{10.3390/instruments5020014}}, \eprint{2101.08558}.

\bibtype{Phdthesis}%
\bibitem{smith2024scaling}
\bibinfo{author}{Jennifer~Pearl Smith}, \bibinfo{title}{Scaling
  Energy-Resolving Microwave Kinetic Inductance Detector Readout},
  \bibinfo{comment}{Ph.D. thesis}, \bibinfo{school}{University of California,
  Santa Barbara} \bibinfo{year}{2024},
  \bibinfo{url}{\urlprefix\url{https://escholarship.org/uc/item/5m22t48n}}.

\bibtype{Article}%
\bibitem{huang2024graphene}
\bibinfo{author}{B. Huang}, \bibinfo{author}{E.G. Arnault}, \bibinfo{author}{W.
  Jung}, \bibinfo{author}{C. Fried}, et al., \bibinfo{title}{Graphene
  calorimetric single-photon detector}, \bibinfo{journal}{arXiv preprint
  arXiv:2410.22433}  (\bibinfo{year}{2024}),
  \bibinfo{url}{\urlprefix\url{https://arxiv.org/abs/2410.22433}}.

\bibtype{Article}%
\bibitem{natarajan2012superconducting}
\bibinfo{author}{Chandrasekara~M. Natarajan}, \bibinfo{author}{Michael~G.
  Tanner}, \bibinfo{author}{Robert~H. Hadfield},
  \bibinfo{title}{Superconducting nanowire single-photon detectors: physics and
  applications}, \bibinfo{journal}{Superconductor Science and Technology}
  \bibinfo{volume}{25} (\bibinfo{number}{6}) (\bibinfo{year}{2012})
  \bibinfo{pages}{063001}, \bibinfo{doi}{\doi{10.1088/0953-2048/25/6/063001}},
  \bibinfo{url}{\urlprefix\url{https://doi.org/10.1088/0953-2048/25/6/063001}}.

\bibtype{Article}%
\bibitem{Flurin:2021smpd}
\bibinfo{author}{E. Flurin}, \bibinfo{author}{R. Lescanne}, \bibinfo{author}{M.
  Villiers}, \bibinfo{author}{P. Bertet}, \bibinfo{author}{D. Vion},
  \bibinfo{author}{D. Esteve}, \bibinfo{title}{Detecting spins by their
  fluorescence with a microwave photon counter}, \bibinfo{journal}{Nature}
  \bibinfo{volume}{599} (\bibinfo{year}{2021}) \bibinfo{pages}{62--67},
  \bibinfo{doi}{\doi{10.1038/s41586-021-04076-z}}.

\bibtype{Article}%
\bibitem{Dixit:2020ymh}
\bibinfo{author}{Akash~V. Dixit}, \bibinfo{author}{Srivatsan Chakram},
  \bibinfo{author}{Kevin He}, \bibinfo{author}{Ankur Agrawal},
  \bibinfo{author}{Ravi~K. Naik}, \bibinfo{author}{David~I. Schuster},
  \bibinfo{author}{Aaron Chou}, \bibinfo{title}{{Searching for Dark Matter with
  a Superconducting Qubit}}, \bibinfo{journal}{Phys. Rev. Lett.}
  \bibinfo{volume}{126} (\bibinfo{number}{14}) (\bibinfo{year}{2021})
  \bibinfo{pages}{141302}, \bibinfo{doi}{\doi{10.1103/PhysRevLett.126.141302}},
  \eprint{2008.12231}.

\bibtype{Article}%
\bibitem{Braggio:2024xed}
\bibinfo{author}{C. Braggio}, et al., \bibinfo{title}{{Quantum-Enhanced Sensing
  of Axion Dark Matter with a Transmon-Based Single Microwave Photon Counter}},
  \bibinfo{journal}{Phys. Rev. X} \bibinfo{volume}{15} (\bibinfo{number}{2})
  (\bibinfo{year}{2025}) \bibinfo{pages}{021031},
  \bibinfo{doi}{\doi{10.1103/PhysRevX.15.021031}}, \eprint{2403.02321}.

\bibtype{Article}%
\bibitem{Ahn:2021fgb}
\bibinfo{author}{Danho Ahn}, \bibinfo{author}{Ohjoon Kwon},
  \bibinfo{author}{Woohyun Chung}, \bibinfo{author}{Wonjun Jang},
  \bibinfo{author}{Doyu Lee}, \bibinfo{author}{Jhinhwan Lee},
  \bibinfo{author}{Sung Woo~Youn}, \bibinfo{author}{Heesu Byun},
  \bibinfo{author}{Dojun Youm}, \bibinfo{author}{Yannis~K. Semertzidis},
  \bibinfo{title}{{Biaxially Textured YBa2Cu3O7\ensuremath{-}x Microwave Cavity
  in a High Magnetic Field for a Dark-Matter Axion Search}},
  \bibinfo{journal}{Phys. Rev. Applied} \bibinfo{volume}{17}
  (\bibinfo{number}{6}) (\bibinfo{year}{2022}) \bibinfo{pages}{L061005},
  \bibinfo{doi}{\doi{10.1103/PhysRevApplied.17.L061005}}, \eprint{2103.14515}.

\bibtype{Article}%
\bibitem{Golm:2021ooj}
\bibinfo{author}{J. Golm}, et al., \bibinfo{title}{{Thin Film (High
  Temperature) Superconducting Radiofrequency Cavities for the Search of Axion
  Dark Matter}}, \bibinfo{journal}{IEEE Trans. Appl. Supercond.}
  \bibinfo{volume}{32} (\bibinfo{number}{4}) (\bibinfo{year}{2022})
  \bibinfo{pages}{1500605}, \bibinfo{doi}{\doi{10.1109/TASC.2022.3147741}},
  \eprint{2110.01296}.

\bibtype{Article}%
\bibitem{Romanov:2020epk}
\bibinfo{author}{Artur Romanov}, \bibinfo{author}{Patrick Krkoti\'c},
  \bibinfo{author}{Guilherme Telles}, \bibinfo{author}{Joan O'Callaghan},
  \bibinfo{author}{Montse Pont}, \bibinfo{author}{Francis Perez},
  \bibinfo{author}{Xavier Granados}, \bibinfo{author}{Sergio Calatroni},
  \bibinfo{author}{Teresa Puig}, \bibinfo{author}{Joffre Gutierrez},
  \bibinfo{title}{{High frequency response of thick REBCO coated conductors in
  the framework of the FCC study}}, \bibinfo{journal}{Sci. Rep.}
  \bibinfo{volume}{10} (\bibinfo{number}{1}) (\bibinfo{year}{2020})
  \bibinfo{pages}{12325}, \bibinfo{doi}{\doi{10.1038/s41598-020-69004-z}}.

\bibtype{Article}%
\bibitem{Schmieden:2024wqp}
\bibinfo{author}{Kristof Schmieden}, \bibinfo{author}{Tim Schneemann},
  \bibinfo{author}{Matthias Schott}, \bibinfo{author}{Malavika Unni},
  \bibinfo{author}{Hendrik Bekker}, \bibinfo{author}{Arne Wickenbrock},
  \bibinfo{author}{Dmitry Budker}, \bibinfo{title}{{Study of NbN as
  superconducting material for the usage in superconducting radio frequency
  cavities}}, \bibinfo{journal}{to appear}  (\bibinfo{year}{2024}),
  \eprint{2412.14958}.

\end{thebibliography*}

\end{document}